\pdfoutput=1
 \documentclass[twocolumn,preprint2]{aastex63}
\usepackage[figuresright]{rotating}
\usepackage{tabularx}
\newcommand{\kms}{{\rm km~s}^{-1}}
\newcommand{\HII}{H\,{\small{II}} }
\newcommand{\HI}{H\,{\small{I}} }

\newcommand{\mjyb}{{\rm mJy~beam}^{-1}}

\newcommand{\mjybkms}{{\rm mJy~beam}^{-1}{\rm km~s}^{-1}}

\newcommand{\Lfir}{L_{\rm FIR}}
\newcommand{\Lsun}{L_{\odot}}

\shorttitle{FASHI OHMs}
\shortauthors{Zhang et al.}

\begin{document}

\title{FASHI: A search for extragalactic OH megamasers with FAST}

\correspondingauthor{Chuan-Peng Zhang}
\email{cpzhang@nao.cas.cn}

\author[0000-0002-4428-3183]{Chuan-Peng Zhang}
\affiliation{National Astronomical Observatories, Chinese Academy of Sciences, Beijing 100101, China}
\affiliation{Guizhou Radio Astronomical Observatory, Guizhou University, Guiyang 550000, China}

\author{Cheng Cheng}
\affiliation{Chinese Academy of Sciences South America Center for Astronomy, National Astronomical Observatories, CAS, Beijing 100101, China}

\author{Ming Zhu}
\author{Jin-Long Xu}
\author{Peng Jiang}
\affiliation{National Astronomical Observatories, Chinese Academy of Sciences, Beijing 100101, China}
\affiliation{Guizhou Radio Astronomical Observatory, Guizhou University, Guiyang 550000, China}




\begin{abstract}
The \textbf{F}AST \textbf{A}ll \textbf{S}ky \textbf{H\,{\footnotesize{I}}} survey (FASHI) is broader in frequency band and sky volume, and deeper in detection sensitivity than the Arecibo Legacy Fast ALFA survey (ALFALFA). To efficiently expand the sample of OH megamasers (OHMs), whose strongest line has a rest frequency of 1667.35903 MHz, we directly matched the IRAS Point Source Catalog Redshift (PSCz) catalog with the corresponding FASHI data cube. From 145 PSCz sources already covered by FASHI, we obtained 27 OHMs with a detection rate of 18.6\%, including 9 previously known and 18 new ones, within a redshift range of $0.14314\lesssim z_{\rm OH} \lesssim0.27656$. We also measured the hyperfine ratio of nine OHMs between the 1667 and 1665\,MHz lines. The ratio ranges from 1.32 to 15.22, with an average of $R_{1667:1665}=4.74$. In a fit to the $L_{\rm OH}$ vs. $L_{\rm FIR}$ relation, we have ${\rm log}L_{\rm OH}= (1.57\pm0.10){\rm log}L_{\rm FIR}-(15.80\pm1.19)$, which is almost the same as derived from previous observations. As expected, since the OHM sample was selected by cross-correlation with the IRAS-selected PSCz, our detected OHMs are [ultra]luminous infrared galaxies ([U]LIRGs). However, not all [U]LIRGs have detectable OH emission, suggesting that the OH emission may be triggered within a specific stage of the merger or can only be seen in specific orientations. In general, FAST, with its 19-beam array and UWB receiver, will be a powerful tool for observing more OHMs and unraveling their mystery in the future. 
\end{abstract}

\section{Introduction} \label{sec:intro}

OH megamasers (OHMs) at rest frequencies of 1667 and 1665\,MHz in extragalactic systems are valuable indicators of galaxy interactions. These megamasers are thought to originate from the starburst nuclei of merging gas-rich galaxy systems, based on both high-resolution interferometer observations \citep{Lo2005} and statistical analysis of luminosity functions \citep{Darling2002a,Darling2006,Giovanelli2015}. OHMs thus offer a distinctive approach to measuring the occurrence of gas-rich galaxy mergers at specific evolutionary stages \citep{Lo2005}. The masing emission or absorption lines are usually broad, with line widths up to $10^3\,\kms$, and exhibit an implied isotropic luminosity in the range of $10-10^4\,\Lsun$ \citep{Darling2002a}. OHMs are connected to high-density molecular gas ($n(\rm H_2) \sim 10^4\,cm^{-3}$) and strong far-infrared radiation. They can offer potential insights into extreme star formation and galactic evolution, particularly high-redshift OHMs \citep{Darling2007,Lockett2008,Roberts2021,Glowacki2022,Jarvis2023}. Therefore, investigating OHMs is a meaningful pursuit.

As of now, approximately 136 OHMs have been discovered, for instance, in studies of \citet{Darling2002a,Willett2012,Haynes2018,Suess2016,Hess2021,Roberts2021,Glowacki2022,Jarvis2023}. Searches with the Arecibo telescope represent the most successful tracking or single-point observation to date, detecting approximately 17\% of OHMs \citep{Darling2000}. All OHMs known to date are exclusively associated with [ultra]luminous infrared galaxies ([U]LIRGs), defined as having far-infrared luminosities $\gtrsim10^{11}\Lsun$ and $\Lfir\gtrsim10^{12}\Lsun$, respectively \citep{Baan1991}. However, only about 20\% of [U]LIRGs are detected with OHM emission, as shown by studies conducted by \citet{Baan1991,Baan1992,Darling2000,Darling2001,Darling2002a}. The production of OHMs may be a transient phenomenon in the evolution of a galaxy, from starburst nucleus to an AGN \citep{Wiggins2016}, or may be subject to beaming and orientation effects, as suggested in \citet{Darling2007}.

The Five-hundred-meter Aperture Spherical radio Telescope (FAST) All Sky \HI survey (FASHI) aims to cover the entire sky visible to FAST, between declinations of $-14^\circ$ and $+66^\circ$, within a frequency range of $1.0-1.5$\,GHz \citep{Nan2011,Jiang2019,Jiang2020}. The survey currently has a typical map rms of $\sim$0.76\,$\mjyb$ and a spectral detection sensitivity of $\sim$1.50\,mJy for a velocity resolution of 6.4\,$\kms$ within a redshift range of $0.11-0.66$ for OHMs \citep{Zhang2023fashi}, while the Arecibo Legacy Fast ALFA (ALFALFA) survey has a lower detection sensitivity of $\sim$1.86\,$\mjyb$ after smoothing to a velocity resolution of 10\,$\kms$ \citep{Haynes2018}. This suggests that the FASHI could detect more OHMs than ALFALFA. Furthermore, the Ultra-Wide Bandwidth (UWB) receiver on FAST can simultaneously cover a frequency range of 500-3300\,MHz (or $z<2.33$), making FAST a powerful instrument for observing OHMs in extragalactic objects \citep{Zhang2023}.

In this paper, we report 18 new and 9 old OHMs detected by FAST. Section\,\ref{sec:data_reduc} shows the FAST observations and source identification. Section\,\ref{sec:result} presents the characterization of the detected OHMs and lists four OHM-related catalogs. Section\,\ref{sec:discu} discusses the hyperfine ratio, host galaxy properties, $L_{\rm OH}$ and $L_{\rm FIR}$ relationship, and FAST OHM detectability. Section\,\ref{sec:summary} is a summary. Throughout this paper, we assume a $\Lambda$CDM cosmology with $H_{0}$ = 75\,$\kms$\,Mpc$^{-1}$, $\Omega_{\rm M} = 0.3$, and $\Omega_{\rm \Lambda} = 0.7$.

\section{Data and identification}
\label{sec:data_reduc}

 \begin{figure*}[htp]
 \centering
 \includegraphics[width=0.95\textwidth, angle=0]{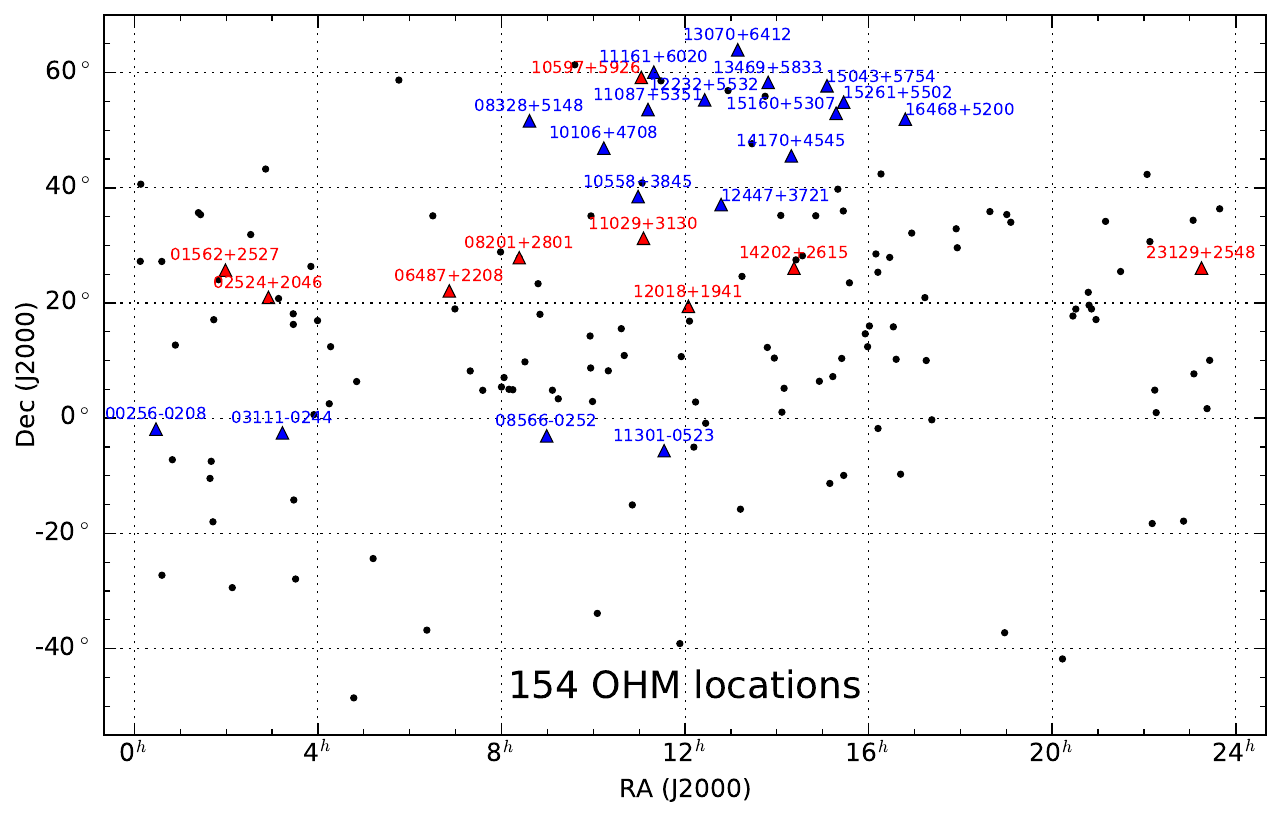}
 \caption{Locations for a total of 154 OHMs, including all previously known \citep{Darling2002a,Darling2006,Willett2012,Suess2016,Hess2021,Glowacki2022,Jarvis2023} and 18 newly discovered OHMs in this work. The triangles with blue IRAS names are the 18 OHMs newly discovered by FAST, while the red ones are 9 known OHMs recovered by FAST. The black dots are OHMs previously discovered by others.}
 \label{Fig:location}
 \end{figure*}

\subsection{Observational data}

Observations were obtained from the FASHI project \citep{Zhang2023fashi}, which uses the FAST 19-beam receiver to efficiently cover the FAST sky. The aperture of FAST is 500\,m and the effective aperture is about 300\,m, resulting in a beam size of $\sim$2.9$'$ at 1.4\,GHz. The full bandwidth of 500\,MHz in the spectral line backend has a frequency coverage of 1000 to 1500\,MHz with a total of 64k channels. The corresponding frequency resolution is 7.63\,kHz or 1.6\,$\kms$ at 1.4\,GHz. The FASHI data are reduced using the FAST spectral data reduction pipeline \texttt{HiFAST} \citep{Jing2024}. The \texttt{HiFAST} pipeline combines data reduction packages including antenna temperature correction, baseline correction, RFI mitigation \citep{Zhang2022rfi}, standing wave correction, gridding, flux correction \citep{Liu2024}, and data cube generation. The data have been smoothed to 6.4\,$\kms$ spectral resolution per channel and the data cubes are gridded to a pixel scale of 1\arcmin. A heliocentric Doppler correction was applied to the data. A detailed observational setup and data reduction procedure is presented in the FASHI paper by \citet{Zhang2023fashi}.

\subsection{Source identification}
\label{sec:identification}

Extragalactic \HI and OHM lines have similar spectral profiles at different redshifts. Spectroscopic redshift information is needed to distinguish them \citep{Hess2021}. The IRAS Point Source Catalog Redshift (PSCz) survey consists of spectroscopic redshifts, infrared and optical photometry, and miscellaneous information for 18351 IRAS sources, mostly selected from the Point Source Catalog \citep{Saunders2000}. The survey covers almost all galaxies with flux brighter than 0.595\,Jy at 60 microns, over 84\% of the sky. It is well known that most previously known OHMs have associated IRAS sources \citep[e.g.,][]{Darling2000,Darling2001,Darling2002a}. To efficiently expand the OHM sample, we directly cross-checked the PSCz catalog \citep{Saunders2000} with the corresponding FASHI data cube to see if there are any emission lines within a beam size of 2.9$'$ and a velocity range of $\sim$600\,$\kms$. The FASHI data covers around 10000 deg$^2$ within $0^h\lesssim{\rm RA}\lesssim17^h$, $22^h\lesssim{\rm RA}\lesssim24^h$, $6^\circ\lesssim{\rm Dec}\lesssim0^\circ$, and $18^\circ\lesssim{\rm Dec}\lesssim64^\circ$ with a typical spectral detection sensitivity of $\sim$1.50\,mJy for a velocity resolution of $\sim$6.4\,$\kms$ at 1.4\,GHz. Because the FASHI project can only operate in schedule-filler mode, parts of the above region are not fully covered. Within the region we found that there are a total of 145 PSCz sources covered by the current FASHI data with spectroscopic redshifts in the frequency range from 1304 to 1461\,MHz. Using the FAST spectral data and the moment 0, 1, 2 images, we check the 145 sources one by one, including the redshift, coordinate, flux density, and the $g$, $r$, and $z$ bands of the SDSS data, and then discard the probable \HI sources as judged by the spectroscopic redshift information. In the end, we discovered and conformed 27 OHMs, 18 of which were first found by FAST. Detailed information on these sources is presented in the following sections.

\section{Detection Results and Analysis}
\label{sec:result}

\subsection{OHM locations}

In Figure\,\ref{Fig:location} we show all the OHM locations, including 18 new OHMs found by FAST and 136 known ones found by others. From Figure\,\ref{Fig:location} we can see that the previously known OHMs are mostly located at $0^\circ\lesssim{\rm Dec}\lesssim30^\circ$, covered by the Arecibo observations. It is expected that the FAST observation will be able to fill the gap of ${\rm Dec}\gtrsim30^\circ$ and ${\rm Dec}\lesssim0^\circ$. Currently, the area covered by FAST has reached $\sim$10000 deg$^2$. We have discovered and identified 27 OHMs, which is 20\% of all known OHMs, and including 18 newly found OHMs, which is 12.4\% of all known OHMs. The 27 OHMs are selected and identified from the 145 PSCz sources, already covered by the current FAST observation. This means that the FAST OHM detection rate is 18.6\%, which is close to the 17\% detection rate of the Arecibo OHM survey \citep{Darling2000}. However, about 20 previously known OHMs with PSCz counterparts are not detected by FAST, although they lie within the volume surveyed by FASHI. The main reason for this is the currently limited detection sensitivity. Furthermore, it also means that about 47 (27+20) OHMs have been successfully detected in the 145 PSCz sources, resulting in a detection rate of 32.4\% for this PSCz sample.

\subsection{Individual sources}
\label{sec:individual}

 \begin{figure*}[htp]
 \centering
 \includegraphics[height=0.25\textwidth, angle=0]{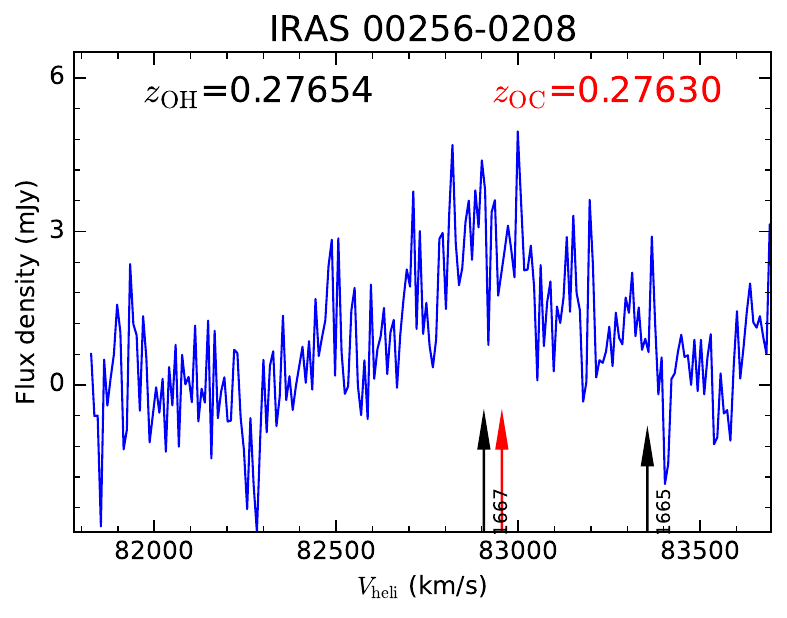}
 \includegraphics[height=0.31\textwidth, angle=0]{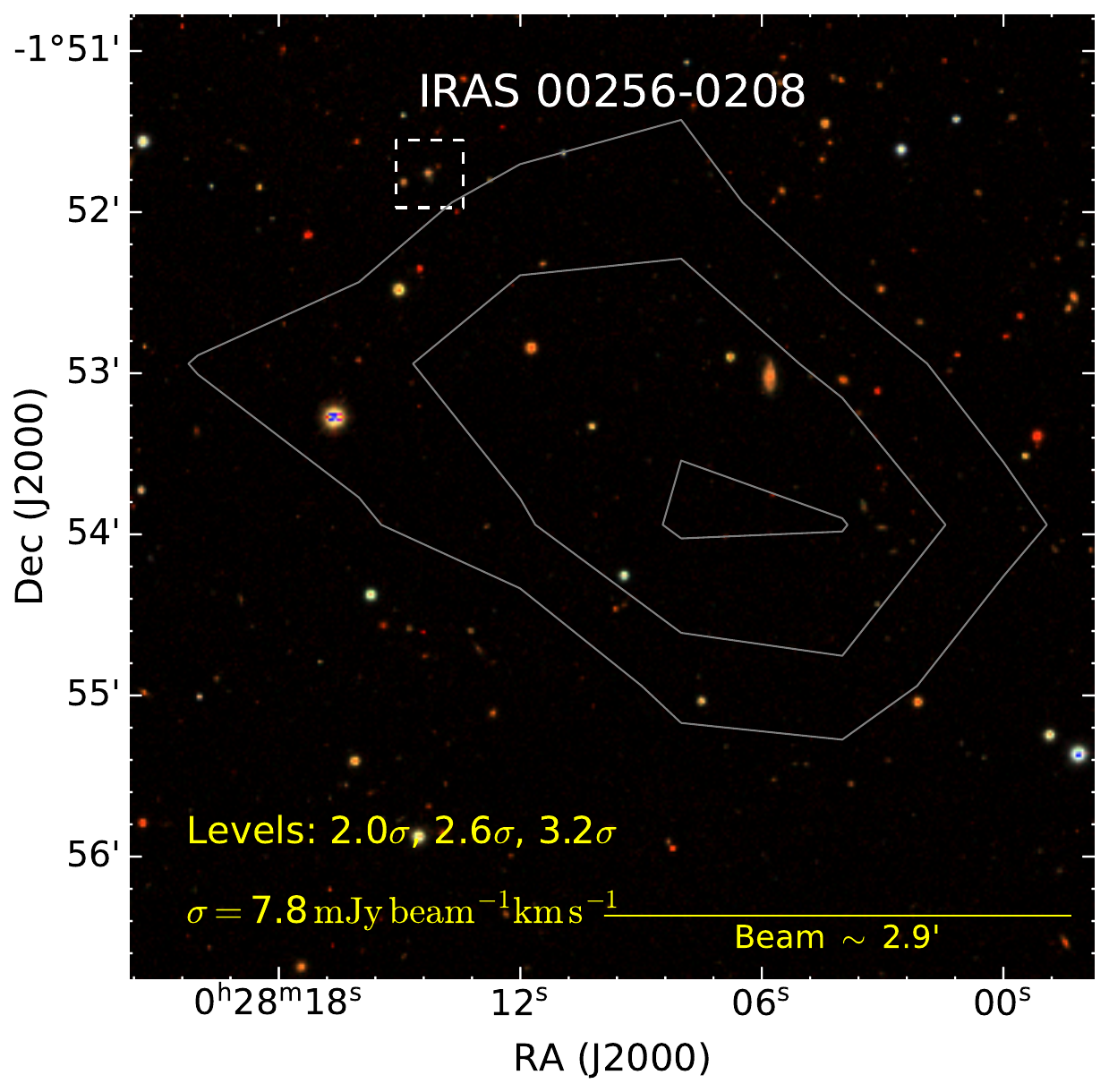}
 \includegraphics[height=0.29\textwidth, angle=0]{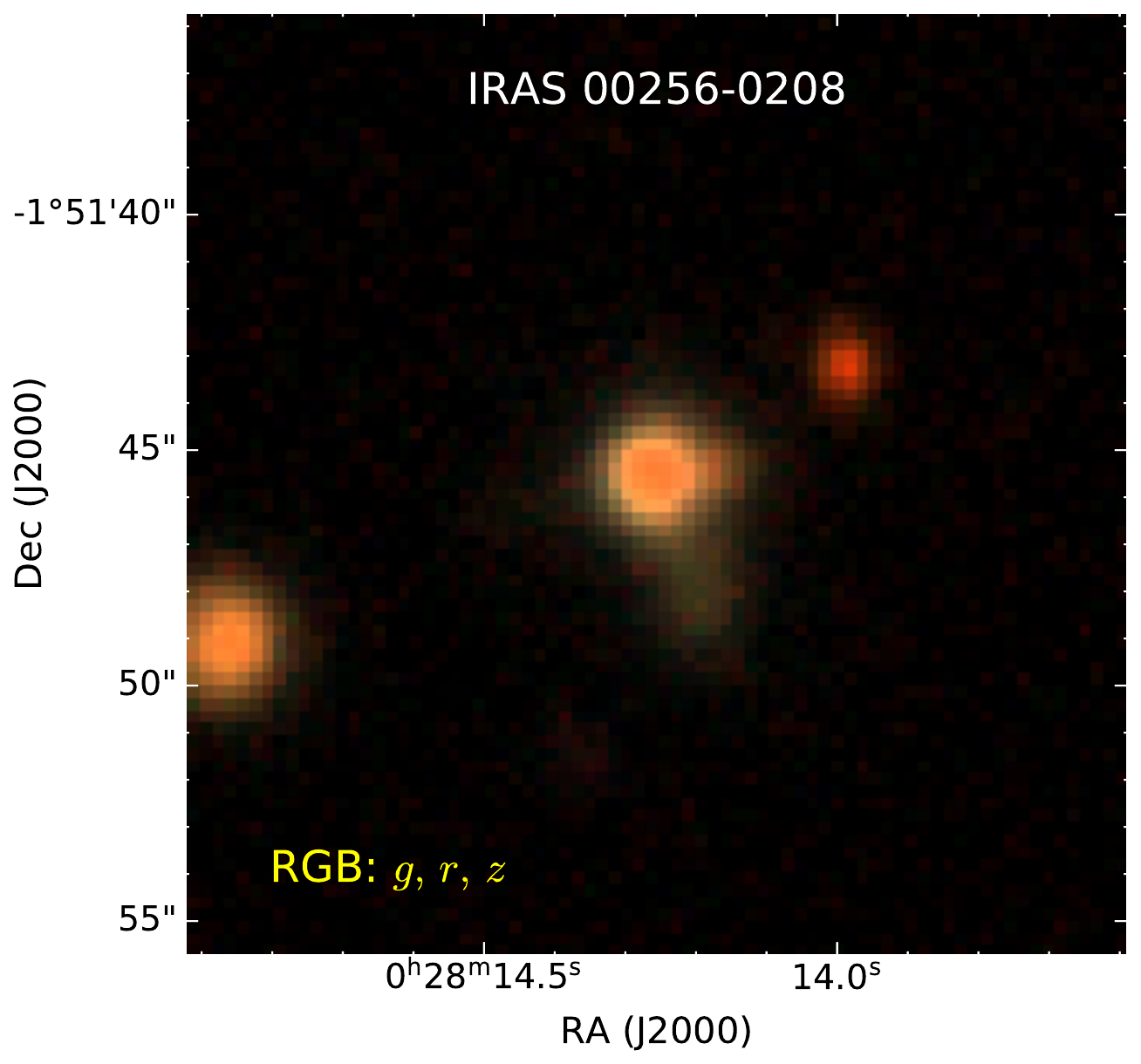}
 \caption{OHM IRAS\,00256-0208. \textit{Left}: FAST OH emission spectrum. The 1667\,MHz OH redshift measured by FAST (in black) and an optical spectroscopic redshift (in red) are presented above and below the spectrum with text and corresponding colored arrows, respectively. The expected location of 1665\,MHz OH line is indicated by a shorter arrow. \textit{Middle}: Gray contours indicate the OH integrated intensity distribution seen by FAST with a beam size of $\sim$2.9$'$. The contour levels are shown in the image. The background shows an RGB image of the $g$, $r$, and $z$ bands of the SDSS data. The white square indicates the OC of the OHM. \textit{Right}: The zoomed-in RGB image of the OC within the white square in the middle panel. The other 26 OHMs are shown in Figure\,\ref{Fig:IRAS-Continued}. }
 \label{Fig:IRAS00256-0208}
 \end{figure*} 

In Tables\,\ref{tab:oh}, \ref{tab:ratio}, and \ref{tab:oc} we list the parameters of 27 OHMs measured by FAST, including 18 newly discovered and 9 previously known sources. Of the latter, eight previously known sources were observed by the Arecibo OHM survey \citep{Darling2000,Darling2001,Darling2002a,Darling2006}, and IRAS\,10597+5926 was first found by \citet{Willett2012}. For all nine previously known OHMs, the positions and redshifts of the optical counterparts (OCs) are in good agreement with those of the FAST OH observations. For all 27 OHMs, the OH spectra, integrated intensity distributions, and RGB images are shown in Figures\,\ref{Fig:IRAS00256-0208} and \ref{Fig:IRAS-Continued}, where both the radio and optical redshifts are given in the spectral panels, and the position of each OC is given in the OH integrated intensity map. In nearly all 27 OHMs, optical RGB images show typical merging systems with elongated tidal tails surrounding the nucleus. For IRAS\,00256-0208, 10558+3845, 12018+1941, 12447+3721, 14170+4545, 14202+2615, 15043+5754, and 15160+5307, the detection sensitivities are low and they require more observation time for further confirmation. IRAS\,08328+5148 is one of four OHMs for which the centroid of the FAST detection is relatively far away on the sky from the position of the putative optical counterpart. (The others, to differing degrees, are IRAS\,03111-0244, IRAS 12447+3721, and IRAS 14202+2615.) These OHMs could be similar to IRAS\,08328+5148, in that the true counterpart for the OHM might be a galaxy that is associated with, but not identical to, the ``optical counterpart'' with the matching spectroscopic redshift. Other notes for some of the OHMs are as follows:

\textbf{IRAS\,00256-0208}
This OHM has the third highest redshift among all known OHMs in the main 1667\,MHz OH emission lines, the highest being J095903.22+025356.1 with a redshift of $z_{\rm OH} \sim 0.7092$, recently discovered by \citet{Jarvis2023}.

\textbf{IRAS\,01562+2527}\footnote{Listed as IRAS\,01562+2528 by \citet{Darling2002a}} 
This has been confirmed as an OHM by the Arecibo OHM survey \citep{Darling2002a}. This survey has a consistent detection result with FAST.

\textbf{IRAS\,02524+2046} 
This OHM was first confirmed by the Arecibo OHM survey \citep{Darling2002a}, which detected an extremely strong peak reaching 40\,mJy. However, the peak detected by FAST is $21.78\pm1.83$\,mJy. This is probably because the OH flux density is time variable. The emission lines are strong and narrow and show an extremely good correspondence between the 1667 and 1665\,MHz lines. This allows us to measure the hyperﬁne line ratio (see Table\,\ref{tab:ratio}).

\textbf{IRAS\,06487+2208} 
This OHM was first found and reported by the Arecibo OHM survey \citep{Darling2000}. The optical spectrum of this source has been described as a composite of \HII and LINER features based on the Osterbrock spectral line ratio classification method \citep{Osterbrock1989,Veilleux1995}. The optical RGB image is unresolved, but there may be another nucleus or tidal tail to the north of the main nucleus.

\textbf{IRAS\,08201+2801} 
This has been confirmed as an OHM by the Arecibo OHM survey \citep{Darling2001}. \citet{Kim1998} classified the nucleus of this OHM as a starburst. The OH spectrum of this source shows multiple velocity components. In Figure\,\ref{Fig:IRAS-Continued}, the optical RGB image shows an obvious interaction morphology with two nuclei connected by arcs and a single kinked tidal tail, clearly indicating an advanced merger \citep[see also][]{Darling2001}.

\textbf{IRAS\,08328+5148}
This is a newly discovered OHM (see Figure\,\ref{Fig:IRAS-Continued}). To the east (OC-1 with RA, Dec = 129.1717$^\circ$, 51.6246$^\circ$) of the OC listed in Table\,\ref{tab:oc}, there is an optical spectroscopic object with a redshift of 0.237 from the SDSS catalog. However, the OC coordinates are 48$''$ away from the OH emission center measured by FASHI. Fortunately, we find another optical object (named OC-2 with RA, Dec = 129.1505$^\circ$, 51.6267$^\circ$) at the center of the OH emission distribution. Although this object has no spectroscopic redshift, its SDSS photometric redshift is $0.234\pm0.070$, which agrees well with the OH redshift of 0.23749. In addition, OC-1 and OC-2 have similar optical colors. Furthermore, the coordinates of IRAS\,08328+5148 are in good agreement with the observed OH line coordinates. Therefore, it is likely that OC-2 is the optical counterpart of this OHM, but its spectroscopic redshift is needed for further confirmation.

\textbf{IRAS\,10106+4708}
The source is bright in all four WISE bands, and the optical data display an evident tidal tail (see Figure\,\ref{Fig:IRAS-Continued}). Despite that, the optical extinction is significant, leading to the absence of the H$\beta$ and [O{\small{III}}] lines. WISE colors are estimated to be [4.6]-[12] = 4.18 and [3.4]-[4.6] = 0.86, leading this object to be described as a dusty quasar \citep{Nikutta2014}.

\textbf{IRAS\,10597+5926}
This OHM was first found and reported by \citet{Willett2012}, where the detected peak flux density is 17.56\,mJy, while it is $32.27\pm2.23$\,mJy measured by FAST. The flux densities are very different, but the line profiles are consistent between the two.

\textbf{IRAS\,11029+3130}\footnote{listed as IRAS\,11028+3130 by \citet{Darling2001}}
This was first found and reported by the Arecibo OHM surveys \citep{Darling2001}. The central core is a LINER-type active galaxy nucleus \citep{Alatalo2016}.

\textbf{IRAS\,11087+5351}
The 1665\,MHz emission line is clearly visible. The measured line ratio is $R_{1667:1665}=1.32\pm0.21$, which is the lowest measured ratio among all sources in this work. In Figure\,\ref{Fig:IRAS-Continued}, the optical RGB image shows an extended morphology with a bright nucleus in the center. The central nucleus is optically identified as a LINER-type active galaxy nucleus by \citet{Toba2014}.

\textbf{IRAS\,11161+6020}
\citet{Toba2014} classified this object spectroscopically as a Seyfert 2 galaxy.

\textbf{IRAS\,11301-0523}
In Figure\,\ref{Fig:IRAS-Continued} the optical RGB image shows an arc-like structure with several close and connected nuclei. The OH spectrum has multiple velocity components, and the 1665\,MHz line is not visible.

\textbf{IRAS\,12447+3721}
\citet{Duarte2017} classified it as a \HII galaxy with a high star formation rate.

\textbf{IRAS\,13469+5833}
The optical RGB image shows a possible merging system with several tidal tails surrounding the nuclei \citep[see also the HST F814W-band images in][]{Murata2017}. The OH spectrum has double velocity components. The IRAS\,13469+5833 is identified as an \HII galaxy \citep{Lin2018}.

\textbf{IRAS\,14202+2615}
This source has been observed by the Arecibo OHM surveys \citep{Darling2000,Darling2006}, but without sufficient sensitivity to see the OH emission line. In the FASHI survey we detected the OHM emission line with a peak flux density of $4.38\pm1.15$\,mJy, at a SNR of only about 3 (see Figure\,\ref{Fig:IRAS-Continued}). The optical redshift from \citet{Saunders2000} is in good agreement with the OH line measured by FAST, while the spatially projected coordinate offset is $\sim$2.0$'$. The optical RGB image shows two close nuclei, one of which has a visible tidal tail. This morphology probably indicates a merging system.

\textbf{IRAS\,15160+5307}
The OH spectrum has a weak peak of $\sim$2.5\,mJy, but the blue wing is broad and the red wing has an absorption feature.

\textbf{IRAS\,15261+5502}
\citet{Hou2009} classified it as a starburst galaxy.

\textbf{IRAS\,16468+5200}
\citet{Weedman2008} classified it as a LINER-type active galaxy nucleus. In Figure\,\ref{Fig:IRAS-Continued} the optical RGB image shows a typical merging system with two interacting nuclei, one of which has an elongated tidal tail. 

\textbf{IRAS\,23129+2548}
This source has been confirmed as an OHM by the Arecibo OHM surveys \citep{Darling2001}. The nucleus of this OHM host is classified as a LINER by \citet{Veilleux1999}.

\subsection{OHM catalog}
\label{sec:catalog_ohm}

\begin{table*}[htp]
\caption{\textbf{FASHI detected OHMs including coordinates, redshift, line width, peak flux.}}
\label{tab:oh}
\centering 
\setlength{\tabcolsep}{1.5mm}{
\begin{tabular}{ccccccccccc}
\hline \hline
[1]  &  [2]  & [3]   & [4]  & [5] & [6]  & [7]  & [8] & [9] & [10]  \\ 
FASHI ID & FASHI name & IRAS name & RA$_{\rm oh}$ & Dec$_{\rm oh}$ & $z_{\odot}$ & $V_{\rm heli}$  & $W_{50}$ & $F_{\rm peak}$ & rms \\
& J2000      & B1950    & deg & deg & &  $\kms$ &   $\kms$  & mJy & mJy    \\
\hline
20240066172 &J002814.52-015310.0 &00256-0208 &7.061 &-1.886 &0.27656 &82912.1 &279.6 &3.05 &0.88 \\
20240066090 &J015902.56+254224.7 &\bf{01562+2527} &29.761 &25.707 &0.16600 &49766.2 &272.8 &6.72 &1.43 \\
20240066093 &J025515.32+205835.5 &\bf{02524+2046} &43.814 &20.977 &0.18067 &54164.0 &86.4 &21.78 &1.83 \\
20240066181 &J031336.48-023255.8 &03111-0244 &48.402 &-2.549 &0.18956 &56827.9 &77.1 &15.27 &1.67 \\
20240066097 &J065143.27+220518.1 &\bf{06487+2208} &102.930 &22.088 &0.14347 &43012.5 &130.4 &5.82 &1.37 \\
20240066055 &J082310.95+275137.6 &\bf{08201+2801} &125.796 &27.860 &0.16793 &50343.5 &242.3 &11.45 &1.50 \\
20240099047 &J083634.98+513745.8 &08328+5148 &129.146 &51.629 &0.23749 &71196.3 &105.9 &10.63 &1.10 \\
20240066185 &J085909.64-030430.6 &08566-0252 &134.790 &-3.075 &0.20262 &60743.1 &190.2 &7.73 &1.27 \\
20240066025 &J101347.44+465356.6 &10106+4708 &153.448 &46.899 &0.20529 &61543.0 &159.4 &24.73 &1.32 \\
20240066062 &J105837.38+382836.6 &10558+3845 &164.656 &38.477 &0.20788 &62319.4 &26.3 &4.35 &1.04 \\
20240066027 &J110245.08+591005.0 &\bf{10597+5926} &165.688 &59.168 &0.19607 &58780.0 &236.2 &32.27 &2.23 \\
20240066060 &J110537.56+311427.4 &\bf{11029+3130} &166.407 &31.241 &0.19885 &59613.8 &130.2 &3.36 &0.97 \\
20240066028 &J111136.57+533514.0 &11087+5351 &167.902 &53.587 &0.14314 &42913.0 &101.9 &3.65 &0.87 \\
20240129741 &J111908.12+600432.5 &11161+6020 &169.784 &60.076 &0.26434 &79246.5 &288.2 &9.07 &2.05 \\
20240066187 &J113239.21-053833.9 &11301-0523 &173.163 &-5.643 &0.22984 &68903.1 &168.0 &6.98 &1.46 \\
20240066105 &J120420.53+192456.7 &\bf{12018+1941} &181.086 &19.416 &0.16783 &50315.5 &254.6 &1.90 &0.83 \\
20240066030 &J122539.49+551528.9 &12232+5532 &186.415 &55.258 &0.23250 &69702.5 &166.1 &2.40 &0.66 \\
20240066069 &J124709.66+370508.6 &12447+3721 &191.790 &37.086 &0.15847 &47506.6 &198.0 &1.40 &0.68 \\
20240066013 &J130854.87+635636.6 &13070+6412 &197.229 &63.944 &0.20523 &61525.8 &154.6 &12.48 &1.97 \\
20240066035 &J134841.23+581817.8 &13469+5833 &207.172 &58.305 &0.15716 &47114.4 &20.6 &9.43 &1.26 \\
20240066034 &J141859.45+453206.3 &14170+4545 &214.748 &45.535 &0.15009 &44997.1 &68.7 &2.43 &0.85 \\
20240066110 &J142227.98+260225.7 &\bf{14202+2615} &215.617 &26.040 &0.15925 &47740.6 &16.0 &4.38 &1.15 \\
20240066037 &J150535.42+574248.3 &15043+5754 &226.398 &57.713 &0.15029 &45054.5 &115.9 &2.17 &1.16 \\
20240066039 &J151732.61+525616.8 &15160+5307 &229.386 &52.938 &0.15227 &45648.2 &195.3 &1.34 &0.51 \\
20240102609 &J152725.70+545139.2 &15261+5502 &231.857 &54.861 &0.22902 &68657.8 &218.0 &8.42 &1.04 \\
20240066042 &J164800.26+515535.2 &16468+5200 &252.001 &51.926 &0.15032 &45065.1 &151.3 &3.98 &1.28 \\
20240066122 &J231519.36+260419.1 &\bf{23129+2548} &348.831 &26.072 &0.17881 &53605.7 &278.4 &3.42 &0.96 \\
\hline
\end{tabular}
}
\begin{flushleft}
\textbf{Notes.} The IRAS names in bold are previously known OHMs, while the others are newly discovered by FAST. \\
\end{flushleft}
\end{table*}

Table\,\ref{tab:oh} presents the main properties of 27 OHMs detected by FASHI, including coordinates, redshift, line width, and flux. Each column in Table\,\ref{tab:oh} is introduced as follows:

\begin{itemize} 

\item {Column\,1: Index number for each FASHI extragalactic source. This index number is unique to each FASHI source.}

\item {Column\,2: Centroid coordinate (J2000) for the OH source in the format of \texttt{Jhhmmss.ss$\pm$ddmmss.s}.}

\item {Column\,3: IRAS source name \citep{IRAS1988}.}

\item {Column\,4-5: Right ascension (RA) and declination (Dec) in units of deg for the FASHI source centroid (J2000).}

\item {Column\,6: Redshift, $z_{\odot}$ of the 1667\,MHz OH line, corresponding to the heliocentric velocity in Column\,7.}

\item {Column\,7: Heliocentric velocity of the OH source, c$z_{\odot}$ in units of $\kms$. All results are presented in this work using the optical helio-centric definition of velocity.}


\item {Column\,8: Velocity width of the OH line profile, $W_{50}$ in $\kms$, measured at the level of 50\% of the peak.}

\item {Column\,9: OH line peak intensity, $F_{\rm peak}$ in mJy.}

\item {Column\,10: OH line noise level for the spatially integrated spectral profile, $\sigma_{\rm{rms}}$ in mJy. It was obtained by measuring over the signal and RFI free portions of the integrated \HI spectrum at a spectral resolution of 6.4\,$\kms$. }

\end{itemize}

\subsection{Hyperfine ratio catalog}
\label{sec:catalog_ratio}

\begin{table*}
\caption{\textbf{Hyperfine ratio between the 1667 and 1665\,MHz lines.}}
\label{tab:ratio}
\centering 
\setlength{\tabcolsep}{2.0mm}{
\begin{tabular}{ccccccccccccc}
\hline \hline
[1]  &  [2]  & [3]   & [4]  & [5] & [6]  & [7]  & [8] & [9] \\ 
FASHI ID & IRAS name & $S_{\rm int}^{1667}$ & $V_{\rm min}^{1667}$ & $V_{\rm max}^{1667}$ & $S_{\rm int}^{1665}$ & $V_{\rm min}^{1665}$ & $V_{\rm max}^{1665}$  & $R_{1667:1665}$ \\
& & mJy$\cdot$$\kms$ & $\kms$ & $\kms$ & mJy$\cdot$$\kms$ & $\kms$ & $\kms$ & ratio      \\
\hline
20240066172 &00256-0208 &1089.2 $\pm$ 65.1 &82246.2 &83457.0 \\
20240066090 &\bf{01562+2527} &2026.8 $\pm$ 93.5 &49363.3 &50148.7 \\
20240066093 &\bf{02524+2046} &1781.4 $\pm$ 59.1 &54028.3 &54243.2 &607.3 $\pm$ 42.6 &54065.6 &54396.2 &2.93 $\pm$ 0.23 \\
20240066181 &03111-0244 &1167.6 $\pm$ 54.0 &56703.8 &56921.8 &424.2 $\pm$ 57.8 &56728.9 &57181.4 &2.75 $\pm$ 0.40 \\
20240066097 &\bf{06487+2208} &848.0 $\pm$ 58.9 &42839.9 &43178.3 &259.7 $\pm$ 46.8 &42869.3 &43121.6 &3.27 $\pm$ 0.63 \\
20240066055 &\bf{08201+2801} &2110.0 $\pm$ 77.8 &50065.3 &50568.3 \\
20240099047 &08328+5148 &1325.4 $\pm$ 50.6 &70921.6 &71376.2 &140.4 $\pm$ 27.1 &71142.6 &71319.7 &9.44 $\pm$ 1.85 \\
20240066185 &08566-0252 &1427.1 $\pm$ 51.1 &60584.7 &60919.2 \\
20240066025 &10106+4708 &4011.7 $\pm$ 68.5 &61332.6 &61876.2 \\
20240066062 &10558+3845 &212.3 $\pm$ 46.7 &62083.1 &62492.4 &130.7 $\pm$ 30.6 &62179.0 &62420.2 &1.62 $\pm$ 0.52 \\
20240066027 &\bf{10597+5926} &7259.4 $\pm$ 112.5 &58535.6 &59039.4 &477.0 $\pm$ 45.2 &58688.7 &59067.3 &15.22 $\pm$ 1.46 \\
20240066060 &\bf{11029+3130} &424.8 $\pm$ 40.6 &59463.6 &59819.7 \\
20240066028 &11087+5351 &289.4 $\pm$ 27.2 &42818.6 &43005.8 &219.1 $\pm$ 28.2 &42797.8 &43035.7 &1.32 $\pm$ 0.21 \\
20240129741 &11161+6020 &2468.3 $\pm$ 114.1 &78901.6 &79578.7 \\
20240066187 &11301-0523 &1014.8 $\pm$ 66.7 &68669.5 &69110.5 \\
20240066105 &\bf{12018+1941} &370.6 $\pm$ 37.4 &50139.7 &50530.3 &80.8 $\pm$ 20.5 &50276.9 &50412.2 &4.59 $\pm$ 1.25 \\
20240066030 &12232+5532 &399.8 $\pm$ 26.0 &69534.9 &69869.2 &259.2 $\pm$ 25.2 &69535.6 &69853.5 &1.54 $\pm$ 0.18 \\
20240066069 &12447+3721 &287.8 $\pm$ 33.0 &47306.6 &47749.7 \\
20240066013 &13070+6412 &2016.0 $\pm$ 91.6 &61324.5 &61764.1 \\
20240066035 &13469+5833 &556.0 $\pm$ 51.5 &46975.8 &47292.6 \\
20240066034 &14170+4545 &197.8 $\pm$ 36.1 &44860.1 &45194.9 \\
20240066110 &\bf{14202+2615} &319.4 $\pm$ 49.4 &47601.5 &47949.1 \\
20240066037 &15043+5754 &211.0 $\pm$ 44.5 &44925.3 &45202.0 \\
20240066039 &15160+5307 &305.5 $\pm$ 28.4 &45368.5 &45923.7 \\
20240102609 &15261+5502 &1857.3 $\pm$ 57.5 &68312.8 &68944.2 \\
20240066042 &16468+5200 &640.3 $\pm$ 57.8 &44889.0 &45267.6 \\
20240066122 &\bf{23129+2548} &845.4 $\pm$ 57.8 &53272.0 &53960.7 \\
\hline
\end{tabular}}
\begin{flushleft}
\textbf{Notes.} The IRAS names in bold are previously known OHMs, while the others are newly discovered by FAST. \\
\end{flushleft}
\end{table*}

Table\,\ref{tab:ratio} presents the integrated flux and the hyperfine ratio between the 1667 and 1665\,MHz lines. Each column in Table\,\ref{tab:ratio} is introduced as follows:

\begin{itemize} 

\item {Column\,1: Index number for each FASHI extragalactic source. This index number is unique to each FASHI source.}

\item {Column\,2: IRAS source name.}

\item {Column\,3: Integrated flux of the 1667\,MHz line, $S_{\rm int}$, in $\mjybkms$, summing all velocity channels containing signal within an appropriate velocity range from each integrated spectrum. The velocity range is listed in Columns\,4 and 5. }

\item {Columns\,4-5: The minimum ($V_{\rm min}^{1667}$) and maximum ($V_{\rm max}^{1667}$) velocities of the integrated flux range for the 1667\,MHz line. }

\item {Column\,6: Integrated flux of the 1665\,MHz line, $S_{\rm int}$, in $\mjybkms$, summing all velocity channels containing signal within an appropriate velocity range from each integrated spectrum. The velocity range is listed in Columns\,7 and 8. }

\item {Columns\,7-8: The minimum ($V_{\rm min}^{1665}$) and maximum ($V_{\rm max}^{1665}$) velocities of the integrated flux range for the 1665\,MHz line. }

\item {Column\,9: Hyperfine line ratio between the 1667 and 1665\,MHz lines. }
\end{itemize}

\subsection{OHM optical counterpart catalog}
\label{sec:catalog_oc}

\begin{table*}
\caption{\textbf{OHM optical counterparts including IRAS flux and FIR properties.}}
\label{tab:oc}
\centering 
\setlength{\tabcolsep}{1.8mm}{
\begin{tabular}{ccccccccccccc}
\hline \hline
[1]  &  [2]  & [3]   & [4]  & [5] & [6]  & [7]  & [8] & [9] & [10] \\ 
FASHI ID &  IRAS name & RA$_{\rm oc}$ & Dec$_{\rm oc}$ & $z_{\odot}$ 
& $f_{60}$ & $f_{100}$ & $D_{\rm L}$ & log$L_{\rm FIR}$ & log$L_{\rm OH}$    \\
& B1950    & deg & deg & PSCz 
& Jy& Jy& $h_{75}^{-1}$Gpc & $h_{75}^{-2}L_{\odot}$ & $h_{75}^{-2}L_{\odot}$     \\
\hline
20240066172 &00256-0208 &7.059 &-1.863 &0.27630 &0.65 $\pm$ 0.07 &1.00 $\pm$ 0.00 &1.320 &12.27 $\pm$ 0.03 &3.40 $\pm$ 0.03 \\
20240066090 &\bf{01562+2527} &29.761 &25.710 &0.16580 &0.80 $\pm$ 0.06 &1.85 $\pm$ 0.15 &0.744 &11.93 $\pm$ 0.02 &3.22 $\pm$ 0.02 \\
20240066093 &\bf{02524+2046} &43.821 &20.983 &0.18153 &0.94 $\pm$ 0.08 &8.48 $\pm$ 0.00 &0.817 &12.46 $\pm$ 0.01 &3.24 $\pm$ 0.01 \\
20240066181 &03111-0244 &48.422 &-2.560 &0.18833 &0.69 $\pm$ 0.06 &1.03 $\pm$ 0.00 &0.862 &11.92 $\pm$ 0.02 &3.09 $\pm$ 0.02 \\
20240066097 &\bf{06487+2208} &102.941 &22.074 &0.14370 &2.07 $\pm$ 0.17 &2.36 $\pm$ 0.26 &0.634 &12.09 $\pm$ 0.03 &2.71 $\pm$ 0.03 \\
20240066055 &\bf{08201+2801} &125.803 &27.861 &0.16750 &1.13 $\pm$ 0.09 &1.60 $\pm$ 0.18 &0.753 &12.01 $\pm$ 0.03 &3.24 $\pm$ 0.02 \\
20240099047 &08328+5148 &129.150 &51.627 &0.23700 &0.37 $\pm$ 0.11 &1.58 $\pm$ 0.00 &1.110 &12.09 $\pm$ 0.05 &3.35 $\pm$ 0.02 \\
20240066185 &08566-0252 &134.789 &-3.070 &0.20201 &0.79 $\pm$ 0.07 &1.10 $\pm$ 0.00 &0.928 &12.03 $\pm$ 0.03 &3.24 $\pm$ 0.02 \\
20240066025 &10106+4708 &153.450 &46.900 &0.20481 &0.77 $\pm$ 0.08 &1.37 $\pm$ 0.30 &0.942 &12.07 $\pm$ 0.05 &3.70 $\pm$ 0.01 \\
20240066062 &10558+3845 &164.664 &38.485 &0.20660 &0.70 $\pm$ 0.06 &1.00 $\pm$ 0.00 &0.955 &12.01 $\pm$ 0.02 &2.44 $\pm$ 0.10 \\
20240066027 &\bf{10597+5926} &165.696 &59.177 &0.19583 &1.02 $\pm$ 0.08 &1.33 $\pm$ 0.17 &0.895 &12.10 $\pm$ 0.03 &3.92 $\pm$ 0.01 \\
20240066060 &\bf{11029+3130} &166.406 &31.242 &0.19890 &1.04 $\pm$ 0.09 &1.38 $\pm$ 0.11 &0.909 &12.12 $\pm$ 0.03 &2.70 $\pm$ 0.04 \\
20240066028 &11087+5351 &167.902 &53.583 &0.14270 &1.00 $\pm$ 0.11 &1.00 $\pm$ 0.13 &0.632 &11.75 $\pm$ 0.04 &2.24 $\pm$ 0.04 \\
20240129741 &11161+6020 &169.779 &60.075 &0.26428 &0.61 $\pm$ 0.05 &1.28 $\pm$ 0.19 &1.253 &12.25 $\pm$ 0.04 &3.72 $\pm$ 0.02 \\
20240066187 &11301-0523 &173.175 &-5.663 &0.23019 &0.70 $\pm$ 0.07 &1.42 $\pm$ 0.16 &1.069 &12.16 $\pm$ 0.03 &3.21 $\pm$ 0.03 \\
20240066105 &\bf{12018+1941} &181.102 &19.419 &0.16865 &1.64 $\pm$ 0.18 &1.86 $\pm$ 0.22 &0.753 &12.14 $\pm$ 0.04 &2.49 $\pm$ 0.04 \\
20240066030 &12232+5532 &186.410 &55.264 &0.23143 &0.62 $\pm$ 0.06 &1.63 $\pm$ 0.00 &1.083 &12.18 $\pm$ 0.02 &2.81 $\pm$ 0.03 \\
20240066069 &12447+3721 &191.782 &37.094 &0.15870 &1.11 $\pm$ 0.12 &1.00 $\pm$ 0.00 &0.707 &11.88 $\pm$ 0.04 &2.33 $\pm$ 0.05 \\
20240066013 &13070+6412 &197.231 &63.941 &0.20492 &0.74 $\pm$ 0.05 &0.71 $\pm$ 0.11 &0.941 &11.96 $\pm$ 0.03 &3.40 $\pm$ 0.02 \\
20240066035 &13469+5833 &207.167 &58.315 &0.15778 &1.34 $\pm$ 0.11 &1.99 $\pm$ 0.14 &0.700 &12.02 $\pm$ 0.02 &2.60 $\pm$ 0.04 \\
20240066034 &14170+4545 &214.745 &45.537 &0.15026 &0.73 $\pm$ 0.06 &1.19 $\pm$ 0.12 &0.666 &11.73 $\pm$ 0.03 &2.11 $\pm$ 0.08 \\
20240066110 &\bf{14202+2615} &215.631 &26.035 &0.15915 &1.45 $\pm$ 0.16 &1.95 $\pm$ 0.23 &0.711 &12.06 $\pm$ 0.04 &2.38 $\pm$ 0.07 \\
20240066037 &15043+5754 &226.415 &57.719 &0.15059 &0.99 $\pm$ 0.07 &1.42 $\pm$ 0.10 &0.667 &11.85 $\pm$ 0.02 &2.14 $\pm$ 0.09 \\
20240066039 &15160+5307 &229.374 &52.947 &0.15199 &0.64 $\pm$ 0.03 &1.38 $\pm$ 0.08 &0.677 &11.74 $\pm$ 0.02 &2.32 $\pm$ 0.04 \\
20240102609 &15261+5502 &231.861 &54.864 &0.22921 &0.53 $\pm$ 0.05 &1.00 $\pm$ 0.00 &1.065 &12.03 $\pm$ 0.03 &3.47 $\pm$ 0.01 \\
20240066042 &16468+5200 &252.006 &51.929 &0.15054 &1.01 $\pm$ 0.07 &1.61 $\pm$ 0.00 &0.667 &11.87 $\pm$ 0.02 &2.63 $\pm$ 0.04 \\
20240066122 &\bf{23129+2548} &348.839 &26.076 &0.17810 &1.80 $\pm$ 0.20 &1.70 $\pm$ 0.17 &0.808 &12.21 $\pm$ 0.04 &2.90 $\pm$ 0.03 \\
\hline
\end{tabular}
}
\begin{flushleft}
\textbf{Notes.} The IRAS names in bold are previously known OHMs, while the others are first discovered by FAST. \\
\end{flushleft}
\end{table*}

Table\,\ref{tab:oc} presents the main properties of the OCs corresponding to the OHMs, including IRAS flux densities, and FIR properties. Each column in Table\,\ref{tab:oc} is introduced as follows:

\begin{itemize} 

\item {Column\,1: Index number for each FASHI extragalactic source. This index number is unique to each FASHI source.}

\item {Column\,2: IRAS source catalog name.}

\item {Column\,3-4: RA and Dec with centroid coordinate (J2000) in units of deg for OHM OCs extracted from the SDSS catalog.}

\item {Column\,5: Heliocentric optical redshift, $z_{\odot}$, extracted from the SDSS catalog.}

\item {Column\,6-7: IRAS 60 and 100\,$\mu$m flux densities $f_{60}$ and $f_{100}$ in units of Jy \citep{IRAS1988}. }

\item {Column\,8: Luminosity distance $D_{\rm L}$ in $h_{75}^{-1}{\rm Gpc}$, estimated with the cosmology calculator\footnote{\url{https://www.astro.ucla.edu/~wright/CosmoCalc.html}} \citep{Wright2006}. }

\item {Column\,9: Logarithm of the FIR luminosity $L_{\rm FIR}$ in units of $h_{75}^{-2}L_{\sun}$, estimated using the following equation from \citet{Sanders1996} with
\begin{equation}
 L_{\rm FIR}=3.96\times10^{11}D_{\rm L}^2(2.58f_{60}+f_{100}),
\end{equation}
where $f_{60}$ and $f_{100}$ are listed in Columns\,6 and 7, and $D_{\rm L}$ in units of $h_{75}^{-1}{\rm Gpc}$ listed in Column\,8. }

\item {Column\,10: OH line luminosity $L_{\rm OH}$ in units of $h_{75}^{-2}L_{\sun}$. From the integrated line flux $S_{\rm int}$ we use the general relation for the spectral line luminosity \citep[e.g.,][]{Briggs1998} with
\begin{equation}
{\frac {L_{\rm OH}}{L_\odot}} = \Big({\frac {1.7064} {1+z_{\rm OH}}}\Big)\,\Big({\frac {S_{\rm int}}{\rm mJy\,km\,s^{-1}}}\Big)\,\Big({\frac {D_L}{\rm Gpc}}\Big)^2,
\end{equation}
where we take the stronger of the two main OH lines with a rest frequency of 1667.35903\,MHz to define $z_{\rm OH}$.
 }

\end{itemize}

\subsection{OHM catalog cross-matching with GSWLC}
\label{sec:catalog_GSWLC}

\begin{table*}
\caption{\textbf{Cross-matched sources between all the known 154 OHMs and GSWLC.}}
\label{tab:GSWLC}
\centering 
\begin{tabular}{ccccccccccc}
\hline \hline
[1]  &  [2]  & [3]   & [4]  & [5] & [6]  & [7]  & [8]  \\   
IRAS name & OBJID & RA & Dec & $z_{\odot}$ & log($M_{\star})$ & log($\rm SFR_{SED})$ & $A_{\rm V}$     \\
B1950 &    & deg & deg &   & $\rm M_{\odot}$ & $\rm M_{\odot}\,yr^{-1}$ & mag  \\
\hline
07556+2859 &1237657629506142647 &119.727 &28.886 &0.1262 &10.70 $\pm$ 0.03 &-0.54 $\pm$0.71 &0.24 $\pm$ 0.10 \\
08122+0505 &1237660668730408974 &123.723 &4.937 &0.1033 &11.04 $\pm$ 0.04 &0.57 $\pm$0.24 &0.43 $\pm$ 0.16 \\
08449+2332 &1237664834854518934 &131.959 &23.353 &0.1517 &10.71 $\pm$ 0.06 &1.47 $\pm$0.05 &0.47 $\pm$ 0.06 \\
09039+0503 &1237658423543595297 &136.642 &4.858 &0.1250 &10.81 $\pm$ 0.03 &1.56 $\pm$0.06 &0.86 $\pm$ 0.06 \\
09116+0334 &1237654605316358298 &138.574 &3.365 &0.1460 &10.91 $\pm$ 0.03 &0.43 $\pm$0.16 &0.23 $\pm$ 0.10 \\
09320+6134 &1237651272966275163 &143.965 &61.353 &0.0394 &11.25 $\pm$ 0.03 &1.16 $\pm$0.01 &0.75 $\pm$ 0.00 \\
09531+1430 &1237671261205495893 &148.958 &14.269 &0.2152 &11.34 $\pm$ 0.04 &1.00 $\pm$0.18 &0.62 $\pm$ 0.12 \\
09540+3521 &1237664669506863314 &149.246 &35.119 &0.1006 &10.72 $\pm$ 0.03 &0.36 $\pm$0.23 &0.59 $\pm$ 0.11 \\
10378+1108 &1237661949170155593 &160.122 &10.888 &0.1363 &11.03 $\pm$ 0.03 &1.57 $\pm$0.02 &0.82 $\pm$ 0.03 \\
10558+3845 &1237664668975562850 &164.664 &38.485 &0.2081 &11.01 $\pm$ 0.09 &1.56 $\pm$0.14 &0.69 $\pm$ 0.14 \\
11011+4107 &1237662194520293391 &165.976 &40.850 &0.0348 &9.97 $\pm$ 0.14 &0.58 $\pm$0.18 &0.49 $\pm$ 0.11 \\
11087+5351 &1237657590319022140 &167.902 &53.583 &0.1429 &10.91 $\pm$ 0.07 &1.44 $\pm$0.20 &0.73 $\pm$ 0.18 \\
11257+5850 &1237655107301277787 &172.140 &58.563 &0.0105 &10.40 $\pm$ 0.04 &0.85 $\pm$0.08 &0.86 $\pm$ 0.09 \\
12018+1941 &1237667914881892435 &181.102 &19.419 &0.1679 &10.89 $\pm$ 0.09 &1.44 $\pm$0.12 &0.60 $\pm$ 0.15 \\
12447+3721 &1237664672714653739 &191.782 &37.094 &0.1582 &10.39 $\pm$ 0.09 &0.96 $\pm$0.18 &0.24 $\pm$ 0.15 \\
13254+4754 &1237661149769760972 &201.900 &47.647 &0.0607 &9.77 $\pm$ 0.07 &0.33 $\pm$0.08 &0.17 $\pm$ 0.08 \\
13428+5608 &1237661387602853939 &206.176 &55.887 &0.0373 &10.77 $\pm$ 0.05 &0.35 $\pm$0.33 &0.39 $\pm$ 0.14 \\
13469+5833 &1237659326015537309 &207.167 &58.314 &0.1575 &11.00 $\pm$ 0.05 &0.90 $\pm$0.14 &0.40 $\pm$ 0.14 \\
14202+2615 &1237665441523957822 &215.631 &26.035 &0.1587 &11.05 $\pm$ 0.05 &1.38 $\pm$0.02 &0.20 $\pm$ 0.02 \\
14312+2825 &1237665351854784676 &218.365 &28.200 &0.1748 &11.30 $\pm$ 0.09 &1.59 $\pm$0.05 &0.73 $\pm$ 0.08 \\
14488+3521 &1237662305655717891 &222.726 &35.144 &0.2057 &11.41 $\pm$ 0.04 &2.23 $\pm$0.02 &0.73 $\pm$ 0.01 \\
15043+5754 &1237671769072206053 &226.415 &57.719 &0.1506 &10.67 $\pm$ 0.06 &1.29 $\pm$0.10 &0.44 $\pm$ 0.14 \\
15160+5307 &1237655463779631232 &229.374 &52.946 &0.1522 &11.21 $\pm$ 0.06 &1.47 $\pm$0.08 &0.37 $\pm$ 0.11 \\
15224+1033 &1237662636908478686 &231.213 &10.380 &0.1346 &10.78 $\pm$ 0.05 &1.13 $\pm$0.05 &0.34 $\pm$ 0.03 \\
15250+3608 &1237662335714852881 &231.748 &35.977 &0.0552 &10.63 $\pm$ 0.04 &0.97 $\pm$0.06 &0.38 $\pm$ 0.06 \\
15327+2340 &1237665537075511390 &233.738 &23.504 &0.0184 &10.87 $\pm$ 0.02 &0.37 $\pm$0.08 &0.69 $\pm$ 0.07 \\
16145+4231 &1237665356696846411 &244.050 &42.400 &0.0232 &9.76 $\pm$ 0.04 &-0.33 $\pm$0.10 &0.19 $\pm$ 0.09 \\
16300+1558 &1237665536008520057 &248.089 &15.863 &0.2418 &11.20 $\pm$ 0.09 &1.71 $\pm$0.13 &0.49 $\pm$ 0.15 \\
\hline
\end{tabular}
\begin{flushleft}
\textbf{Notes.} The listed Columns [2]-[8] in this catalog are extracted from GSWLC \citep{Salim2016}. \\
\end{flushleft}
\end{table*}

Table\,\ref{tab:GSWLC} shows the 28 sources recovered when all known 154 OHMs are cross-matched with the GALEX-SDSS-WISE Legacy Catalog (GSWLC) catalog \citep{Salim2016} under the condition $\rm\delta_{RA} \leq 3'$, $\rm\delta_{Dec} \leq 3'$ and $\rm\delta_{HI\,velocity} \leq 300\,\kms$. Nine of these 28 OHMs are detected are detected by FAST. Each column in Table\,\ref{tab:oc} is introduced as follows:

\begin{itemize} 

\item {Column\,1: IRAS source catalog name.}

\item {Column\,2: OBJID, SDSS photometric identification number, from GSWLC.}

\item {Column\,3-4: RA and Dec with centroid coordinate (J2000) in units deg, from GSWLC.}

\item {Column\,5: Redshift, $z_{\odot}$, from GSWLC.}

\item {Column\,6: Stellar mass, log($M_{\star})$ with its error, from GSWLC.}

\item {Column\,7: UV/optical (SED) star formation rate, log($\rm SFR_{SED})$ with its error, from GSWLC.}

\item {Column\,8: Dust attenuation, $A_{\rm V}$, in rest-frame $V$, from GSWLC.}

\end{itemize}

\section{Discussion}
\label{sec:discu}

\subsection{Hyperfine ratio}

The 1667 and 1665\,MHz lines refer to the main lines of OH at the radio rest frequencies of 1667.359 and 1665.402\,MHz, while two satellite lines are located at 1720.530 and 1612.231\,MHz \citep{Radford1964}. In local thermodynamic equilibrium (LTE), the OH line ratios are approximately 1:5:9:1 for 1612, 1665, 1667, and 1720\,MHz lines \citep{Henkel1991,Hess2021}. The 1667\,MHz line is relatively common in extragalactic observations, but the 1665\,MHz line is very rare because of its weakness. \citet{McBride2013} reported that the hyperfine line ratio is $1<R_{1667:1665}<10$ with an average of $R_{1667:1665}\approx5$. In previous observations \citep[e.g.,][]{Darling2000,Darling2001,Darling2002a}, the 1667 and 1665\,MHz lines often blend into each other, resulting in a barely measurable line ratio. The main reason for this is the inherently large line width ($>500\,\kms$) of the 1667\,MHz line. In Table\,\ref{tab:ratio} we present the hyperfine line ratio between the integrated fluxes of 1667 and 1665\,MHz. In total we have detected 27 OHMs, 9 of which show a visible 1665\,MHz line. The detection rate is 33.3\% for the 1665\,MHz line. For detected sources, the line ratio $R_{1667:1665}$ ranges from 1.32 to 15.22, with an average of $R_{1667:1665}=4.74$, in agreement with the results reported by \citet{McBride2013}. Although the hyperfine ratio is often consistent with the thermal emission condition \citep{Baan1982}, e.g. for IRAS\,10558+3845 with $R_{1667:1665}=1.62\pm0.52$, 11087+5351 with $R_{1667:1665}=1.32\pm0.21$, and 12232+5532 with $R_{1667:1665}=1.54\pm0.18$, which have relatively low hyperfine ratios with $R_{1667:1665}<2$, \citet{McBride2013} suggested that the OH emission probably includes a significant non-thermal contribution. Furthermore, the wide range from 1.32 to 15.22 is indicative of complex maser pumping mechanisms.

\subsection{Host galaxy properties}

\begin{figure}[htp]
\centering
\includegraphics[width=0.47\textwidth, angle=0]{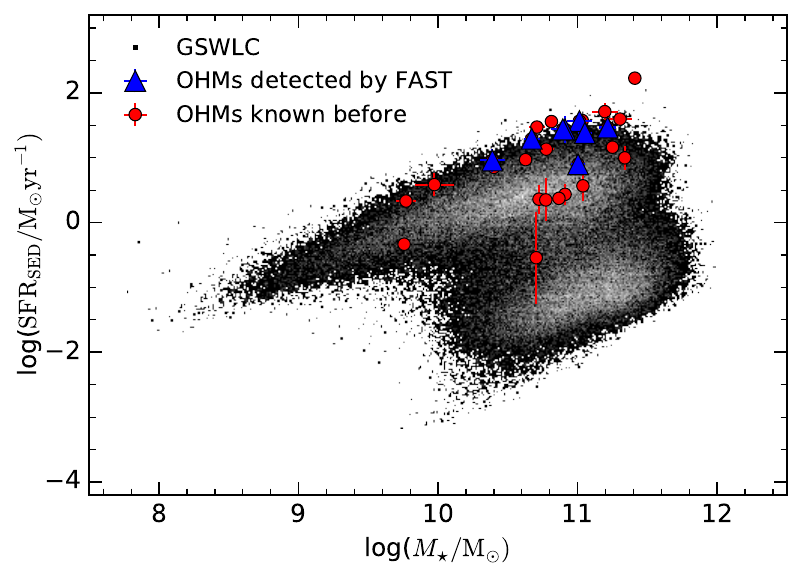}
\includegraphics[width=0.47\textwidth, angle=0]{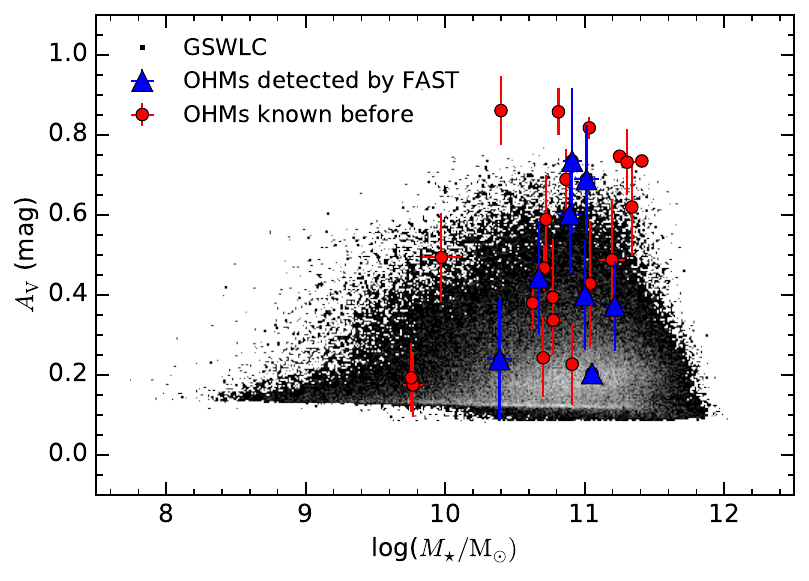}
\includegraphics[width=0.47\textwidth, angle=0]{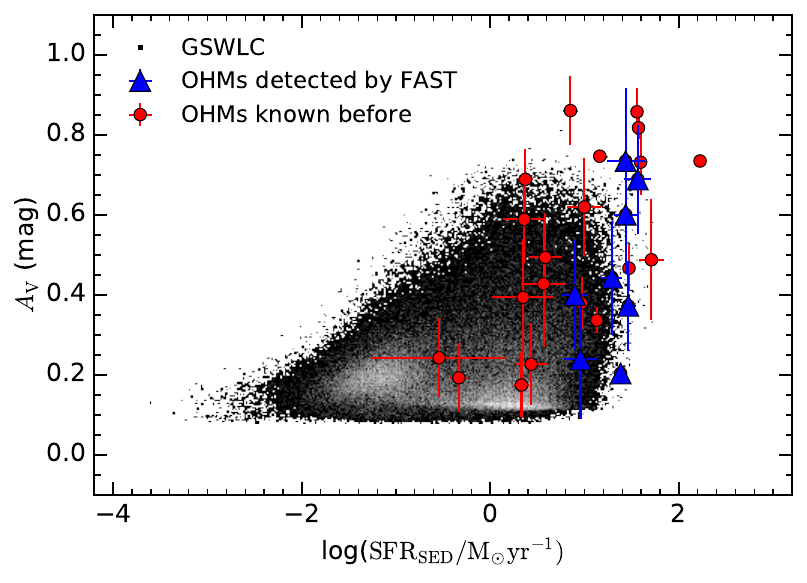}
\caption{Two-dimensional distributions of log($M_{\star})$, log($\rm SFR_{SED})$, and $A_{\rm V}$. The background shows all the data in the GSWLC. The blue triangles indicate cross-matched OHMs detected by FAST. The red dots indicate an additional 19 cross-matched OHMs not detected by FAST among the 154 known OHMs.}
\label{Fig:GSWLC}
\end{figure}

As expected, since the OHM sample was selected by cross-correlation with the IRAS-selected PSCz, which contains only objects with 60\,$\mu$m flux densities $>0.595$\,Jy, all of our detected OHMs are hosted by Luminous Infrared Galaxies (LIRGs) with $L_{\rm FIR} > 10^{11} \Lsun$, or ultra-luminous infrared galaxies (ULIRGs) with FIR luminosities $L_{\rm FIR} > 10^{12} \Lsun$ \citep[e.g.,][]{Darling2002a}. 13 OHMs are ULIRGs and 14 OHMs are LIRGs based on the measured $L_{\rm FIR}$, which ranges from $10^{11.70\pm0.03}$ to $10^{12.43\pm0.01} \Lsun$ with a median of $10^{12.00\pm0.02} \Lsun$. This means that all of the 27 OHMs have relatively high FIR luminosities. The strong FIR emission suggest that such 27 OHMs has undergone a very recent and strong starburst \citep[e.g.,][]{Sanders1996}.

The majority of [U]LIRGs or OHMs show evidence of interaction with other galaxies or have recently undergone a galaxy merger \citep{Andreasian1994,Darling2002a}. Mergers could be a way to funnel molecular gas into the core of the [U]LIRG, producing high molecular densities and stimulating the high star formation rates \citep{Burdiuzha1990}. \citet{Pihlstrom2007} suggested that the maser emission may originate in thick, circumnuclear structures, based on the results of high-resolution imaging.

OHMs are likely a disturbed system in a galaxy merger, as evidenced by tidal tails \citep{Clements1996}. We found that almost all of the 27 OHMs have recently merged or interacted with another galaxy (see Figures\,\ref{Fig:IRAS00256-0208} and \ref{Fig:IRAS-Continued}). This could be an indication that such galaxy system is in the process of merging as seen in the optical morphology (see Figures\,\ref{Fig:IRAS00256-0208} and \ref{Fig:IRAS-Continued}). In general, OHM line emission is always detected in [U]LIRGs \citep[e.g.,][]{Darling2007,Henkel1990,Suess2016,Willett2012}. It is likely that OHMs occur during a particular state or stage of the merger, which are consequences of tidal density enhancements that accompany galaxy interactions \citep{Darling2007}. In fact, very-long-baseline interferometry (VLBI) radio images of Arp220 show that the OHM emission arises in regions only a few parsecs in size in its core, and have been used to support active galactic nucleus (AGN) models for this galaxy \citep{Skinner1997}.

The multi-wavelength SED fitting results in Figure\,\ref{Fig:GSWLC} also show that the OH host galaxies are massive ($M_\star \simeq 10^{11}M_\odot$) galaxies with the SFR above the star-forming main sequence. The higher $A_V$ value of the OH galaxies from the SED fit is also consistent with the high IR luminosities. On the other hand, not all ULIRGs have detectable OH emission, suggesting that the OH emission may be triggered within a specific stage of the merger or may be seen by orientation effects. High-resolution OH mapping observations, e.g. with the Very Large Array (VLA), are still needed to better constrain the physical origin of the OH emission.

\subsection{$L_{\rm OH}$ and $L_{\rm FIR}$ Relationship}

\begin{figure}[htp]
\centering
\includegraphics[width=0.47\textwidth, angle=0]{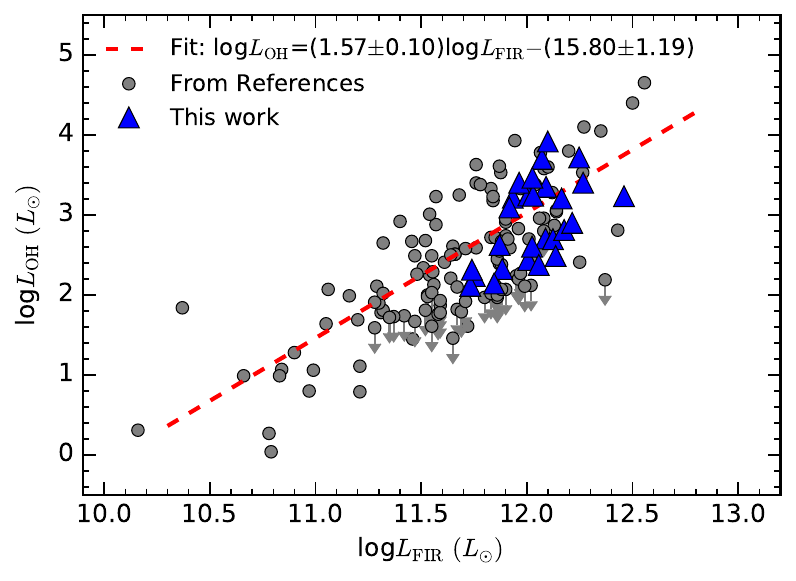}
\caption{$L_{\rm OH}$ vs. $L_{\rm FIR}$ distribution including 154 sources. The blue triangles indicate OHMs discovered by FASHI. The grey dots indicate previously known OHMs \citep{Darling2002a,Darling2006,Willett2012,Suess2016,Hess2021,Glowacki2022,Jarvis2023}, where the sources with $L_{\rm OH}$ having only an upper limit are excluded from the current fitting statistics. The red dashed line shows the fitting result of the $L_{\rm OH}$-$L_{\rm FIR}$ relation.}
\label{Fig:Loh_Lfir}
\end{figure}

OHMs are thought to be radiatively pumped by the FIR radiation ﬁeld at 35 and 53\,$\mu$m, with masing stimulated by 18\,cm continuum emission from the surrounding environment \citep{Henkel1987}. In a simple scenario of low-gain unsaturated masing, the maser power is proportional to the pumping rate and the stimulated emission rate \citep{Baan1989}. If the observed OHM represents an ensemble of many individual masing regions with different saturation states, then the relationship between OH and FIR luminosity is $L_{\rm OH}\propto L_{\rm FIR}^{1<\gamma<2}$ \citep{Baan1989,Darling2002a,Yun2001}. The measurement of $\gamma$ has been frustrated by small samples, survey bias, and theoretical bias. \citet{Darling2002a} derived the relationship between OH and far IR luminosity with 95 data points as ${\rm log}L_{\rm OH}= (1.57\pm0.11){\rm log}L_{\rm FIR}-(15.76\pm1.22)$. In this work we detect 27 OHMs and 18 new OHMs, plus 127 other previously known OHMs with OH and FIR luminosities from the literature \citep{Darling2002a,Darling2006,Willett2012,Suess2016,Hess2021,Glowacki2022}, where 31 sources with $L_{\rm OH}$ having only an upper limit are excluded from the current fitting statistics. Using the least-squares fitting method, the samples produce a fitting result with ${\rm log}L_{\rm OH}= (1.57\pm0.10){\rm log}L_{\rm FIR}-(15.80\pm1.19)$ in Figure\,\ref{Fig:Loh_Lfir}. The $\gamma=1.57\pm0.10$ agrees well with the result of \citet{Darling2002a} with $\gamma = 1.57\pm0.11$. Therefore, $\gamma$ remains almost the same when the OHM samples reach up to 154, which is $\sim$1.6 times the number of previously known OHMs. 

\subsection{FAST OHM detectability}

\begin{figure}[htp]
\centering
\includegraphics[width=0.47\textwidth, angle=0]{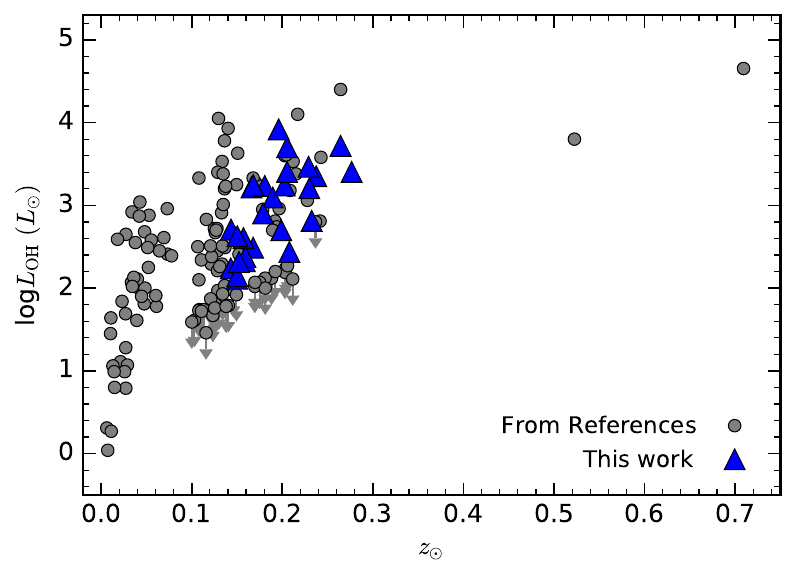}
\caption{$L_{\rm OH}$ vs. $z_{\odot}$ distribution including 154 sources. The blue triangles indicate 18 new found OHMs by FASHI. The grey dots indicate detections of previously known OHMs \citep{Darling2002a,Darling2006,Willett2012,Suess2016,Hess2021,Glowacki2022,Jarvis2023}. Each grey dot with a down arrow indicates the source with $L_{\rm OH}$ having only an upper limit. }
\label{Fig:Loh_z}
\end{figure}

Currently, the highest redshift detection of such an OHM in the main 1667\,MHz OH emission line is J095903.22+025356.1 with a redshift of $z_{\rm OH} = 0.7092$, recently found by \citet{Jarvis2023}. The second highest is LADUMA\,J033046.20-275518.1 with a redshift of $z_{\rm OH} = 0.5225$, found by \citet{Glowacki2022}. However, the third highest redshift of OHMs is IRAS\,00256-0208 with $z_{\rm OH} = 0.27656$, found in this paper. The large gap (see Figure\,\ref{Fig:Loh_z}) between the three highest redshifts of OHMs requires more observatories to complete the OHM picture. The 19-beam receiver with FAST is capable of simultaneously covering a frequency range of 1000-1500\,MHz, and the Ultra-Wide Bandwidth (UWB) receiver with FAST could simultaneously cover 500-3300\,MHz. FAST will be a powerful instrument for observing highly redshifted OH emission in extragalactic objects \citep{Zhang2023}.

Furthermore, radio frequency interference (RFI) always affects the search results of astronomical observations for all radio telescopes on Earth. At present, the redshifts of all known OHMs are mainly below $z\approx0.27$ or $f\approx1310$\,MHz \citep{Darling2002a,Willett2012,Haynes2018,Suess2016,Hess2021,Roberts2021}, which is mainly due to the existence of strong RFI above $z\approx0.27$. From the statistics of RFI in the FAST 19-beam array \citep{Zhang2023}, we can see that there is a lot of seriously strong RFI from communication and navigation satellites, over two broad frequency bands in 1150-1310 and 1465-1500\,MHz, corresponding to a redshift of $0.27\lesssim z\lesssim0.45$ and $0.11\lesssim z\lesssim0.14$, respectively. It is because the frequency band in 1310-1465\,MHz is very clear that FAST has already detected 27 OHMs in this band. In addition, the RFI is bearable in the frequency band of 1050-1150\,MHz or $0.45\lesssim z\lesssim0.55$. Therefore, it is feasible to search for potential OHMs within $0.45\lesssim z\lesssim0.55$ using FASHI data in the future.

Moreover, the FASHI data have detected 27 OHMs in the frequency range 1310-1465\,MHz with a typical spectral detection sensitivity of $\sim$1.50\,mJy at a spectral resolution of $\sim$6.4\,$\kms$ at 1.4\,GHz \citep{Zhang2023fashi}. The coverage of the current FASHI data is $\sim$10000\,deg$^2$. If FASHI observations cover $\sim$20000\,deg$^2$ in the future, FASHI is expected to obtain a total of 54 OHMs at the current detection sensitivity. If a longer integration time than the current FASHI drift survey is applied to each PSCz source, more OHMs would be extracted. Please note that the method of source identification in this work is to directly compare the PSCz catalog with the corresponding FASHI data cube to see if there are any emission lines (see Section\,\ref{sec:identification}). A blind and complete OHM identification throughout the FASHI data cube would expand the sample again, especially if OHMs can be hosted by galaxies with sub-LIRG far-IR luminosities. Thus, FASHI or FAST could provide great opportunities to detect more samples.

\section{Summary}
\label{sec:summary}

The \textbf{F}AST \textbf{A}ll \textbf{S}ky \textbf{H\,{\footnotesize{I}}} survey (FASHI) aims to cover the entire sky observable by the Five-hundred-meter Aperture Spherical radio Telescope (FAST), covering $\sim$22000 square degrees of declination between $-14^\circ$ and $+66^\circ$, and in the frequency range of 1000-1500\,MHz, with an expectation of detecting more than 100000 \HI sources finally. At present, FASHI has observed more than 7600 square degrees. It has a typical spectral detection sensitivity of $\sim$1.50\,mJy for a velocity resolution of $\sim$6.4\,$\kms$ at a frequency of $\sim$1.4\,GHz. As of now, a total of 41741 extragalactic \HI sources have been detected and released by FASHI.

To efficiently expand the OHM sample, we have directly cross-checked the PSCz catalog with the corresponding FASHI data cube. From 145 PSCz sources already covered by FASHI, we obtained 27 OHMs with a detection rate of 18.6\%, including 9 previously known and 18 new ones, within a redshift range of $0.14314\lesssim z_{\rm OH} \lesssim0.27656$. For 9 of the detected 27 OHMs, we also measured the hyperfine line ratio between the integrated fluxes of 1667 and 1665\,MHz. The detection rate is 33.3\% for the 1665\,MHz line. The line ratio $R_{1667:1665}$ ranges from 1.32 to 15.22, with an average of $R_{1667:1665}=4.74$, in agreement with the results reported by \citet{McBride2013}.

We now have a total of 154 OHMs, including the 18 OHMs newly detected by FASHI. The full sample yields an $L_{\rm OH}$ and $L_{\rm FIR}$ relationship with ${\rm log}L_{\rm OH}= (1.57\pm0.10){\rm log}L_{\rm FIR}-(15.80\pm1.19)$, which agrees well with the result of \citet{Darling2002a} that $\gamma = 1.57\pm0.11$ from 95 data points. This indicates that $\gamma$ remains almost the same when the OHM samples reach up to $\sim$1.6 times the previously known OHMs.

Of the detected 27 OHMs of this work, 13 OHMs are ULIRGs and 14 OHMs are LIRGs, ranging from $10^{11.70\pm0.03}$ to $10^{12.43\pm0.01} \Lsun$ with a median of $10^{12.00\pm0.02} \Lsun$. As expected, all 27 OHMs have relatively high FIR luminosities. This is because the OHM sample was selected by cross-correlation with the IRAS-selected PSCz. The multi-wavelength SED fitting results also show that the OH host galaxies for such 27 OHMs are massive ($M_\star \simeq 10^{11}M_\odot$) galaxies with {SFRs that place them above} the star-forming main sequence. The higher $A_V$ values for the OH hosts from the SED fit are also consistent with the high IR luminosities. On the other hand, not all ULIRGs have detectable OH emission, suggesting that the OH emission may be triggered within a specific stage of the merger or may have its visibility limited by orientation effects.

FAST, with its 19-beam array and UWB receiver, will be a powerful tool for observing more OHMs and unraveling their mystery in the future.

\section*{Acknowledgements}
\addcontentsline{toc}{section}{Acknowledgements}

This work is supported by the National Key R\&D Program of China (2022YFA1602901). CPZ acknowledges support by the West Light Foundation of the Chinese Academy of Sciences (CAS). CC and JLX are supported by the National Natural Science Foundation of China, Nos.\,11803044, 11933003, 12173045, and 11933011. This work is sponsored (in part) by the Chinese Academy of Sciences (CAS), through a grant to the CAS South America Center for Astronomy (CASSACA). We acknowledge the science research grants from the China Manned Space Project with No.\,CMS-CSST-2021-A05. We also wish to thank the anonymous referee for comments that improved the clarity of the paper.

FAST is a Chinese national mega-science facility, operated by the National Astronomical Observatories of Chinese Academy of Sciences (NAOC). SDSS acknowledges support and resources from the Center for High-Performance Computing at the University of Utah. The SDSS web site is www.sdss4.org. SDSS is managed by the Astrophysical Research Consortium for the Participating Institutions of the SDSS Collaboration including the Brazilian Participation Group, the Carnegie Institution for Science, Carnegie Mellon University, Center for Astrophysics | Harvard \& Smithsonian (CfA), the Chilean Participation Group, the French Participation Group, Instituto de Astrofísica de Canarias, The Johns Hopkins University, Kavli Institute for the Physics and Mathematics of the Universe (IPMU) / University of Tokyo, the Korean Participation Group, Lawrence Berkeley National Laboratory, Leibniz Institut für Astrophysik Potsdam (AIP), Max-Planck-Institut für Astronomie (MPIA Heidelberg), Max-Planck-Institut für Astrophysik (MPA Garching), Max-Planck-Institut für Extraterrestrische Physik (MPE), National Astronomical Observatories of China, New Mexico State University, New York University, University of Notre Dame, Observatório Nacional / MCTI, The Ohio State University, Pennsylvania State University, Shanghai Astronomical Observatory, United Kingdom Participation Group, Universidad Nacional Autónoma de México, University of Arizona, University of Colorado Boulder, University of Oxford, University of Portsmouth, University of Utah, University of Virginia, University of Washington, University of Wisconsin, Vanderbilt University, and Yale University.

\bibliography{references}{}

\begin{thebibliography}{}
\expandafter\ifx\csname natexlab\endcsname\relax\def\natexlab#1{#1}\fi
\providecommand{\url}[1]{\href{#1}{#1}}
\providecommand{\dodoi}[1]{doi:~\href{http://doi.org/#1}{\nolinkurl{#1}}}
\providecommand{\doeprint}[1]{\href{http://ascl.net/#1}{\nolinkurl{http://ascl.net/#1}}}
\providecommand{\doarXiv}[1]{\href{https://arxiv.org/abs/#1}{\nolinkurl{https://arxiv.org/abs/#1}}}

\bibitem[{{Alatalo} {et~al.}(2016){Alatalo}, {Cales}, {Rich}, {Appleton},
  {Kewley}, {Lacy}, {Lanz}, {Medling}, \& {Nyland}}]{Alatalo2016}
{Alatalo}, K., {Cales}, S.~L., {Rich}, J.~A., {et~al.} 2016, \apjs, 224, 38,
  \dodoi{10.3847/0067-0049/224/2/38}

\bibitem[{{Andreasian} \& {Alloin}(1994)}]{Andreasian1994}
{Andreasian}, N., \& {Alloin}, D. 1994, \aaps, 107, 23

\bibitem[{{Baan}(1989)}]{Baan1989}
{Baan}, W.~A. 1989, \apj, 338, 804, \dodoi{10.1086/167237}

\bibitem[{{Baan}(1991)}]{Baan1991}
{Baan}, W.~A. 1991, in Astronomical Society of the Pacific Conference Series,
  Vol.~16, Atoms, Ions and Molecules: New Results in Spectral Line
  Astrophysics, ed. A.~D. {Haschick} \& P.~T.~P. {Ho}, 45

\bibitem[{{Baan} {et~al.}(1992){Baan}, {Rhoads}, {Fisher}, {Altschuler}, \&
  {Haschick}}]{Baan1992}
{Baan}, W.~A., {Rhoads}, J., {Fisher}, K., {Altschuler}, D.~R., \& {Haschick},
  A. 1992, \apjl, 396, L99, \dodoi{10.1086/186526}

\bibitem[{{Baan} {et~al.}(1982){Baan}, {Wood}, \& {Haschick}}]{Baan1982}
{Baan}, W.~A., {Wood}, P.~A.~D., \& {Haschick}, A.~D. 1982, \apjl, 260, L49,
  \dodoi{10.1086/183868}

\bibitem[{{Briggs}(1998)}]{Briggs1998}
{Briggs}, F.~H. 1998, \aap, 336, 815, \dodoi{10.48550/arXiv.astro-ph/9710143}

\bibitem[{{Burdiuzha} \& {Vikulov}(1990)}]{Burdiuzha1990}
{Burdiuzha}, V.~V., \& {Vikulov}, K.~A. 1990, \mnras, 244, 86

\bibitem[{{Clements} {et~al.}(1996){Clements}, {Sutherland}, {McMahon}, \&
  {Saunders}}]{Clements1996}
{Clements}, D.~L., {Sutherland}, W.~J., {McMahon}, R.~G., \& {Saunders}, W.
  1996, \mnras, 279, 477, \dodoi{10.1093/mnras/279.2.477}

\bibitem[{{Darling}(2007)}]{Darling2007}
{Darling}, J. 2007, \apjl, 669, L9, \dodoi{10.1086/523756}

\bibitem[{{Darling} \& {Giovanelli}(2000)}]{Darling2000}
{Darling}, J., \& {Giovanelli}, R. 2000, \aj, 119, 3003, \dodoi{10.1086/301403}

\bibitem[{{Darling} \& {Giovanelli}(2001)}]{Darling2001}
---. 2001, \aj, 121, 1278, \dodoi{10.1086/319413}

\bibitem[{{Darling} \& {Giovanelli}(2002)}]{Darling2002a}
---. 2002, \aj, 124, 100, \dodoi{10.1086/341166}

\bibitem[{{Darling} \& {Giovanelli}(2006)}]{Darling2006}
---. 2006, \aj, 132, 2596, \dodoi{10.1086/508513}

\bibitem[{{Duarte Puertas} {et~al.}(2017){Duarte Puertas}, {Vilchez},
  {Iglesias-P{\'a}ramo}, {Kehrig}, {P{\'e}rez-Montero}, \&
  {Rosales-Ortega}}]{Duarte2017}
{Duarte Puertas}, S., {Vilchez}, J.~M., {Iglesias-P{\'a}ramo}, J., {et~al.}
  2017, \aap, 599, A71, \dodoi{10.1051/0004-6361/201629044}

\bibitem[{{Giovanelli} \& {Haynes}(2015)}]{Giovanelli2015}
{Giovanelli}, R., \& {Haynes}, M.~P. 2015, \aapr, 24, 1,
  \dodoi{10.1007/s00159-015-0085-3}

\bibitem[{{Glowacki} {et~al.}(2022){Glowacki}, {Collier}, {Kazemi-Moridani},
  {Frank}, {Roberts}, {Darling}, {Kl{\"o}ckner}, {Adams}, {Baker}, {Bershady},
  {Blecher}, {Blyth}, {Bowler}, {Catinella}, {Chemin}, {Crawford}, {Cress},
  {Dav{\'e}}, {Deane}, {de Blok}, {Delhaize}, {Duncan}, {Elson}, {February},
  {Gawiser}, {Hatfield}, {Healy}, {Henning}, {Hess}, {Heywood}, {Holwerda},
  {Hoosain}, {Hughes}, {Hutchens}, {Jarvis}, {Kannappan}, {Katz},
  {Kere{\v{s}}}, {Korsaga}, {Kraan-Korteweg}, {Lah}, {Lochner}, {Maddox},
  {Makhathini}, {Meurer}, {Meyer}, {Obreschkow}, {Oh}, {Oosterloo}, {Oppor},
  {Pan}, {Pisano}, {Randriamiarinarivo}, {Ravindranath}, {Schr{\"o}der},
  {Skelton}, {Smirnov}, {Smith}, {Somerville}, {Srianand}, {Staveley-Smith},
  {Tanaka}, {Vaccari}, {van Driel}, {Verheijen}, {Walter}, {Wu}, \&
  {Zwaan}}]{Glowacki2022}
{Glowacki}, M., {Collier}, J.~D., {Kazemi-Moridani}, A., {et~al.} 2022, \apjl,
  931, L7, \dodoi{10.3847/2041-8213/ac63b0}

\bibitem[{{Haynes} {et~al.}(2018){Haynes}, {Giovanelli}, {Kent}, {Adams},
  {Balonek}, {Craig}, {Fertig}, {Finn}, {Giovanardi}, {Hallenbeck}, {Hess},
  {Hoffman}, {Huang}, {Jones}, {Koopmann}, {Kornreich}, {Leisman}, {Miller},
  {Moorman}, {O'Connor}, {O'Donoghue}, {Papastergis}, {Troischt}, {Stark}, \&
  {Xiao}}]{Haynes2018}
{Haynes}, M.~P., {Giovanelli}, R., {Kent}, B.~R., {et~al.} 2018, \apj, 861, 49,
  \dodoi{10.3847/1538-4357/aac956}

\bibitem[{Helou \& Walker(1988)}]{IRAS1988}
Helou, G., \& Walker, D.~W. 1988, 0.
\newblock \url{https://www.osti.gov/biblio/6458812}

\bibitem[{{Henkel} {et~al.}(1991){Henkel}, {Baan}, \&
  {Mauersberger}}]{Henkel1991}
{Henkel}, C., {Baan}, W.~A., \& {Mauersberger}, R. 1991, \aapr, 3, 47,
  \dodoi{10.1007/BF00873457}

\bibitem[{{Henkel} {et~al.}(1987){Henkel}, {Guesten}, \& {Baan}}]{Henkel1987}
{Henkel}, C., {Guesten}, R., \& {Baan}, W.~A. 1987, \aap, 185, 14

\bibitem[{{Henkel} \& {Wilson}(1990)}]{Henkel1990}
{Henkel}, C., \& {Wilson}, T.~L. 1990, \aap, 229, 431

\bibitem[{{Hess} {et~al.}(2021){Hess}, {Roberts}, {D{\'e}nes}, {Adebahr},
  {Darling}, {Adams}, {de Blok}, {Kutkin}, {Lucero}, {Morganti}, {Moss},
  {Oosterloo}, {Schulz}, {van der Hulst}, {Coolen}, {Damstra}, {Ivashina},
  {Loose}, {Maan}, {Mika}, {Mulder}, {Norden}, {Oostrum}, {Ruiter}, {van
  Leeuwen}, {Vermaas}, {Vohl}, {Wijnholds}, \& {Ziemke}}]{Hess2021}
{Hess}, K.~M., {Roberts}, H., {D{\'e}nes}, H., {et~al.} 2021, \aap, 647, A193,
  \dodoi{10.1051/0004-6361/202040019}

\bibitem[{{Hou} {et~al.}(2009){Hou}, {Wu}, \& {Han}}]{Hou2009}
{Hou}, L.~G., {Wu}, X.-B., \& {Han}, J.~L. 2009, \apj, 704, 789,
  \dodoi{10.1088/0004-637X/704/1/789}

\bibitem[{{Jarvis} {et~al.}(2023){Jarvis}, {Heywood}, {Jewell}, {Deane},
  {Kl{\"o}ckner}, {Ponomareva}, {Maddox}, {Baker}, {Bianchetti}, {Hess},
  {Roberts}, {Rodighiero}, {Ruffa}, {Sinigaglia}, {Varadaraj}, {Whittam},
  {Adams}, {Baes}, {Murphy}, {Pan}, \& {Vaccari}}]{Jarvis2023}
{Jarvis}, M.~J., {Heywood}, I., {Jewell}, S.~M., {et~al.} 2023, arXiv e-prints,
  arXiv:2312.04345.
\newblock \doarXiv{2312.04345}

\bibitem[{{Jiang} {et~al.}(2019){Jiang}, {Yue}, {Gan}, {Yao}, {Li}, {Pan},
  {Sun}, {Yu}, {Liu}, {Tang}, {Qian}, {Lu}, {Yan}, {Peng}, {Zhang}, {Wang},
  {Li}, \& {Li}}]{Jiang2019}
{Jiang}, P., {Yue}, Y., {Gan}, H., {et~al.} 2019, Science China Physics,
  Mechanics, and Astronomy, 62, 959502, \dodoi{10.1007/s11433-018-9376-1}

\bibitem[{{Jiang} {et~al.}(2020){Jiang}, {Tang}, {Hou}, {Liu}, {Kr{\v{c}}o},
  {Qian}, {Sun}, {Ching}, {Liu}, {Duan}, {Yue}, {Gan}, {Yao}, {Li}, {Pan},
  {Yu}, {Liu}, {Li}, {Peng}, {Yan}, \& {FAST Collaboration}}]{Jiang2020}
{Jiang}, P., {Tang}, N.-Y., {Hou}, L.-G., {et~al.} 2020, Research in Astronomy
  and Astrophysics, 20, 064, \dodoi{10.1088/1674-4527/20/5/64}

\bibitem[{{Jing} {et~al.}(2024){Jing}, {Wang}, {Xu}, {Liu}, {Chen}, {Liang},
  {Xu}, {Cao}, {Wang}, {Hu}, {Zhang}, {Guo}, {Gao}, {Ai}, {Gan}, {Gao}, {Han},
  {Hou}, {Hou}, {Jiang}, {Kong}, {Li}, {Liu}, {Shao}, {Pan}, {Pan}, {Qian},
  {Sun}, {Tang}, {Yang}, {Zhang}, {Zhang}, \& {Zhu}}]{Jing2024}
{Jing}, Y., {Wang}, J., {Xu}, C., {et~al.} 2024, Science China Physics,
  Mechanics, and Astronomy, 67, 259514, \dodoi{10.1007/s11433-023-2333-8}

\bibitem[{{Kim} {et~al.}(1998){Kim}, {Veilleux}, \& {Sanders}}]{Kim1998}
{Kim}, D.~C., {Veilleux}, S., \& {Sanders}, D.~B. 1998, \apj, 508, 627,
  \dodoi{10.1086/306409}

\bibitem[{{Lin} {et~al.}(2018){Lin}, {Huang}, \& {Chen}}]{Lin2018}
{Lin}, Y.-T., {Huang}, H.-J., \& {Chen}, Y.-C. 2018, \aj, 155, 188,
  \dodoi{10.3847/1538-3881/aab5b4}

\bibitem[{{Liu} {et~al.}(2024){Liu}, {Wang}, {Jing}, {Zhang}, {Xu}, {Liang},
  {Chen}, {Tang}, \& {Yang}}]{Liu2024}
{Liu}, Z., {Wang}, J., {Jing}, Y., {et~al.} 2024, arXiv e-prints,
  arXiv:2406.08278, \dodoi{10.48550/arXiv.2406.08278}

\bibitem[{{Lo}(2005)}]{Lo2005}
{Lo}, K.~Y. 2005, \araa, 43, 625,
  \dodoi{10.1146/annurev.astro.41.011802.094927}

\bibitem[{{Lockett} \& {Elitzur}(2008)}]{Lockett2008}
{Lockett}, P., \& {Elitzur}, M. 2008, \apj, 677, 985, \dodoi{10.1086/533429}

\bibitem[{{McBride} {et~al.}(2013){McBride}, {Heiles}, \&
  {Elitzur}}]{McBride2013}
{McBride}, J., {Heiles}, C., \& {Elitzur}, M. 2013, \apj, 774, 35,
  \dodoi{10.1088/0004-637X/774/1/35}

\bibitem[{{Murata} {et~al.}(2017){Murata}, {Yamada}, {Oyabu}, {Kaneda},
  {Ishihara}, {Yamagishi}, {Kokusho}, \& {Takeuchi}}]{Murata2017}
{Murata}, K.~L., {Yamada}, R., {Oyabu}, S., {et~al.} 2017, \mnras, 472, 39,
  \dodoi{10.1093/mnras/stx1902}

\bibitem[{{Nan} {et~al.}(2011){Nan}, {Li}, {Jin}, {Wang}, {Zhu}, {Zhu},
  {Zhang}, {Yue}, \& {Qian}}]{Nan2011}
{Nan}, R., {Li}, D., {Jin}, C., {et~al.} 2011, International Journal of Modern
  Physics D, 20, 989, \dodoi{10.1142/S0218271811019335}

\bibitem[{{Nikutta} {et~al.}(2014){Nikutta}, {Hunt-Walker}, {Nenkova},
  {Ivezi{\'c}}, \& {Elitzur}}]{Nikutta2014}
{Nikutta}, R., {Hunt-Walker}, N., {Nenkova}, M., {Ivezi{\'c}}, {\v{Z}}., \&
  {Elitzur}, M. 2014, \mnras, 442, 3361, \dodoi{10.1093/mnras/stu1087}

\bibitem[{{Osterbrock}(1989)}]{Osterbrock1989}
{Osterbrock}, D.~E. 1989, {Astrophysics of gaseous nebulae and active galactic
  nuclei}

\bibitem[{{Pihlstr{\"o}m}(2007)}]{Pihlstrom2007}
{Pihlstr{\"o}m}, Y.~M. 2007, in Astrophysical Masers and their Environments,
  ed. J.~M. {Chapman} \& W.~A. {Baan}, Vol. 242, 446--451,
  \dodoi{10.1017/S1743921307013579}

\bibitem[{{Radford}(1964)}]{Radford1964}
{Radford}, H.~E. 1964, \prl, 13, 534, \dodoi{10.1103/PhysRevLett.13.534}

\bibitem[{{Roberts} {et~al.}(2021){Roberts}, {Darling}, \&
  {Baker}}]{Roberts2021}
{Roberts}, H., {Darling}, J., \& {Baker}, A.~J. 2021, \apj, 911, 38,
  \dodoi{10.3847/1538-4357/abe944}

\bibitem[{{Salim} {et~al.}(2016){Salim}, {Lee}, {Janowiecki}, {da Cunha},
  {Dickinson}, {Boquien}, {Burgarella}, {Salzer}, \& {Charlot}}]{Salim2016}
{Salim}, S., {Lee}, J.~C., {Janowiecki}, S., {et~al.} 2016, \apjs, 227, 2,
  \dodoi{10.3847/0067-0049/227/1/2}

\bibitem[{{Sanders} \& {Mirabel}(1996)}]{Sanders1996}
{Sanders}, D.~B., \& {Mirabel}, I.~F. 1996, \araa, 34, 749,
  \dodoi{10.1146/annurev.astro.34.1.749}

\bibitem[{{Saunders} {et~al.}(2000){Saunders}, {Sutherland}, {Maddox},
  {Keeble}, {Oliver}, {Rowan-Robinson}, {McMahon}, {Efstathiou}, {Tadros},
  {White}, {Frenk}, {Carrami{\~n}ana}, \& {Hawkins}}]{Saunders2000}
{Saunders}, W., {Sutherland}, W.~J., {Maddox}, S.~J., {et~al.} 2000, \mnras,
  317, 55, \dodoi{10.1046/j.1365-8711.2000.03528.x}

\bibitem[{{Skinner} {et~al.}(1997){Skinner}, {Smith}, {Sturm}, {Barlow},
  {Cohen}, \& {Stacey}}]{Skinner1997}
{Skinner}, C.~J., {Smith}, H.~A., {Sturm}, E., {et~al.} 1997, \nat, 386, 472,
  \dodoi{10.1038/386472a0}

\bibitem[{{Suess} {et~al.}(2016){Suess}, {Darling}, {Haynes}, \&
  {Giovanelli}}]{Suess2016}
{Suess}, K.~A., {Darling}, J., {Haynes}, M.~P., \& {Giovanelli}, R. 2016,
  \mnras, 459, 220, \dodoi{10.1093/mnras/stw666}

\bibitem[{{Toba} {et~al.}(2014){Toba}, {Oyabu}, {Matsuhara}, {Malkan},
  {Gandhi}, {Nakagawa}, {Isobe}, {Shirahata}, {Oi}, {Ohyama}, {Takita},
  {Yamauchi}, \& {Yano}}]{Toba2014}
{Toba}, Y., {Oyabu}, S., {Matsuhara}, H., {et~al.} 2014, \apj, 788, 45,
  \dodoi{10.1088/0004-637X/788/1/45}

\bibitem[{{Veilleux} {et~al.}(1999){Veilleux}, {Kim}, \&
  {Sanders}}]{Veilleux1999}
{Veilleux}, S., {Kim}, D.~C., \& {Sanders}, D.~B. 1999, \apj, 522, 113,
  \dodoi{10.1086/307634}

\bibitem[{{Veilleux} {et~al.}(1995){Veilleux}, {Kim}, {Sanders}, {Mazzarella},
  \& {Soifer}}]{Veilleux1995}
{Veilleux}, S., {Kim}, D.~C., {Sanders}, D.~B., {Mazzarella}, J.~M., \&
  {Soifer}, B.~T. 1995, \apjs, 98, 171, \dodoi{10.1086/192158}

\bibitem[{{Weedman} \& {Houck}(2008)}]{Weedman2008}
{Weedman}, D.~W., \& {Houck}, J.~R. 2008, \apj, 686, 127,
  \dodoi{10.1086/591123}

\bibitem[{{Wiggins} {et~al.}(2016){Wiggins}, {Migenes}, \&
  {Smidt}}]{Wiggins2016}
{Wiggins}, B.~K., {Migenes}, V., \& {Smidt}, J.~M. 2016, \apj, 816, 55,
  \dodoi{10.3847/0004-637X/816/2/55}

\bibitem[{{Willett}(2012)}]{Willett2012}
{Willett}, K.~W. 2012, in Cosmic Masers - from OH to H0, ed. R.~S. {Booth},
  W.~H.~T. {Vlemmings}, \& E.~M.~L. {Humphreys}, Vol. 287, 345--349,
  \dodoi{10.1017/S1743921312007284}

\bibitem[{{Wright}(2006)}]{Wright2006}
{Wright}, E.~L. 2006, \pasp, 118, 1711, \dodoi{10.1086/510102}

\bibitem[{{Yun} {et~al.}(2001){Yun}, {Reddy}, \& {Condon}}]{Yun2001}
{Yun}, M.~S., {Reddy}, N.~A., \& {Condon}, J.~J. 2001, \apj, 554, 803,
  \dodoi{10.1086/323145}

\bibitem[{{Zhang} {et~al.}(2022){Zhang}, {Xu}, {Wang}, {Jing}, {Liu}, {Zhu}, \&
  {Jiang}}]{Zhang2022rfi}
{Zhang}, C.-P., {Xu}, J.-L., {Wang}, J., {et~al.} 2022, Research in Astronomy
  and Astrophysics, 22, 025015, \dodoi{10.1088/1674-4527/ac3f2d}

\bibitem[{{Zhang} {et~al.}(2023){Zhang}, {Jiang}, {Zhu}, {Pan}, {Cheng}, {Liu},
  {Zhu}, {Sun}, \& {FAST Collaboration}}]{Zhang2023}
{Zhang}, C.-P., {Jiang}, P., {Zhu}, M., {et~al.} 2023, Research in Astronomy
  and Astrophysics, 23, 075016, \dodoi{10.1088/1674-4527/acd58e}

\bibitem[{Zhang {et~al.}(2024)Zhang, Zhu, Jiang, Cheng, Wang, Wang, Xu, Liu,
  Yu, Qian, Yu, Ai, Jing, Xu, Liu, Guan, Sun, Yang, Huang, Hao, \&
  Collaboration}]{Zhang2023fashi}
Zhang, C.-P., Zhu, M., Jiang, P., {et~al.} 2024, SCPMA, 67, 219511,
  \dodoi{10.1007/s11433-023-2219-7}

\end{thebibliography}
\bibliographystyle{aasjournal}



  
 
  \begin{figure*}[htp]
 \centering
 \renewcommand{\thefigure}{\arabic{figure}}
 \addtocounter{figure}{-0}
 \includegraphics[height=0.25\textwidth, angle=0]{./figures/spec_172.pdf}
 \includegraphics[height=0.31\textwidth, angle=0]{./figures/overlay_172.pdf}
 \includegraphics[height=0.29\textwidth, angle=0]{./figures/overlay_172_small.pdf}
 \includegraphics[height=0.25\textwidth, angle=0]{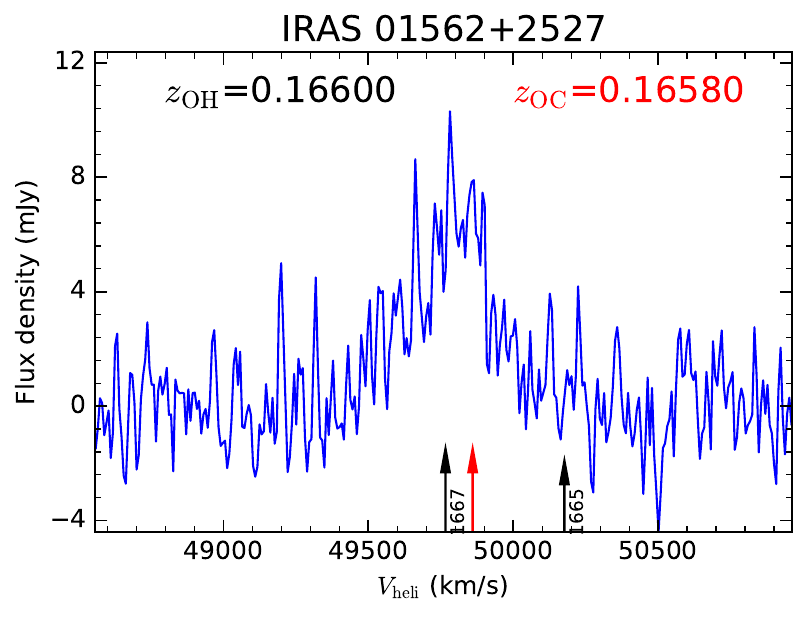}
 \includegraphics[height=0.31\textwidth, angle=0]{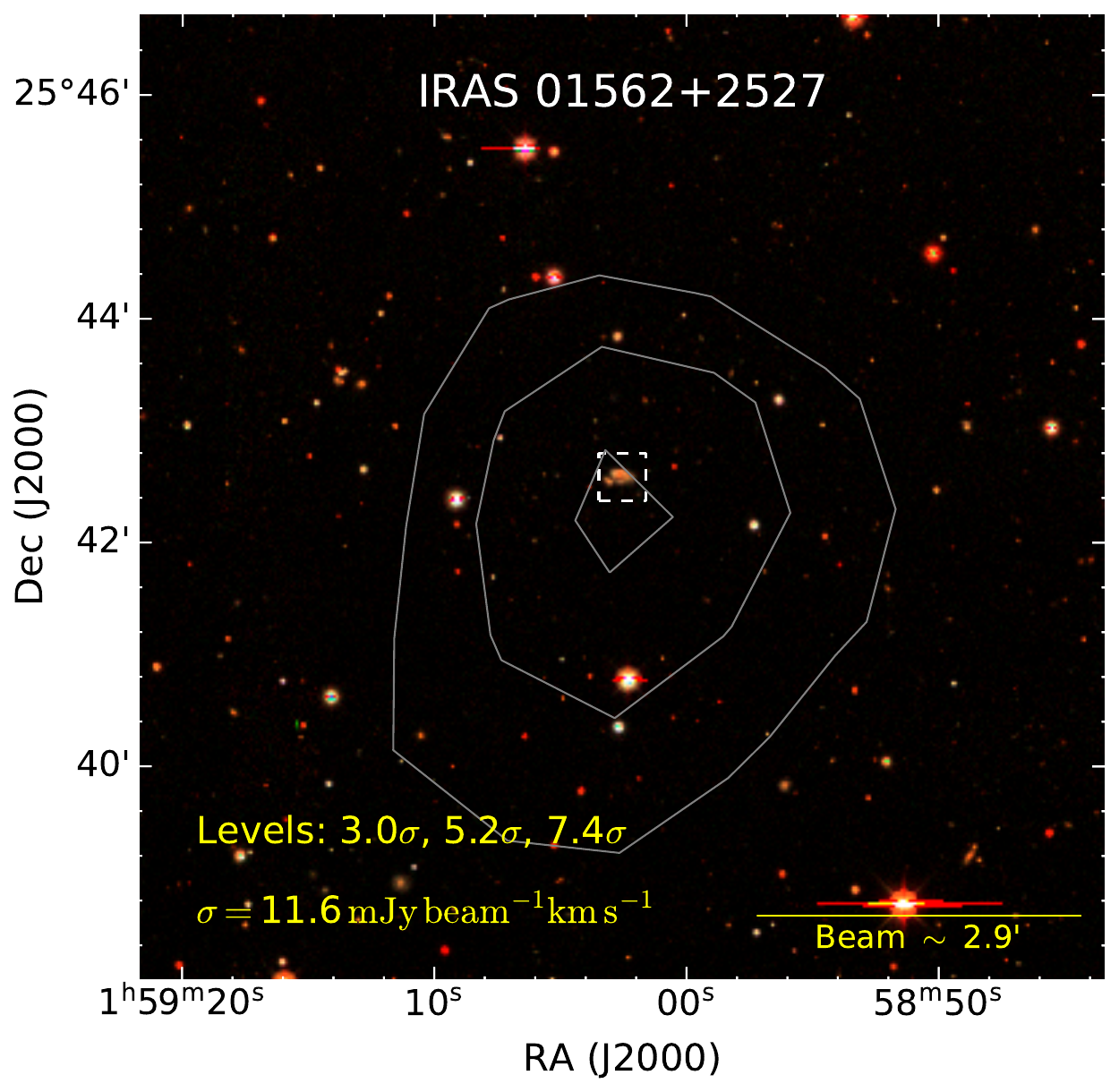}
 \includegraphics[height=0.29\textwidth, angle=0]{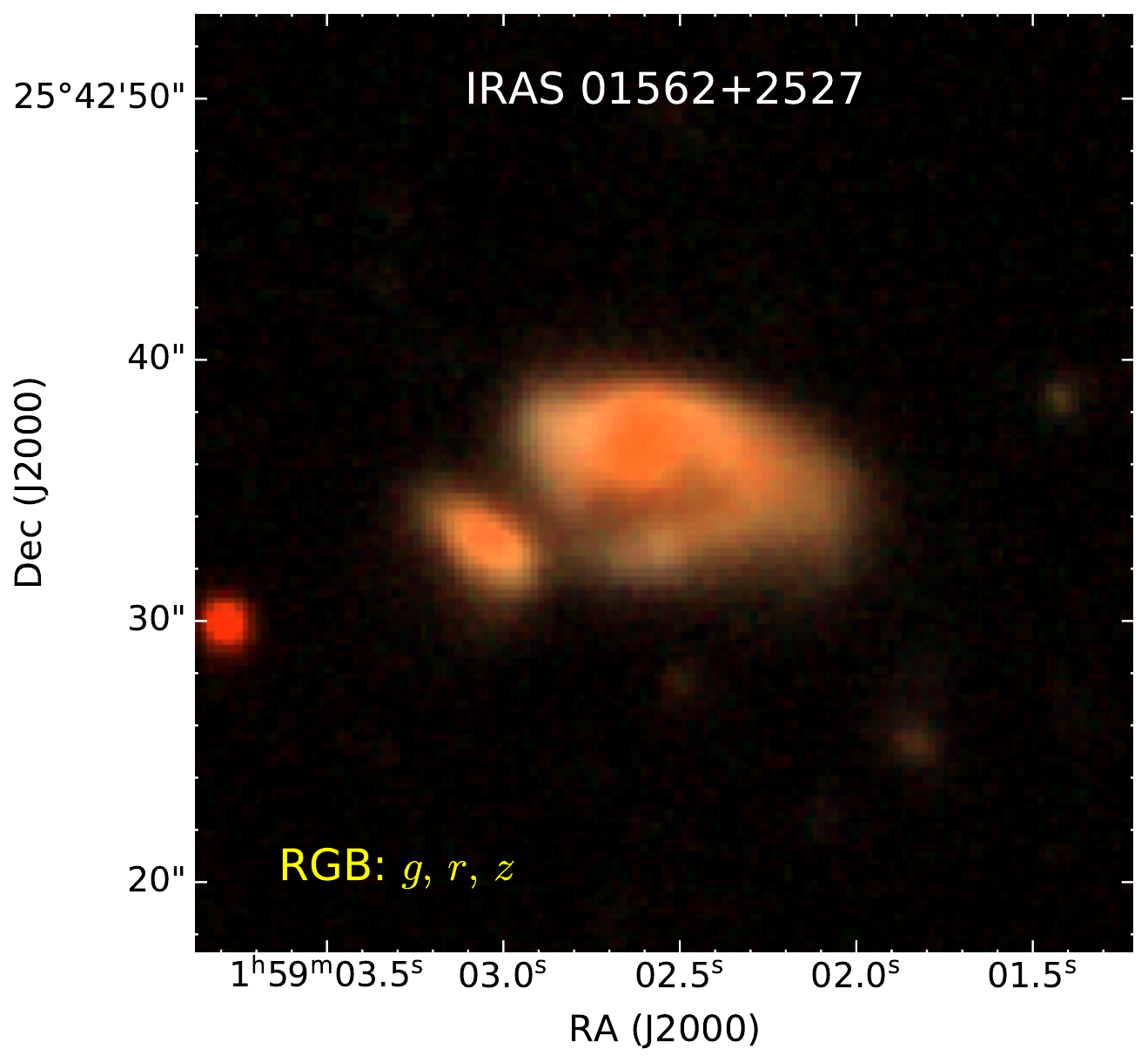}
 \includegraphics[height=0.25\textwidth, angle=0]{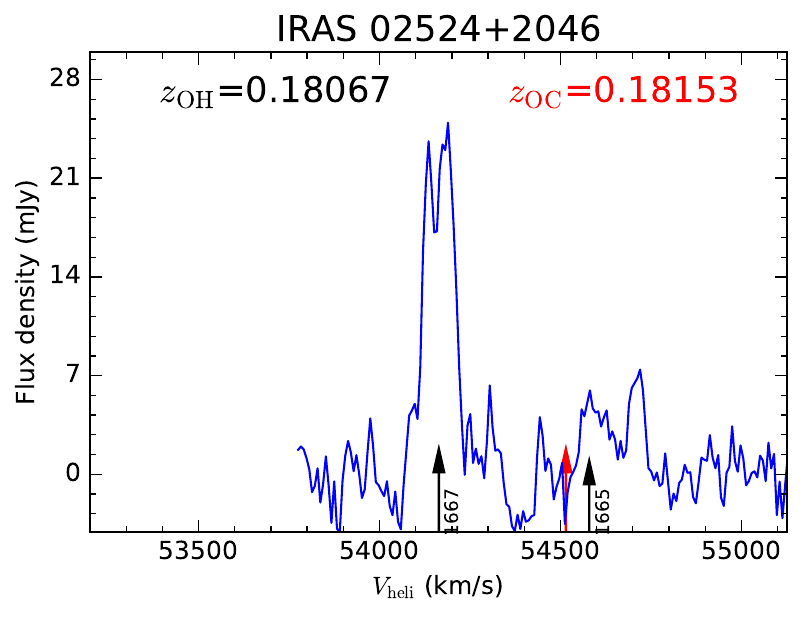}
 \includegraphics[height=0.31\textwidth, angle=0]{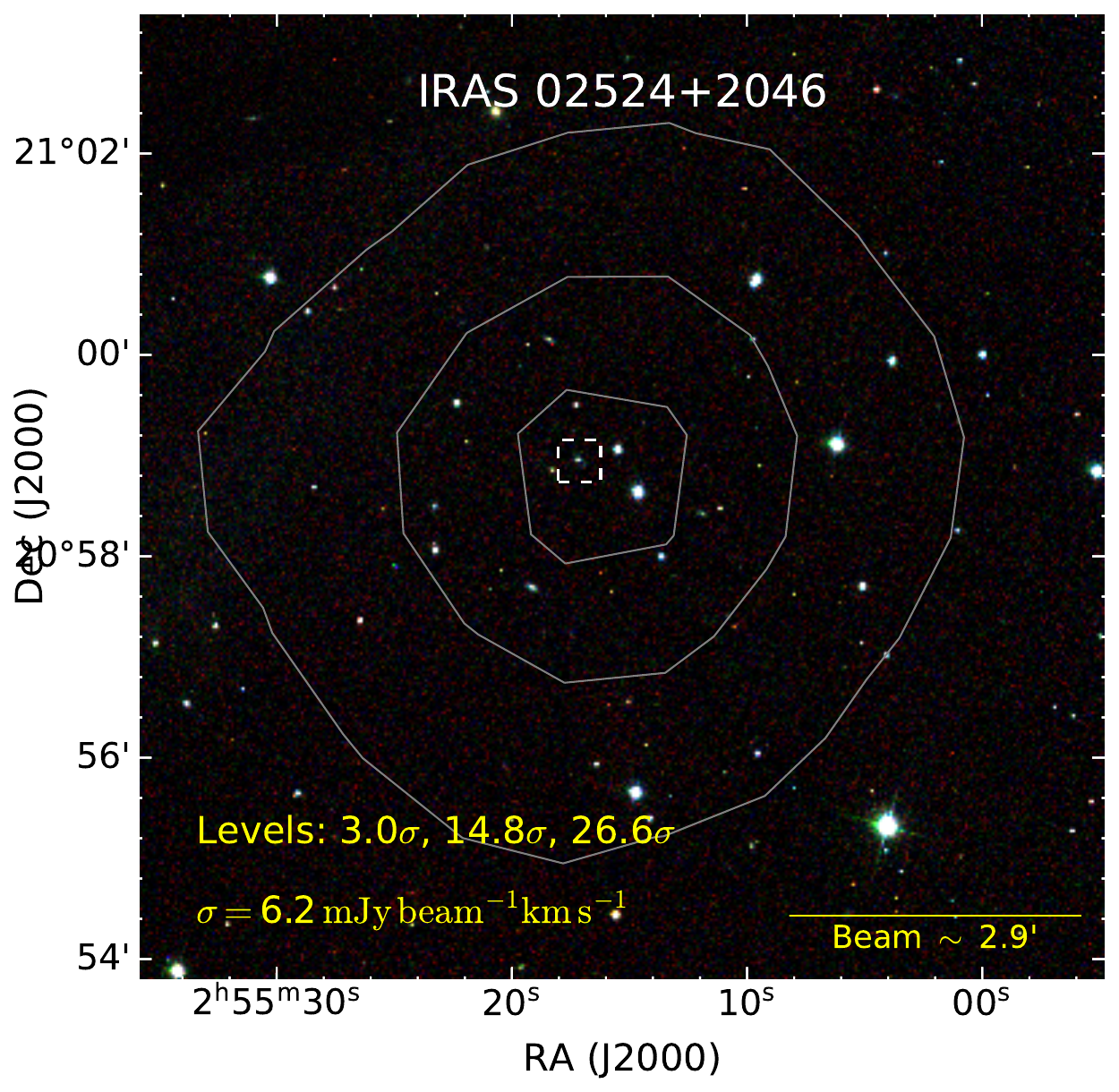}
 \includegraphics[height=0.29\textwidth, angle=0]{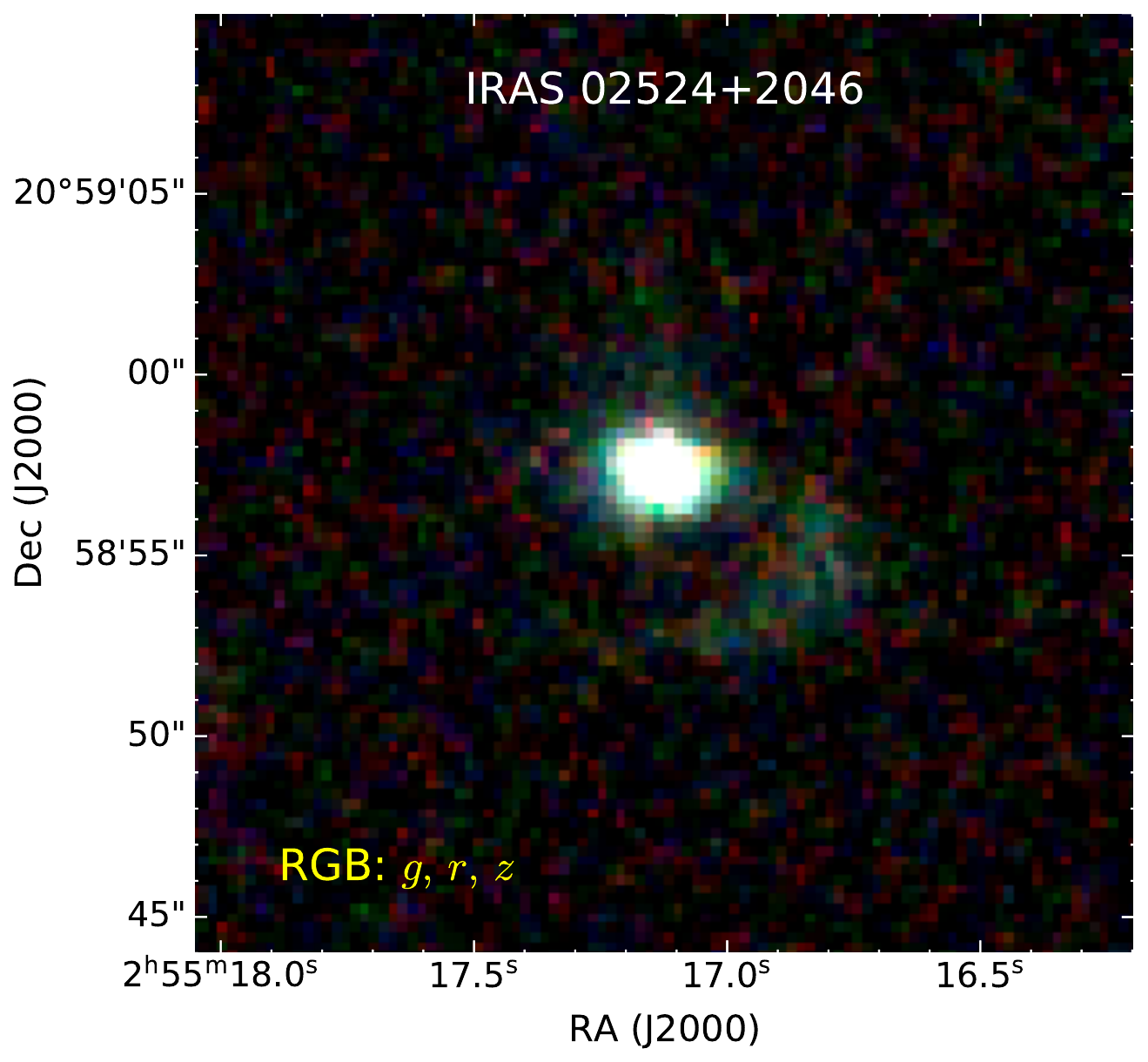}
 \includegraphics[height=0.25\textwidth, angle=0]{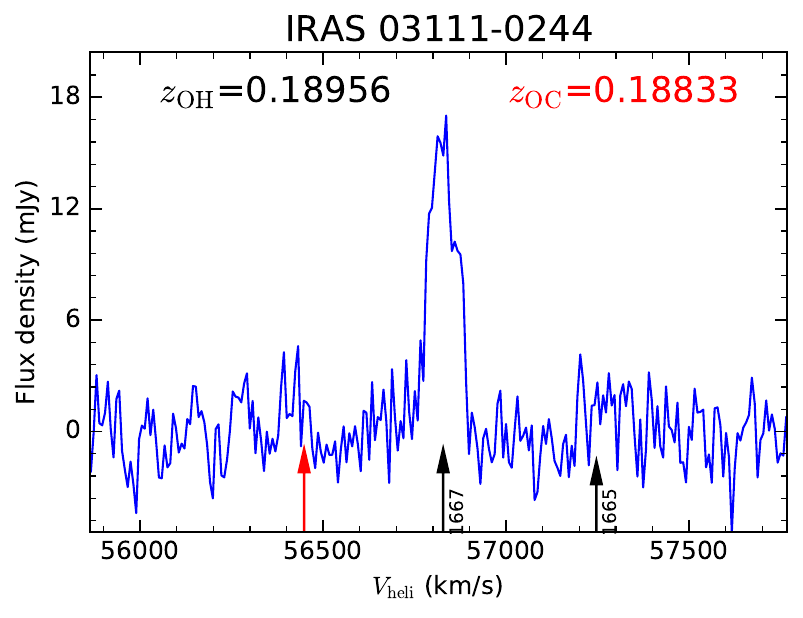}
 \includegraphics[height=0.31\textwidth, angle=0]{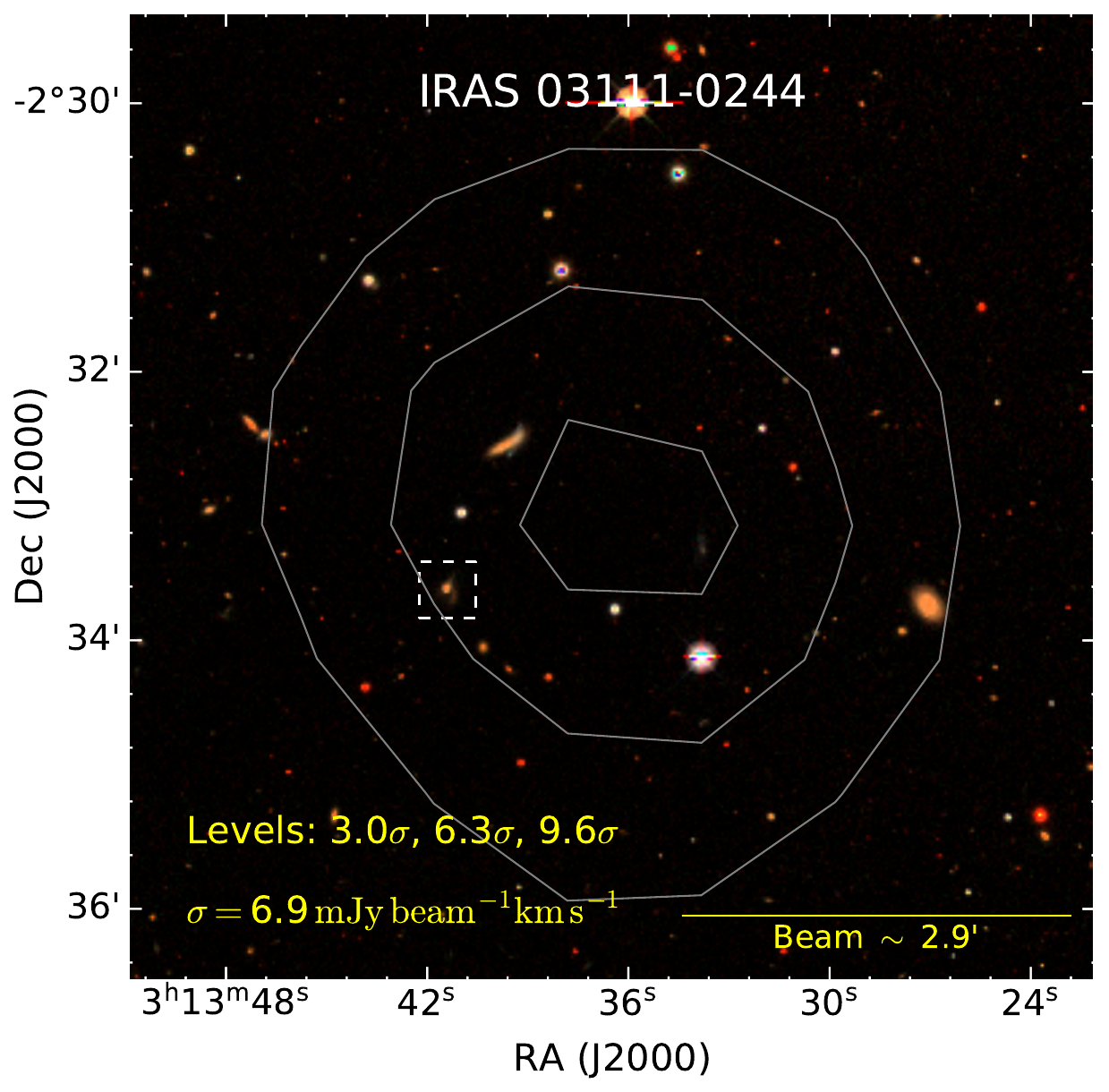}
 \includegraphics[height=0.29\textwidth, angle=0]{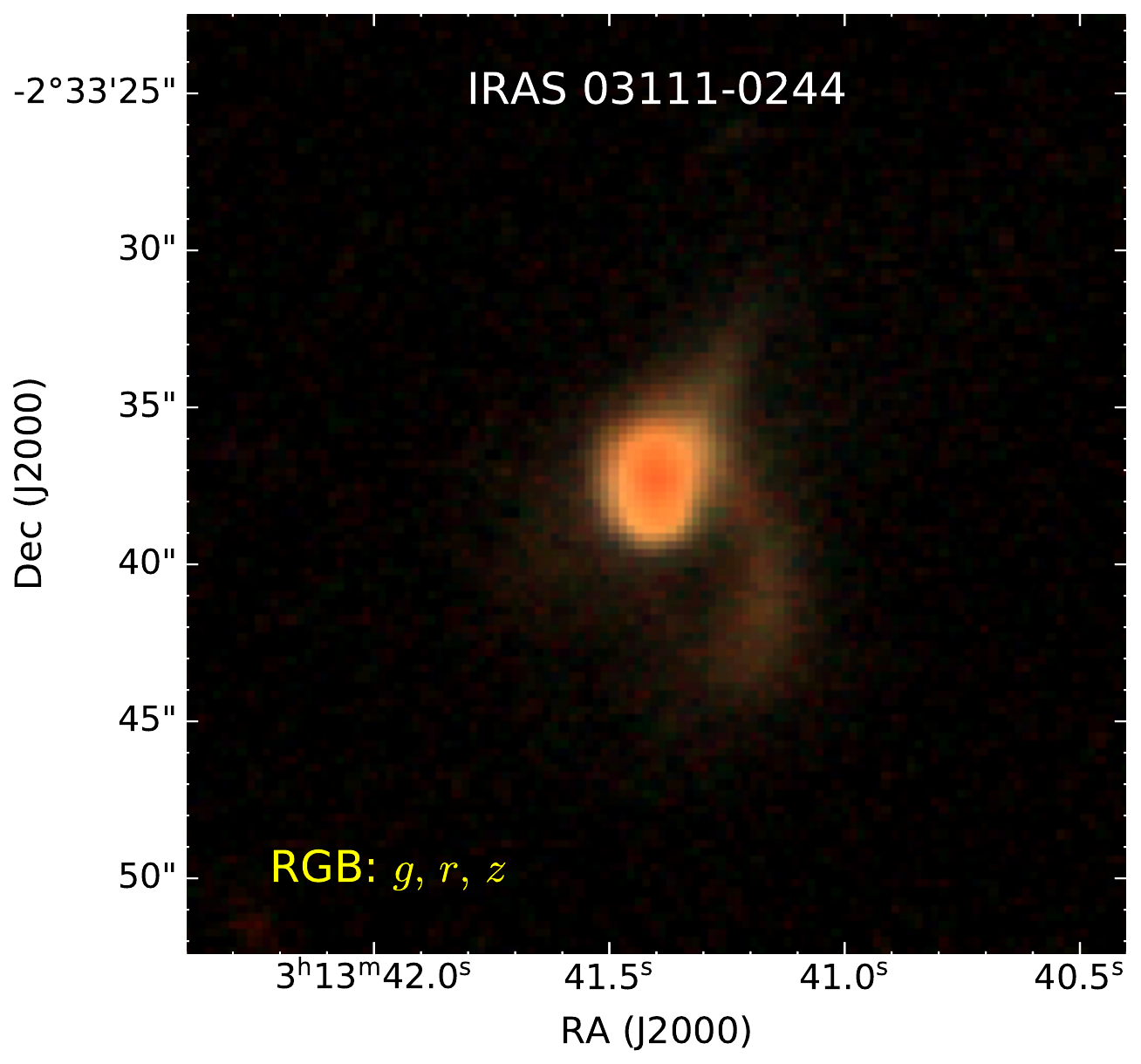}
 \caption{See caption in Figure\,\ref{Fig:IRAS00256-0208}}
 \label{Fig:IRAS-Continued}
 \end{figure*} 

 \begin{figure*}[htp]
 \centering
 \renewcommand{\thefigure}{\arabic{figure} (Continued)}
 \addtocounter{figure}{-1}
 \includegraphics[height=0.25\textwidth, angle=0]{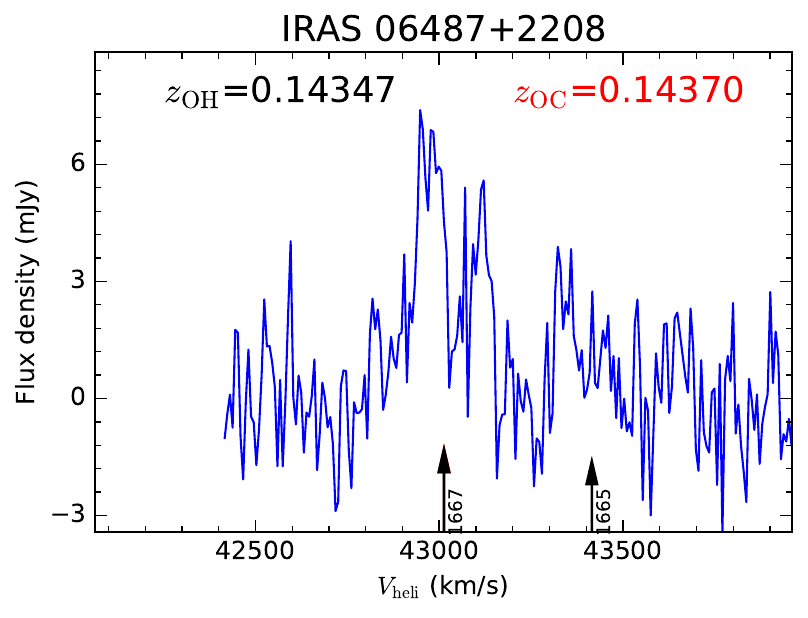}
 \includegraphics[height=0.31\textwidth, angle=0]{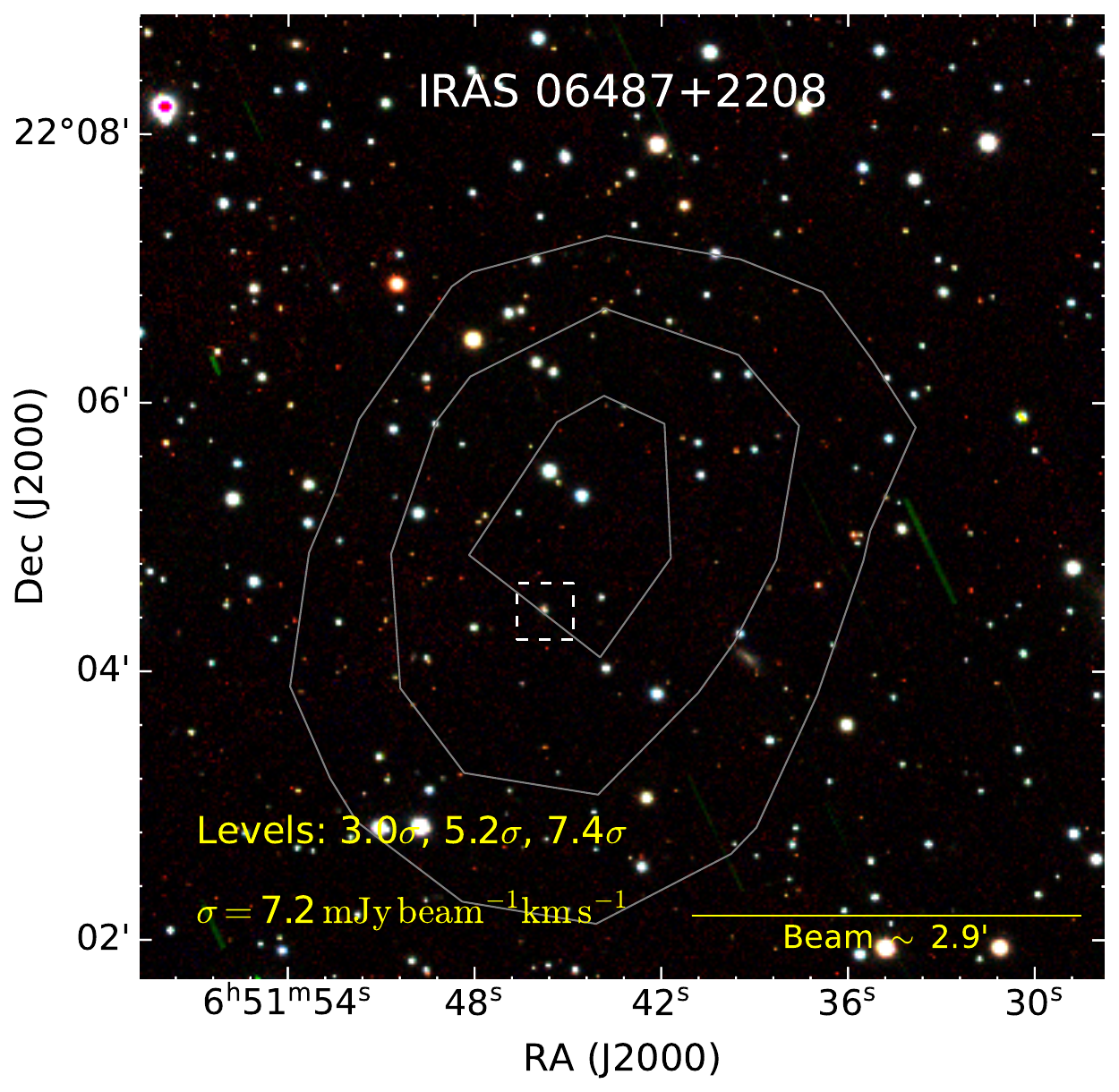}
 \includegraphics[height=0.29\textwidth, angle=0]{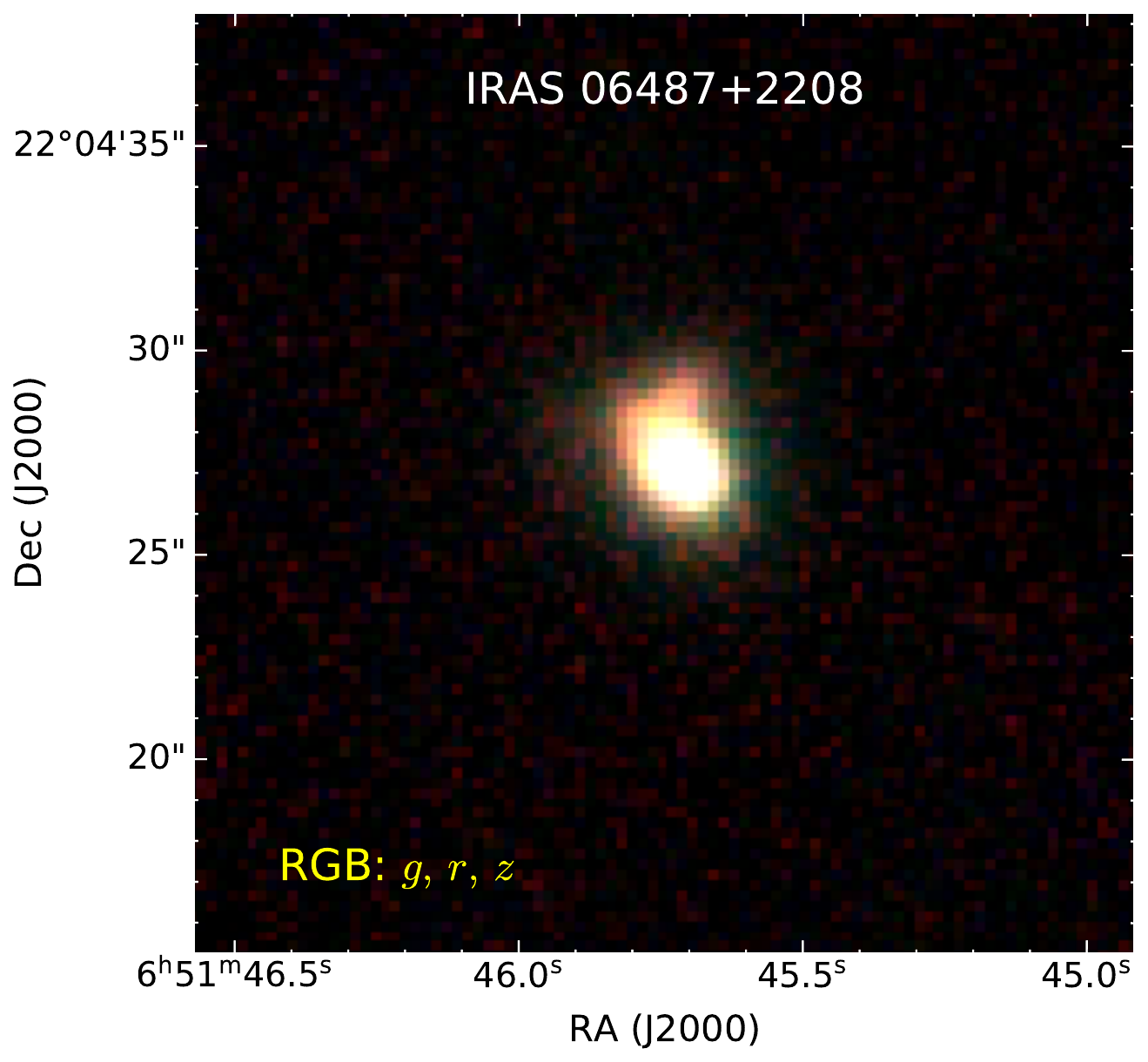}
 \includegraphics[height=0.25\textwidth, angle=0]{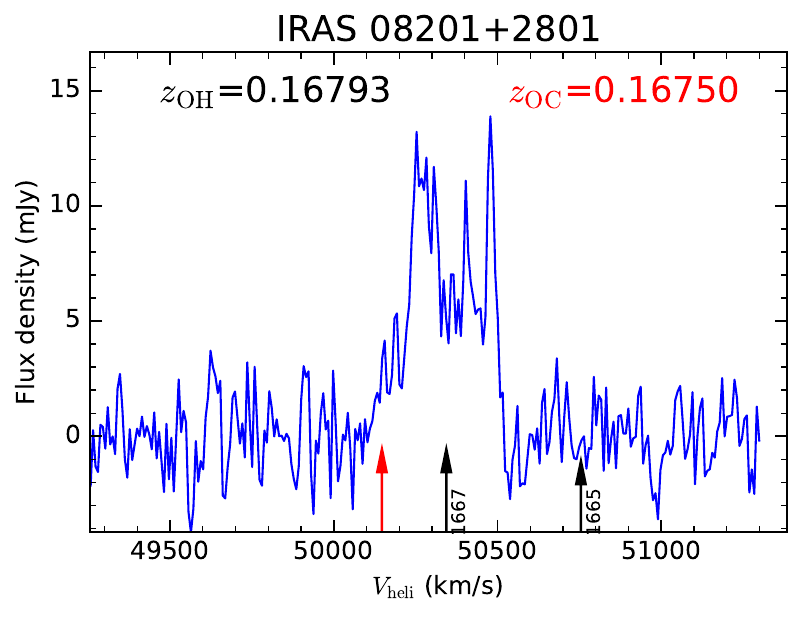}
 \includegraphics[height=0.31\textwidth, angle=0]{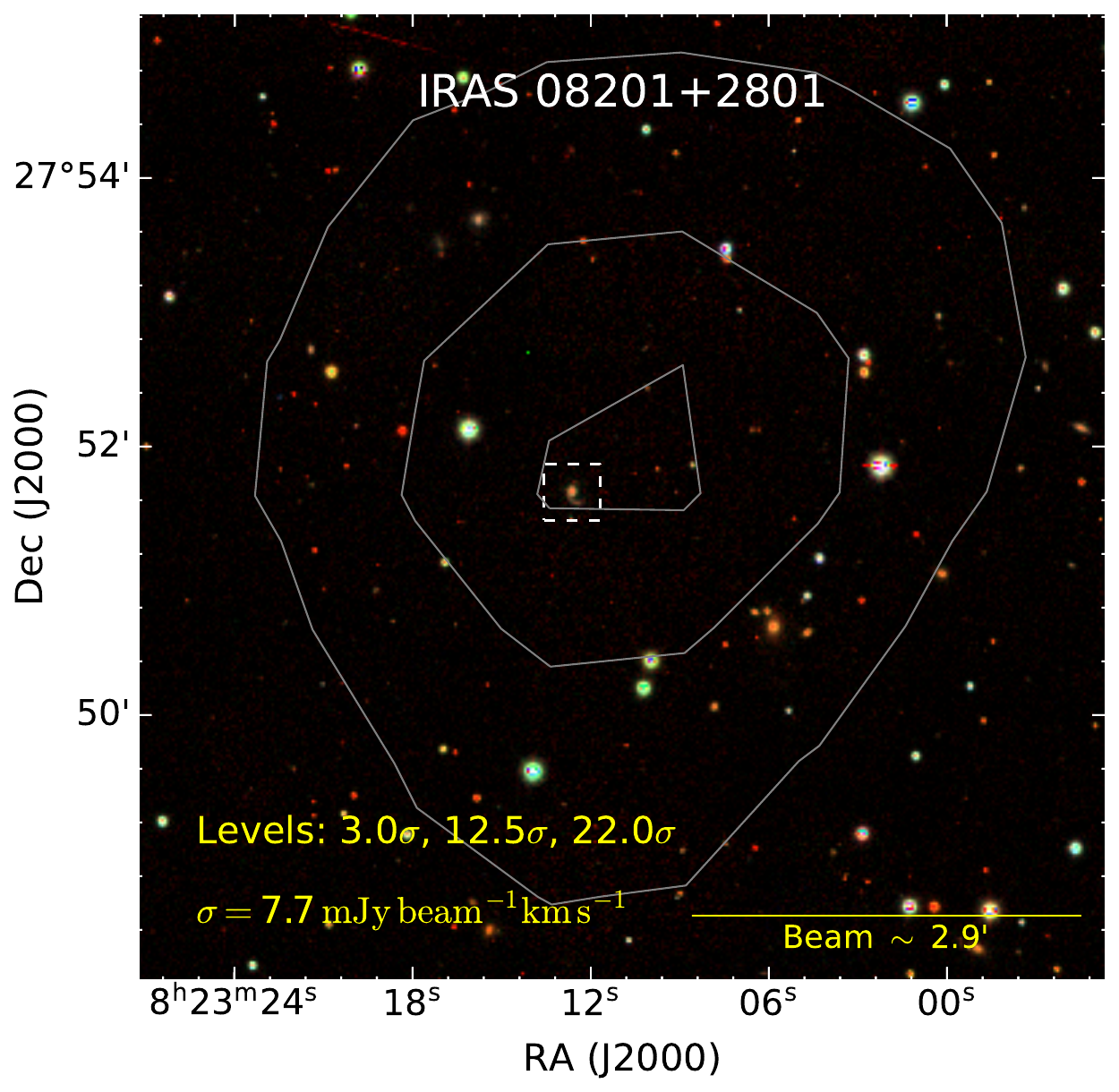}
 \includegraphics[height=0.29\textwidth, angle=0]{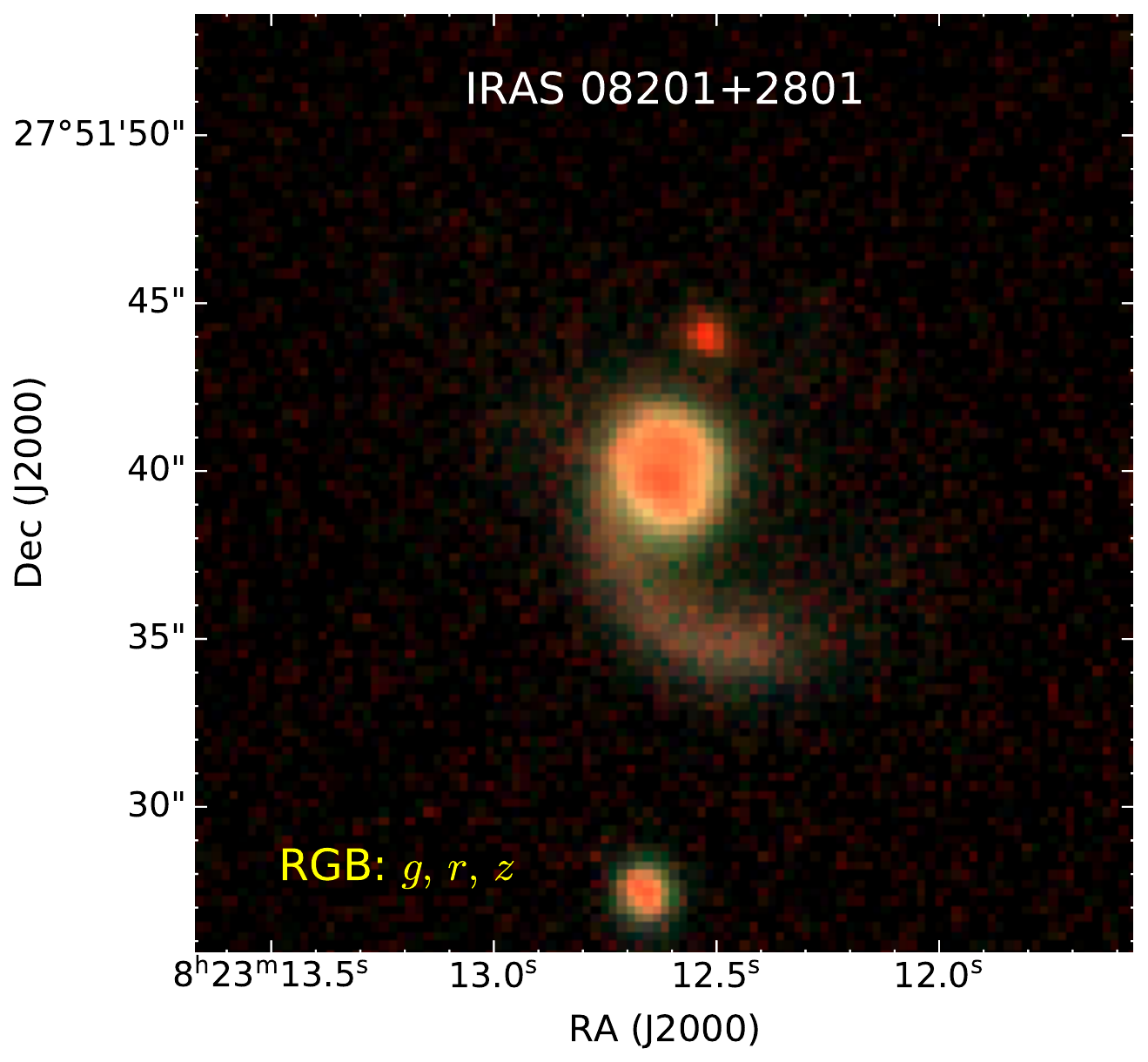}
 \includegraphics[height=0.25\textwidth, angle=0]{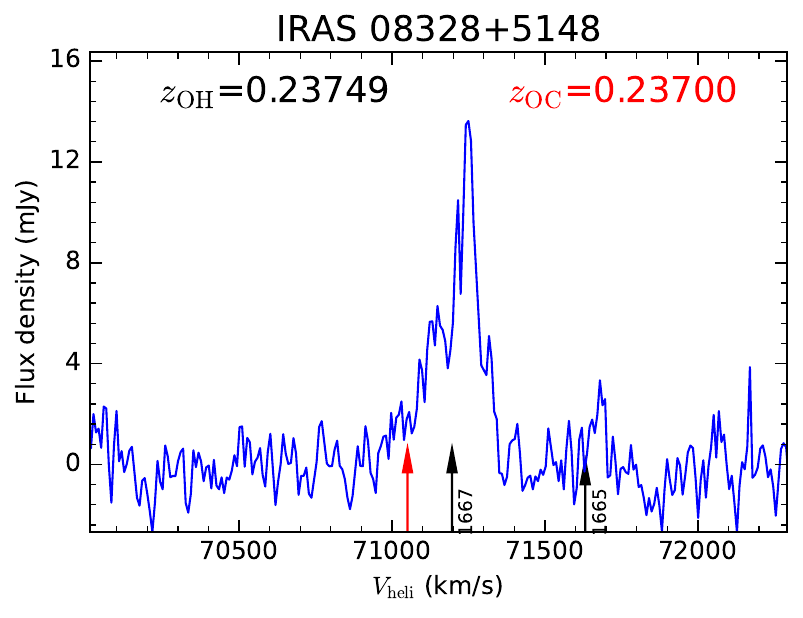}
 \includegraphics[height=0.31\textwidth, angle=0]{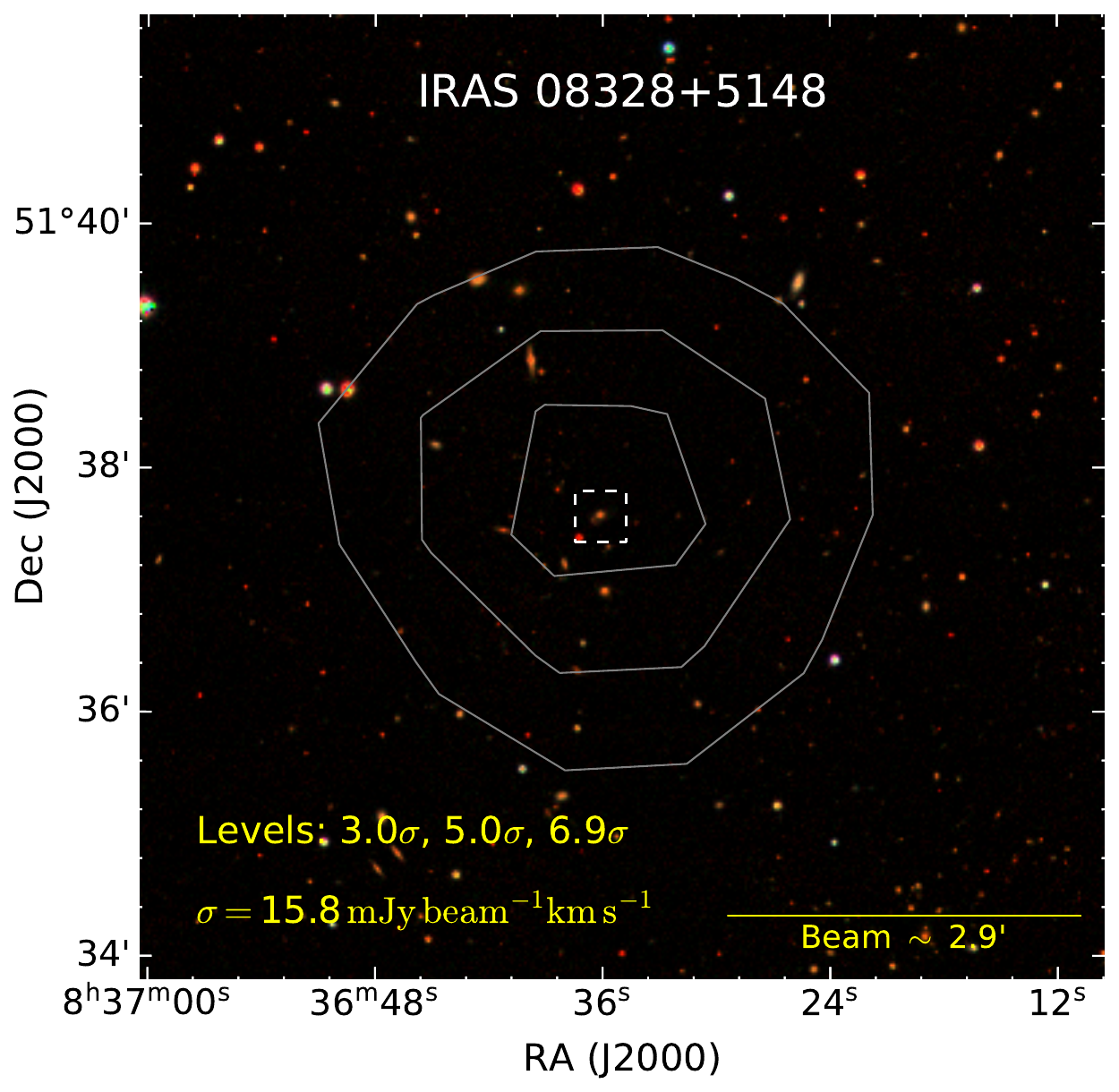}
 \includegraphics[height=0.29\textwidth, angle=0]{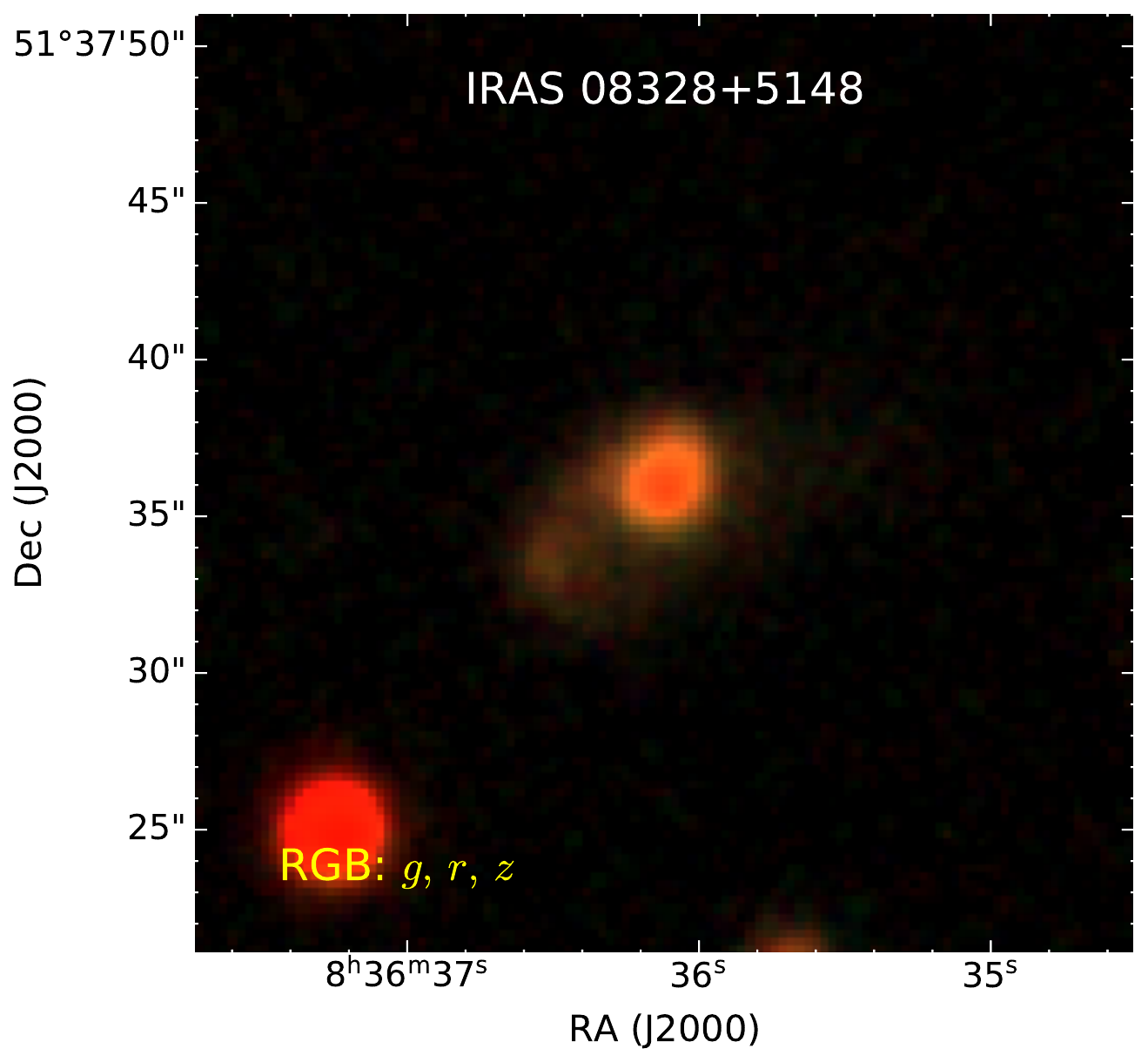}
 \includegraphics[height=0.25\textwidth, angle=0]{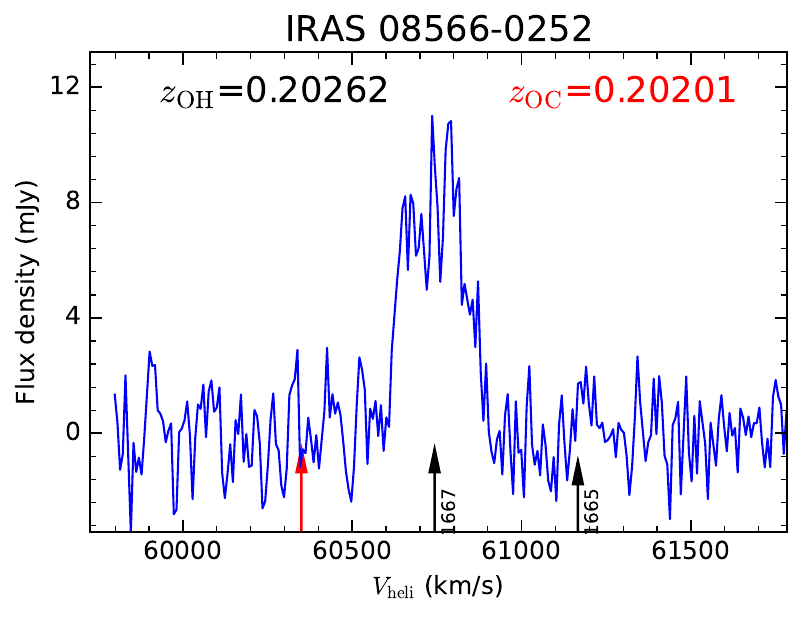}
 \includegraphics[height=0.31\textwidth, angle=0]{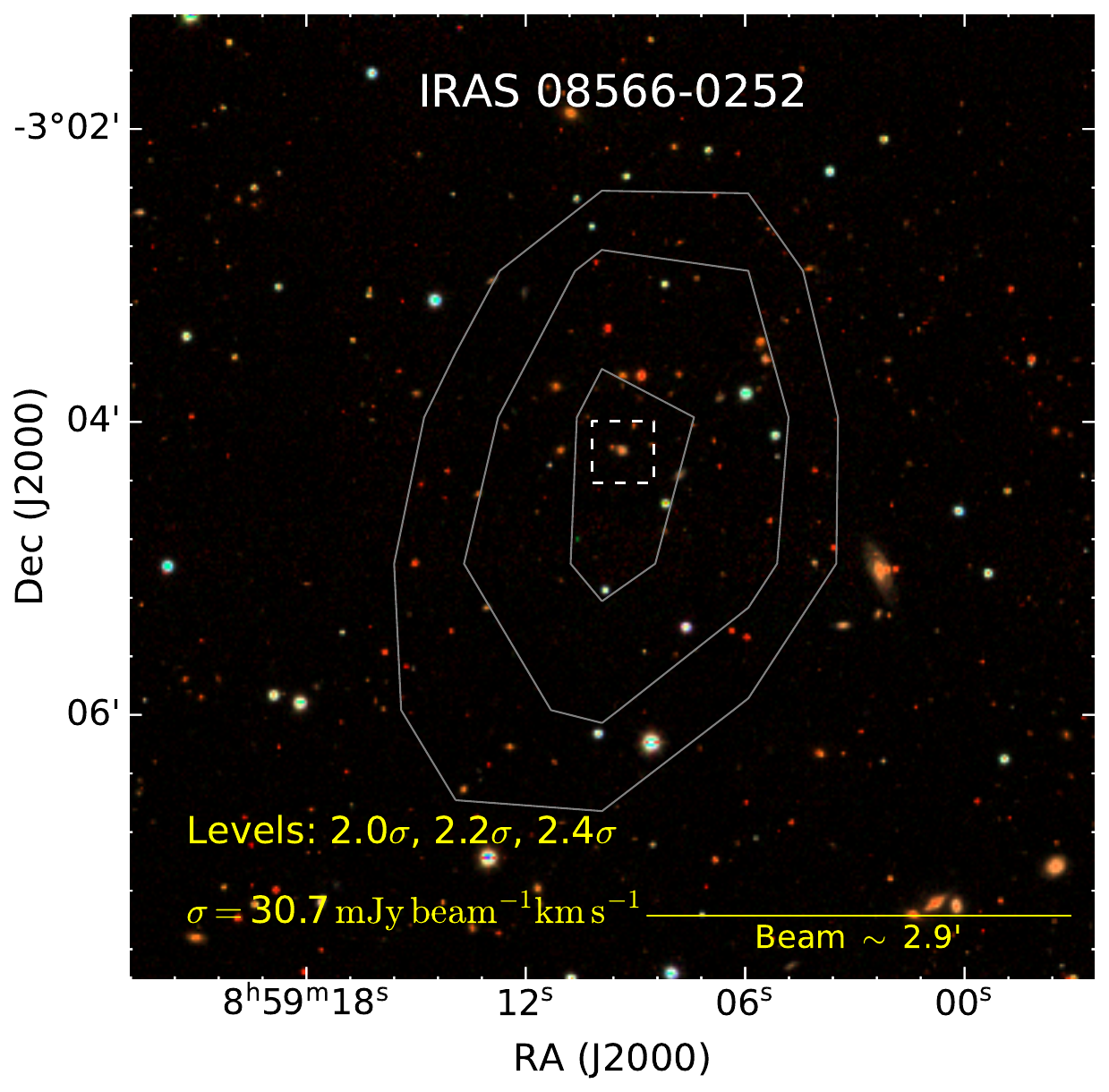}
 \includegraphics[height=0.29\textwidth, angle=0]{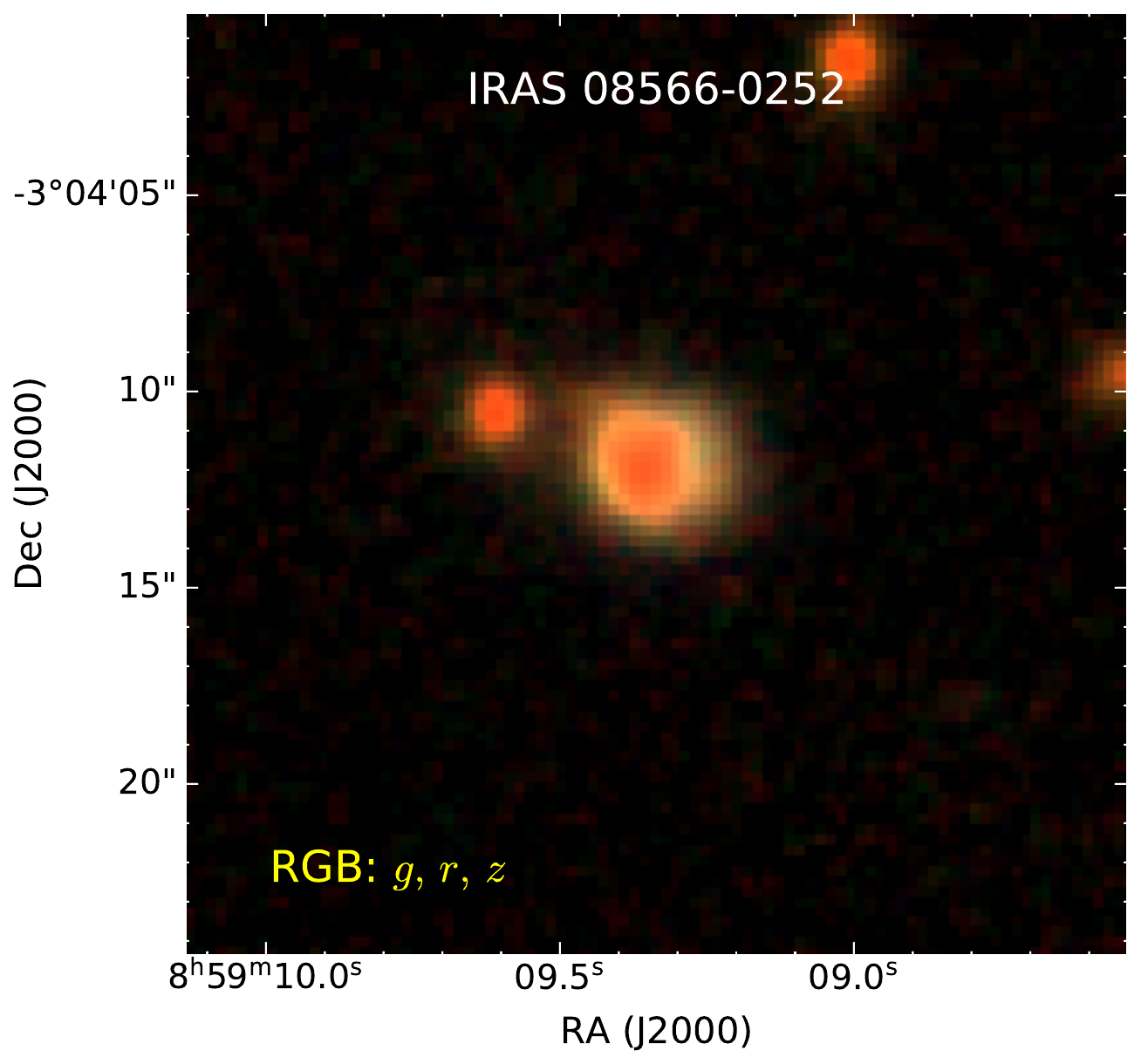}
 \caption{See caption in Figure\,\ref{Fig:IRAS00256-0208}}
 \end{figure*} 

 \begin{figure*}[htp]
 \centering
 \renewcommand{\thefigure}{\arabic{figure} (Continued)}
 \addtocounter{figure}{-1}
 \includegraphics[height=0.25\textwidth, angle=0]{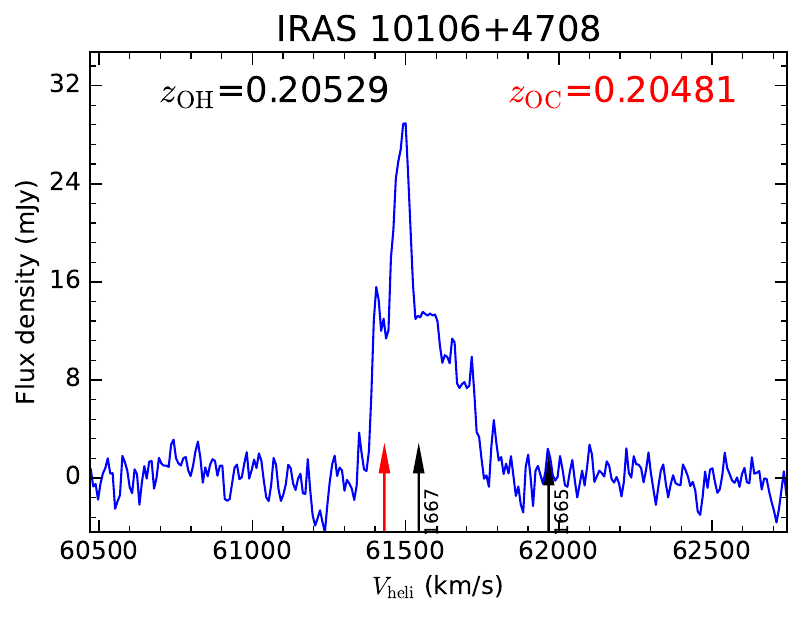}
 \includegraphics[height=0.31\textwidth, angle=0]{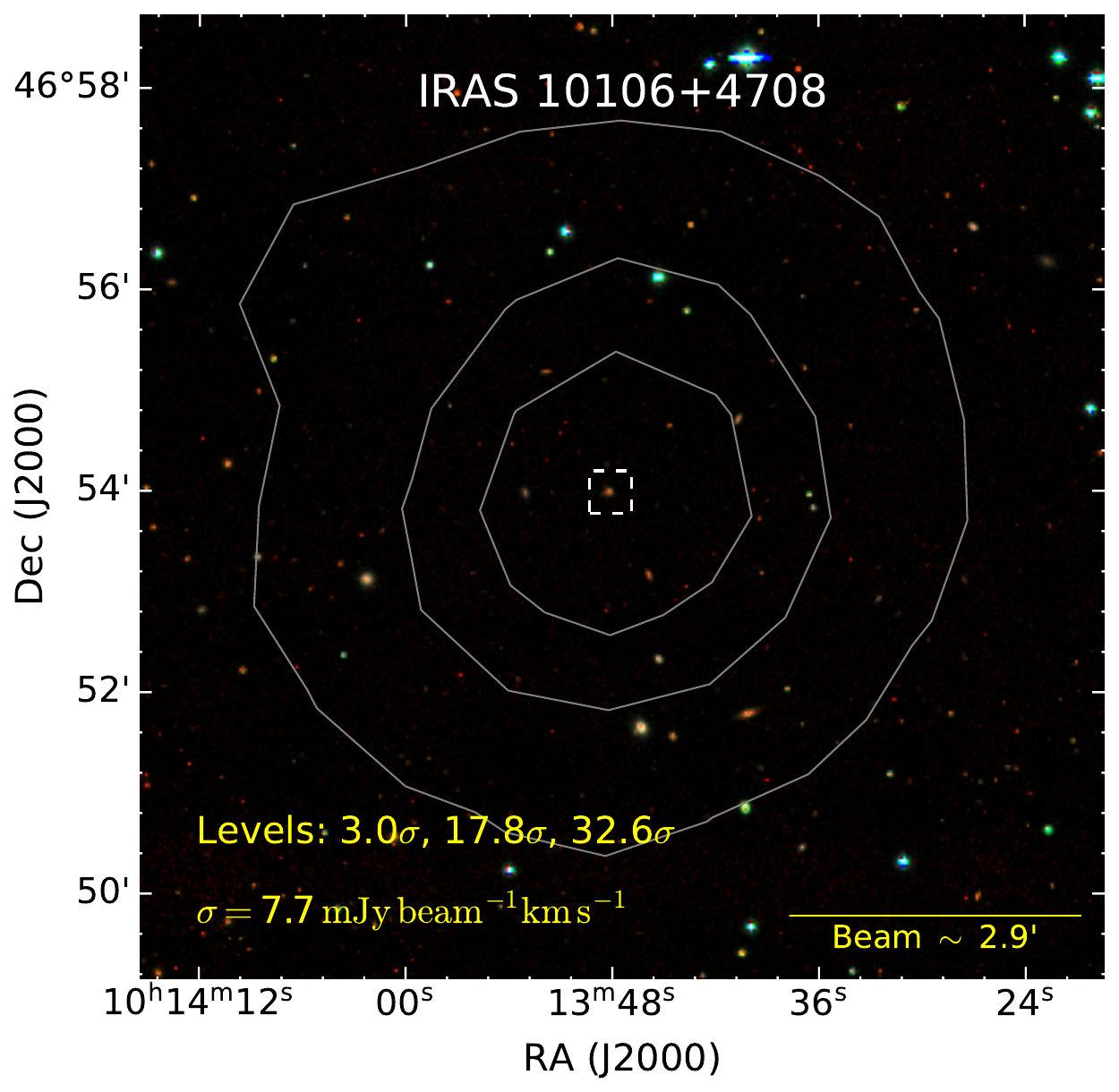}
 \includegraphics[height=0.29\textwidth, angle=0]{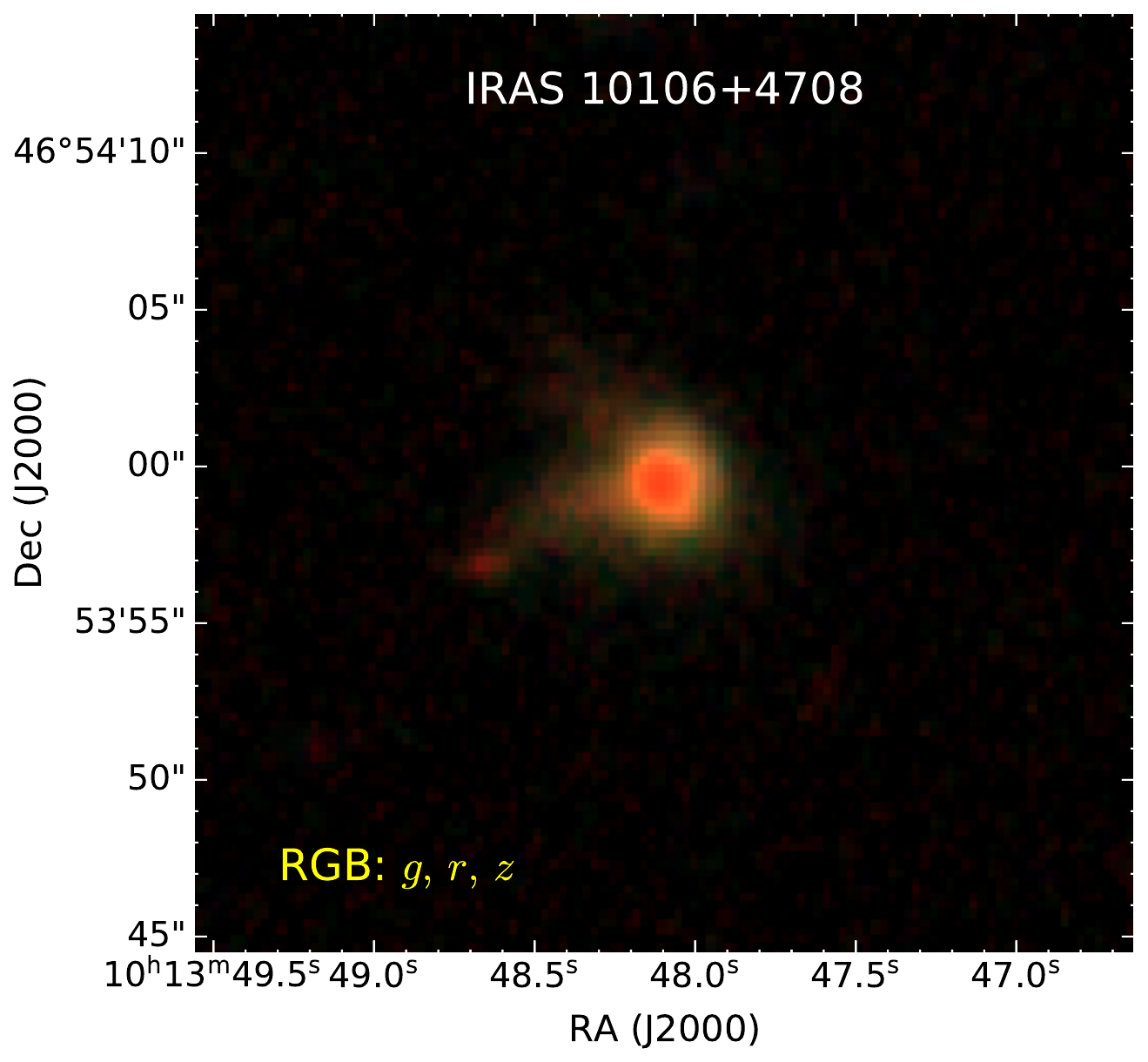}
 \includegraphics[height=0.25\textwidth, angle=0]{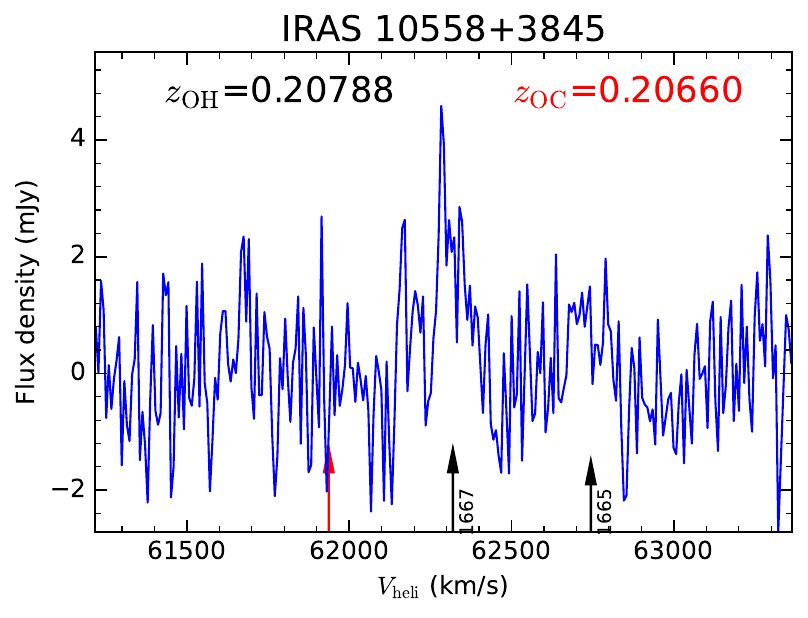}
 \includegraphics[height=0.31\textwidth, angle=0]{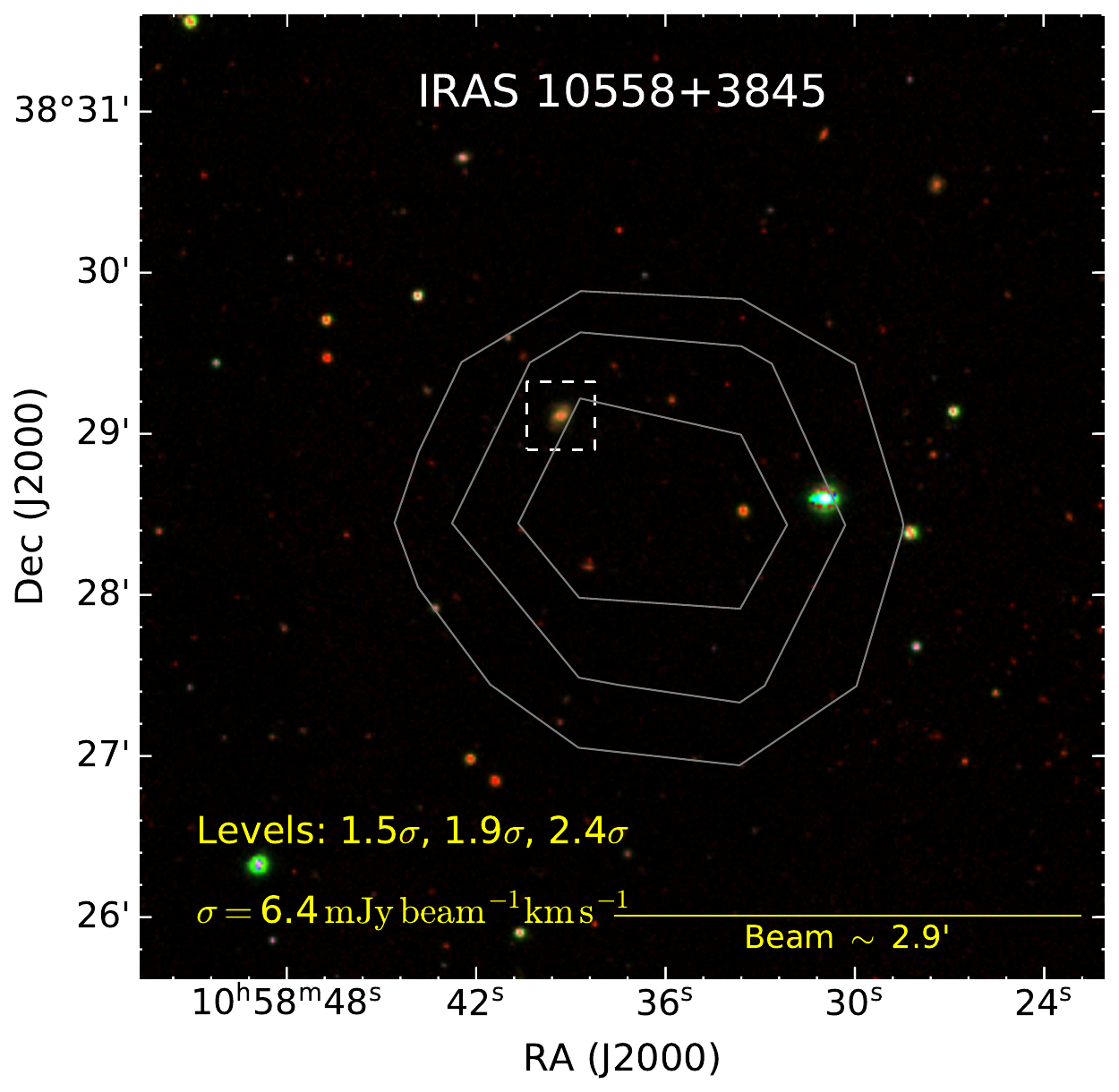}
 \includegraphics[height=0.29\textwidth, angle=0]{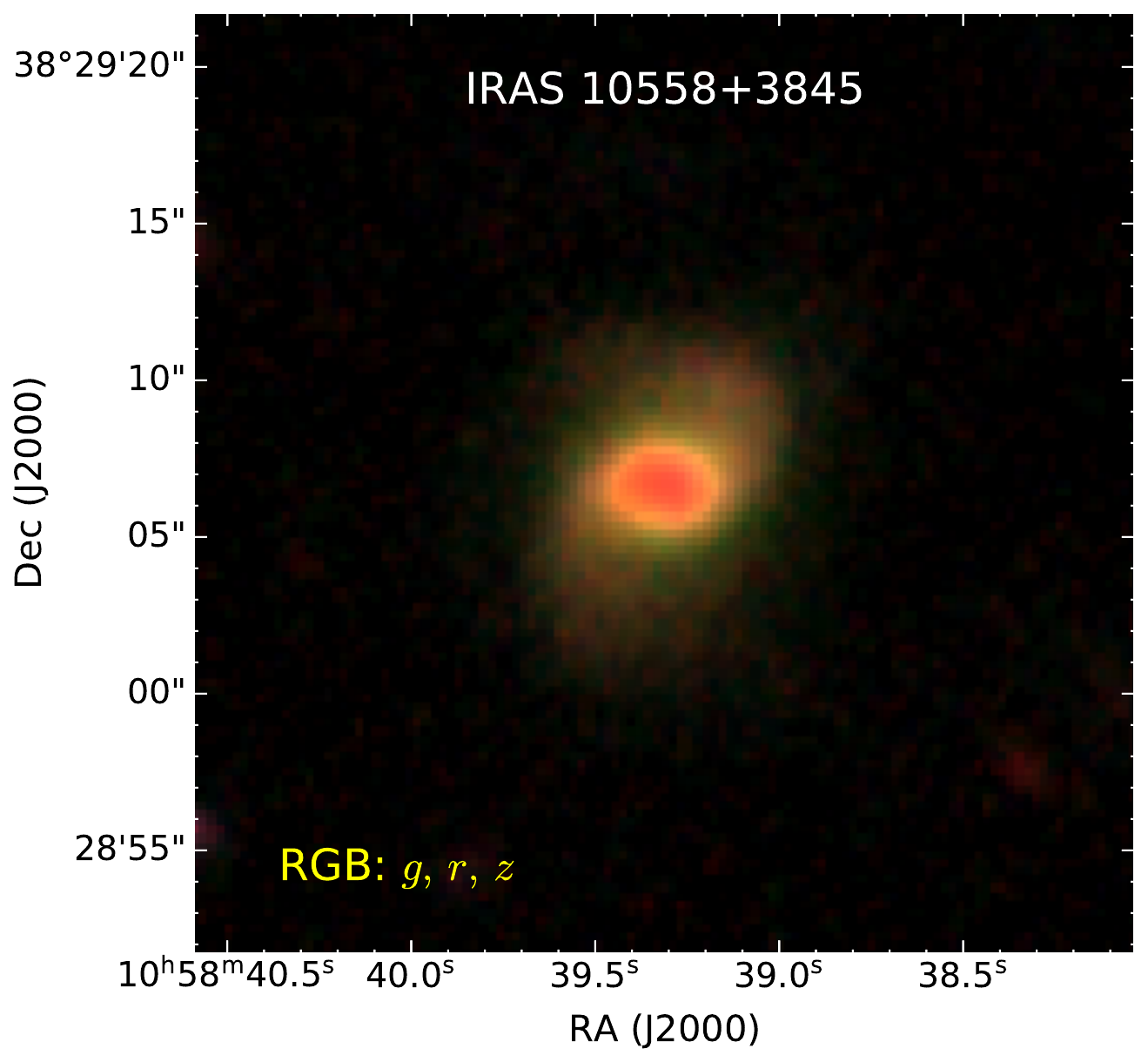}
 \includegraphics[height=0.25\textwidth, angle=0]{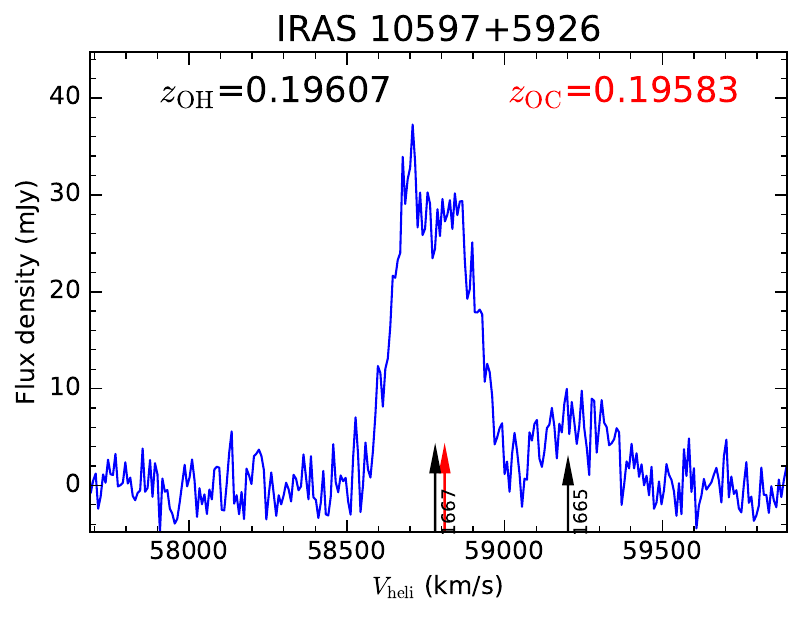}
 \includegraphics[height=0.31\textwidth, angle=0]{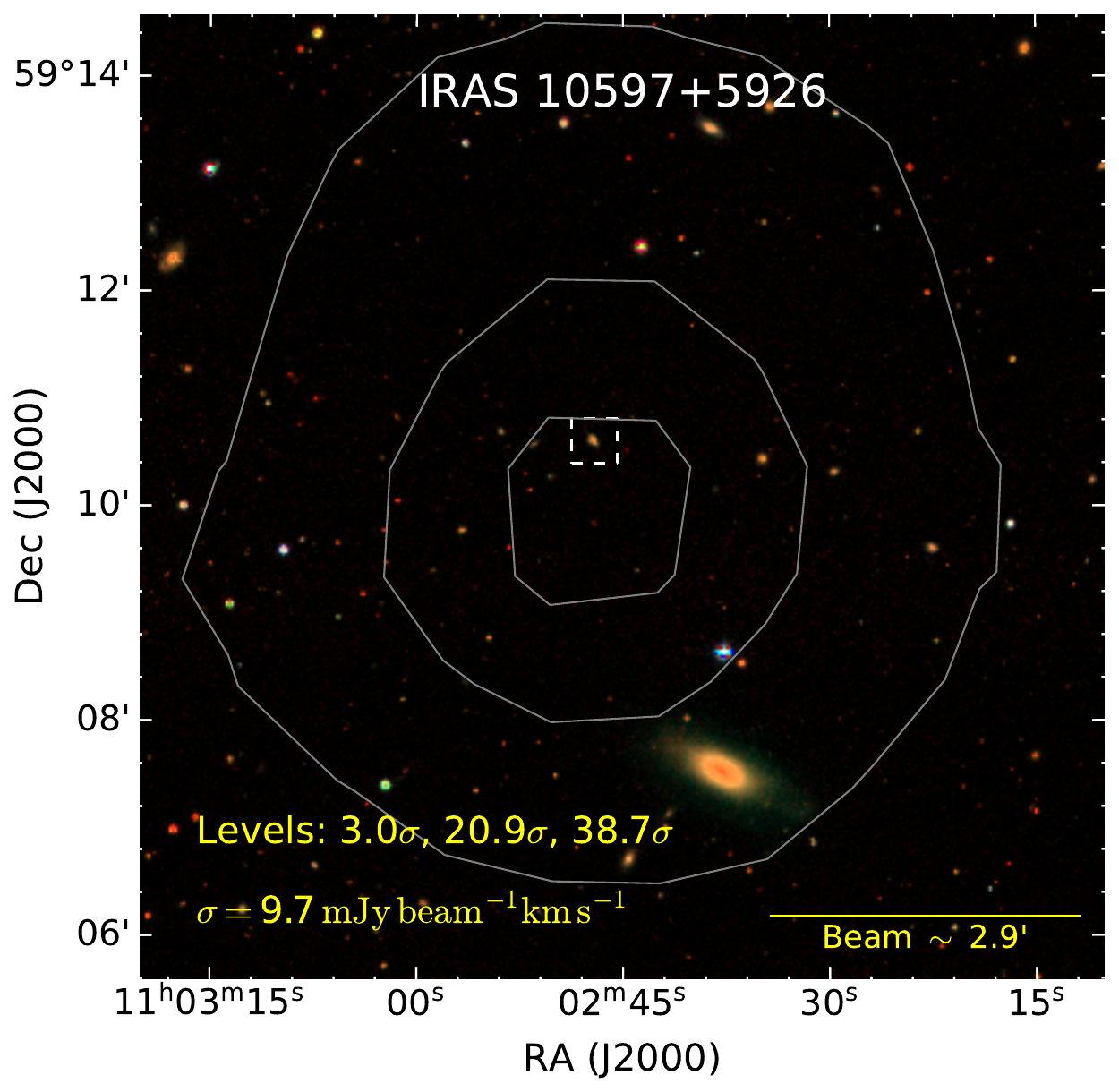}
 \includegraphics[height=0.29\textwidth, angle=0]{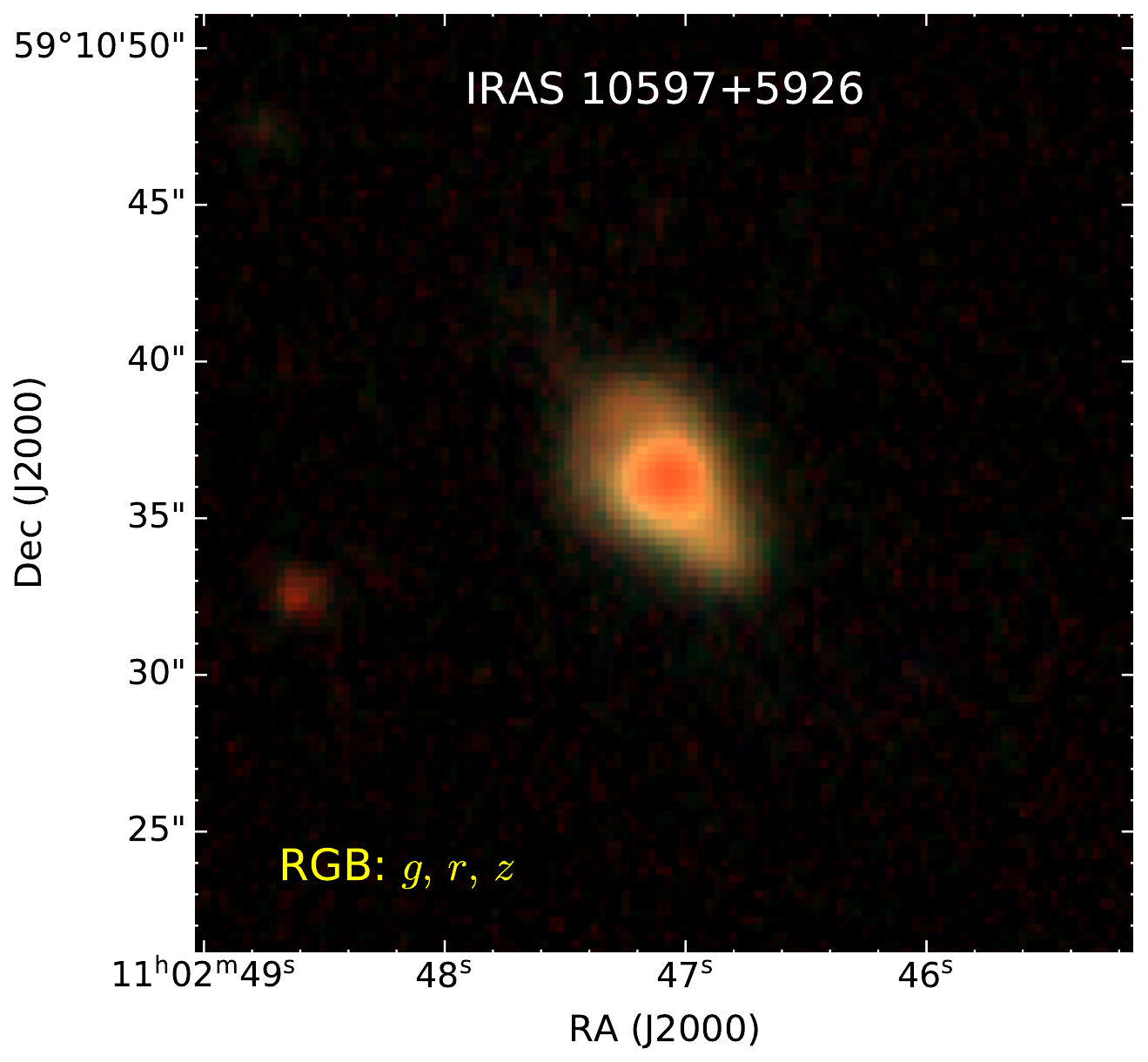}
 \includegraphics[height=0.25\textwidth, angle=0]{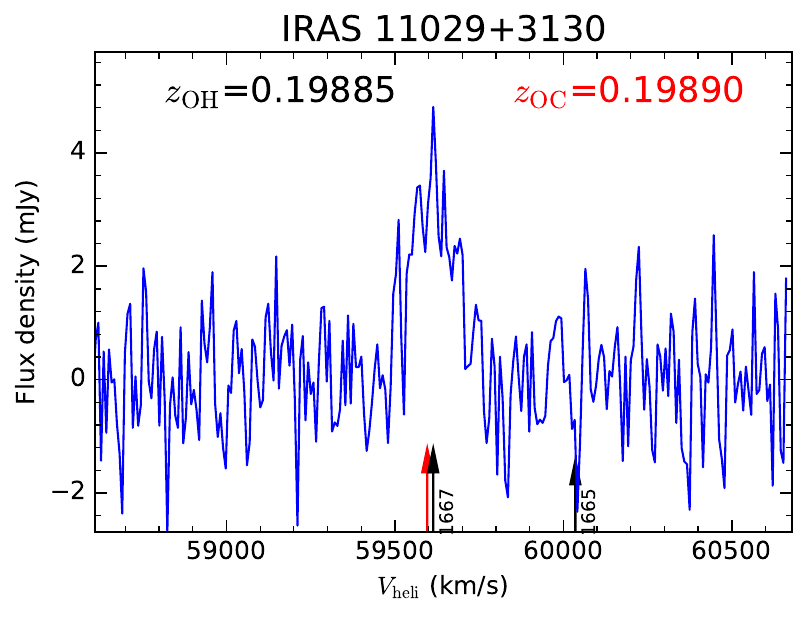}
 \includegraphics[height=0.31\textwidth, angle=0]{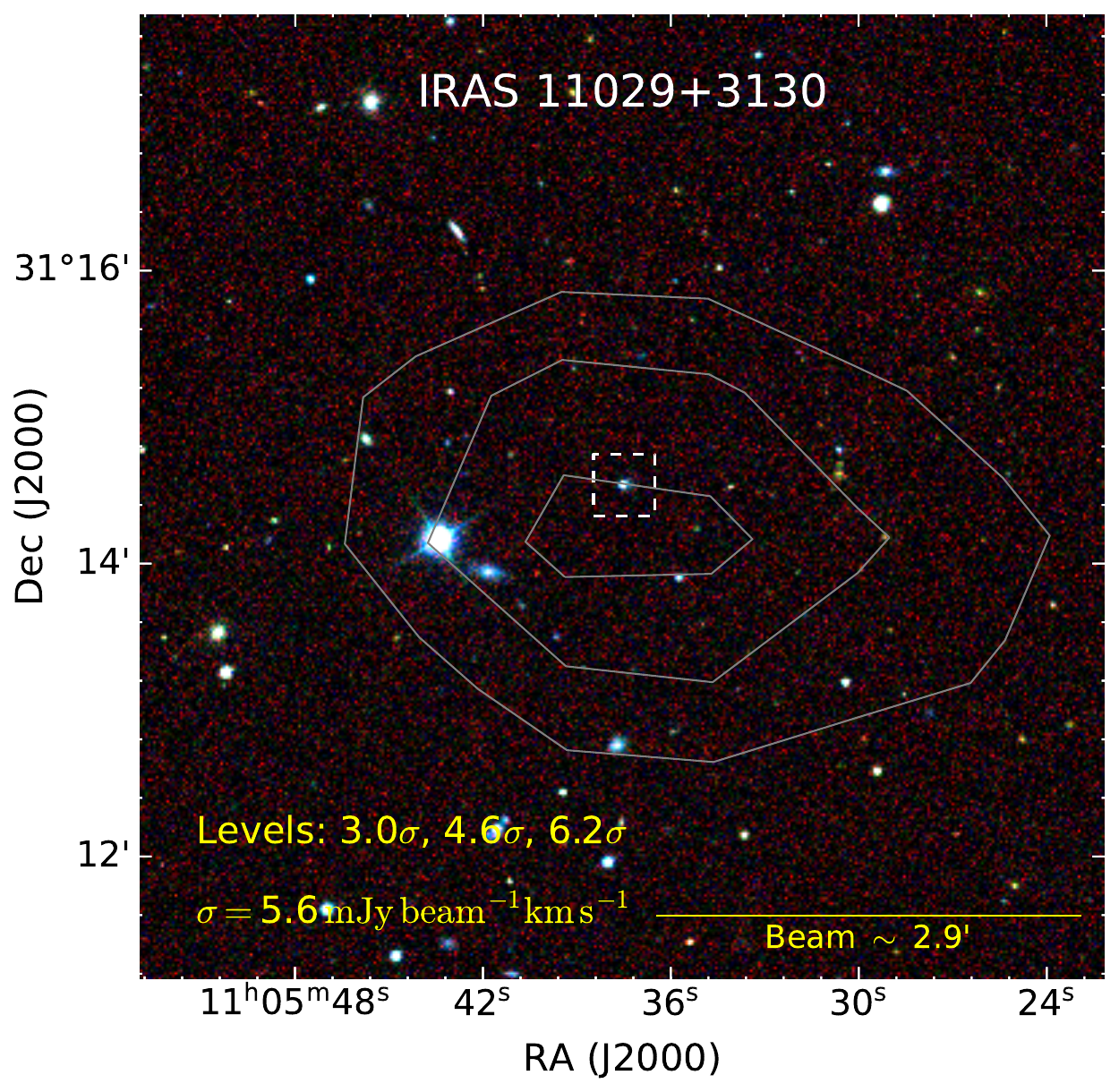}
 \includegraphics[height=0.29\textwidth, angle=0]{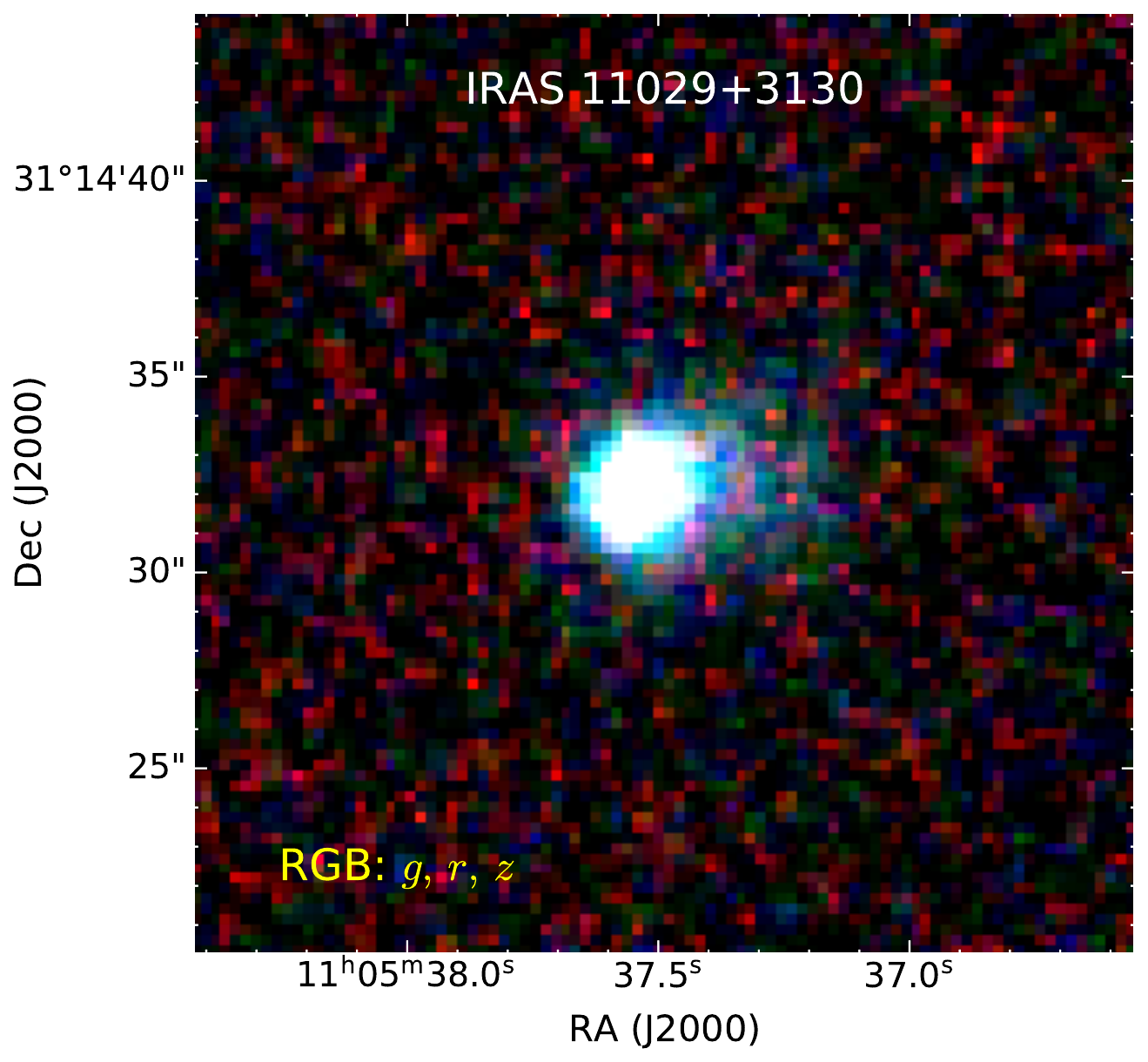}
 \caption{See caption in Figure\,\ref{Fig:IRAS00256-0208}}
 \end{figure*} 

 \begin{figure*}[htp]
 \centering
 \renewcommand{\thefigure}{\arabic{figure} (Continued)}
 \addtocounter{figure}{-1}
 \includegraphics[height=0.25\textwidth, angle=0]{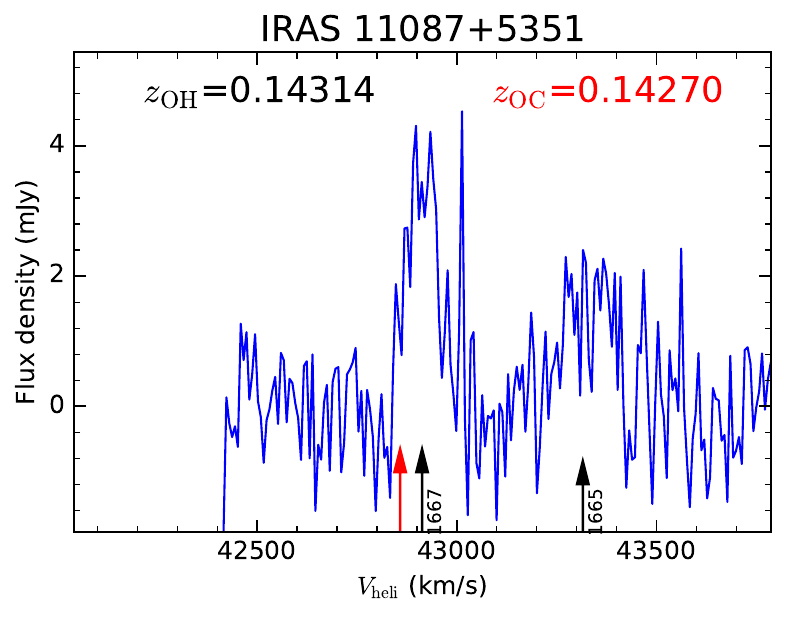}
 \includegraphics[height=0.31\textwidth, angle=0]{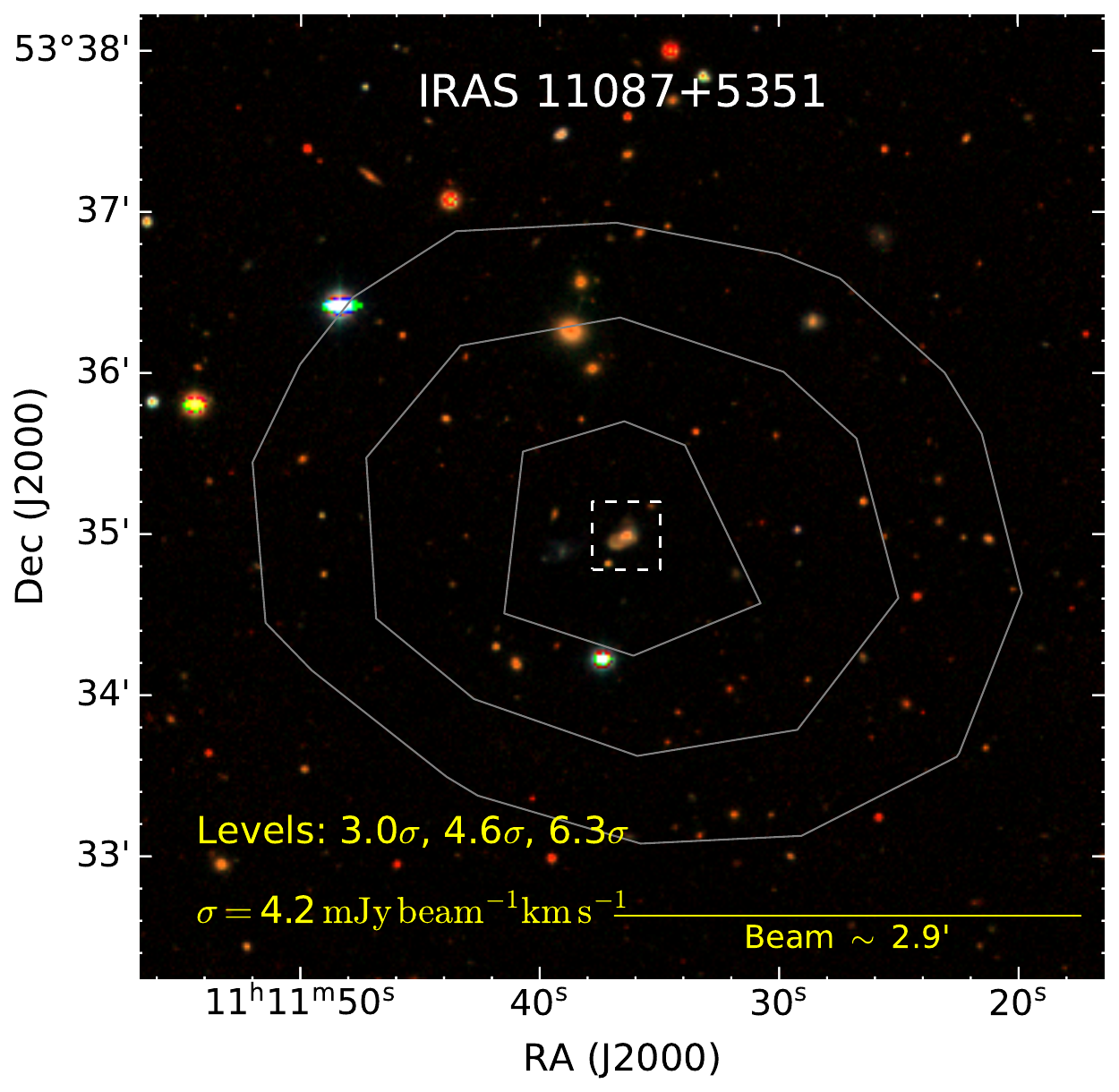}
 \includegraphics[height=0.29\textwidth, angle=0]{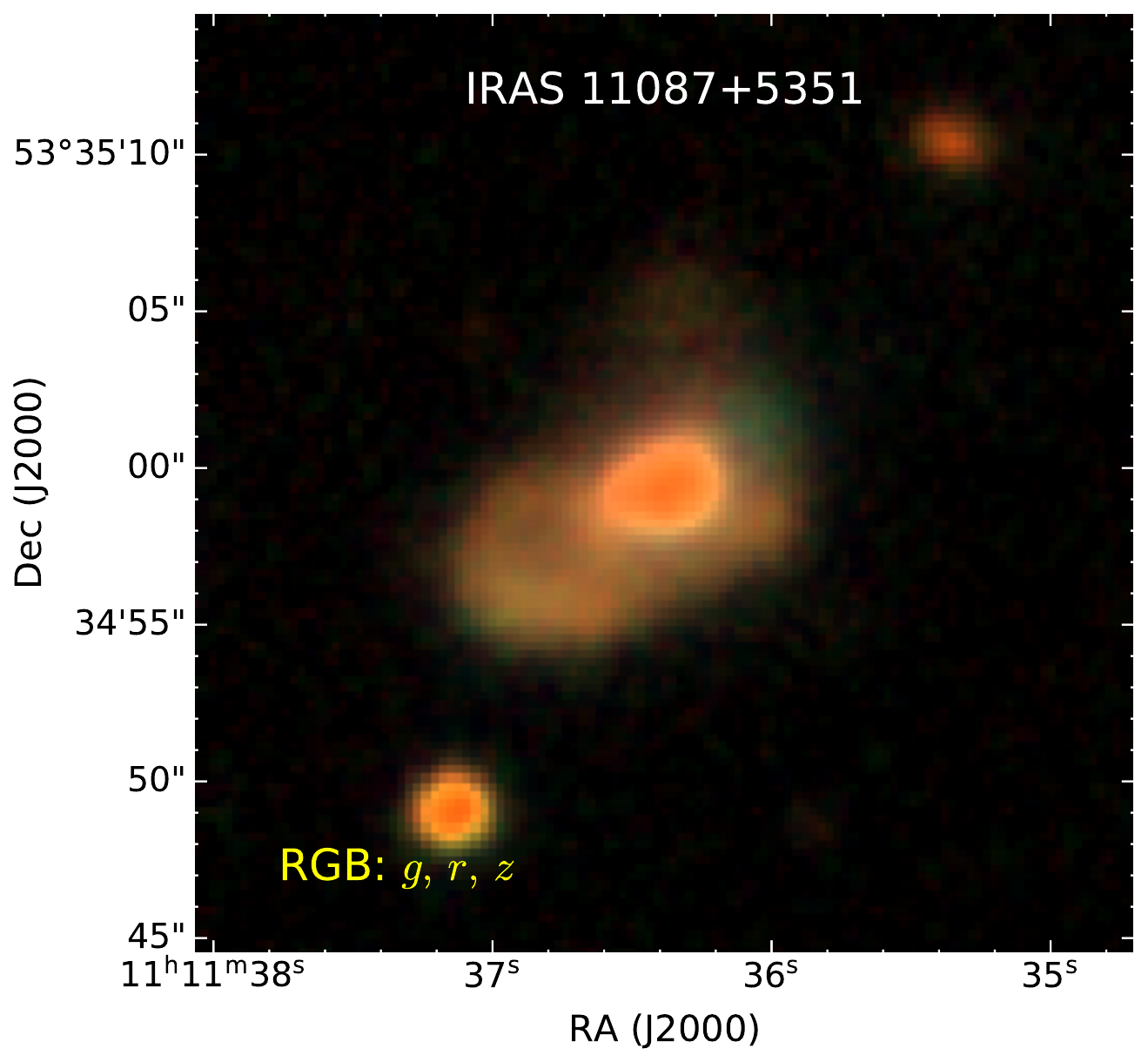}
 \includegraphics[height=0.25\textwidth, angle=0]{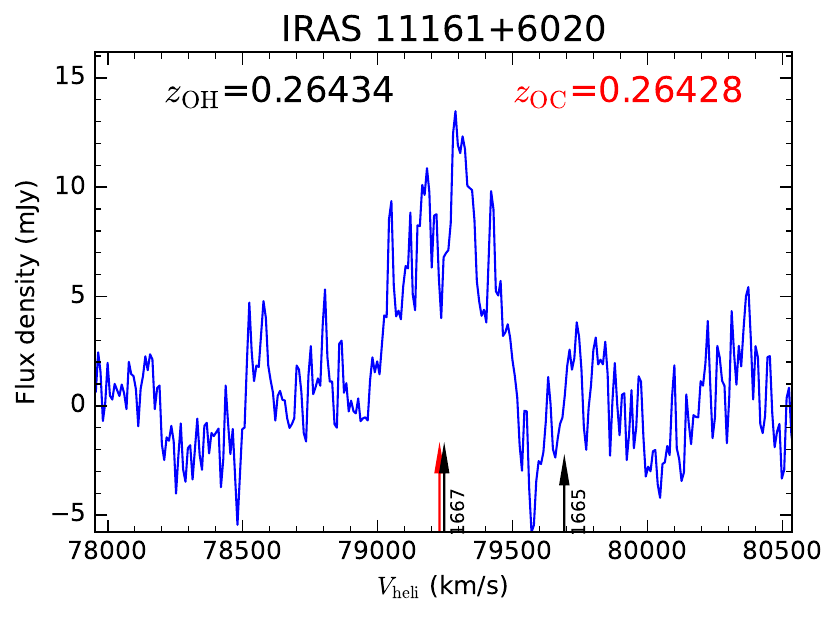}
 \includegraphics[height=0.31\textwidth, angle=0]{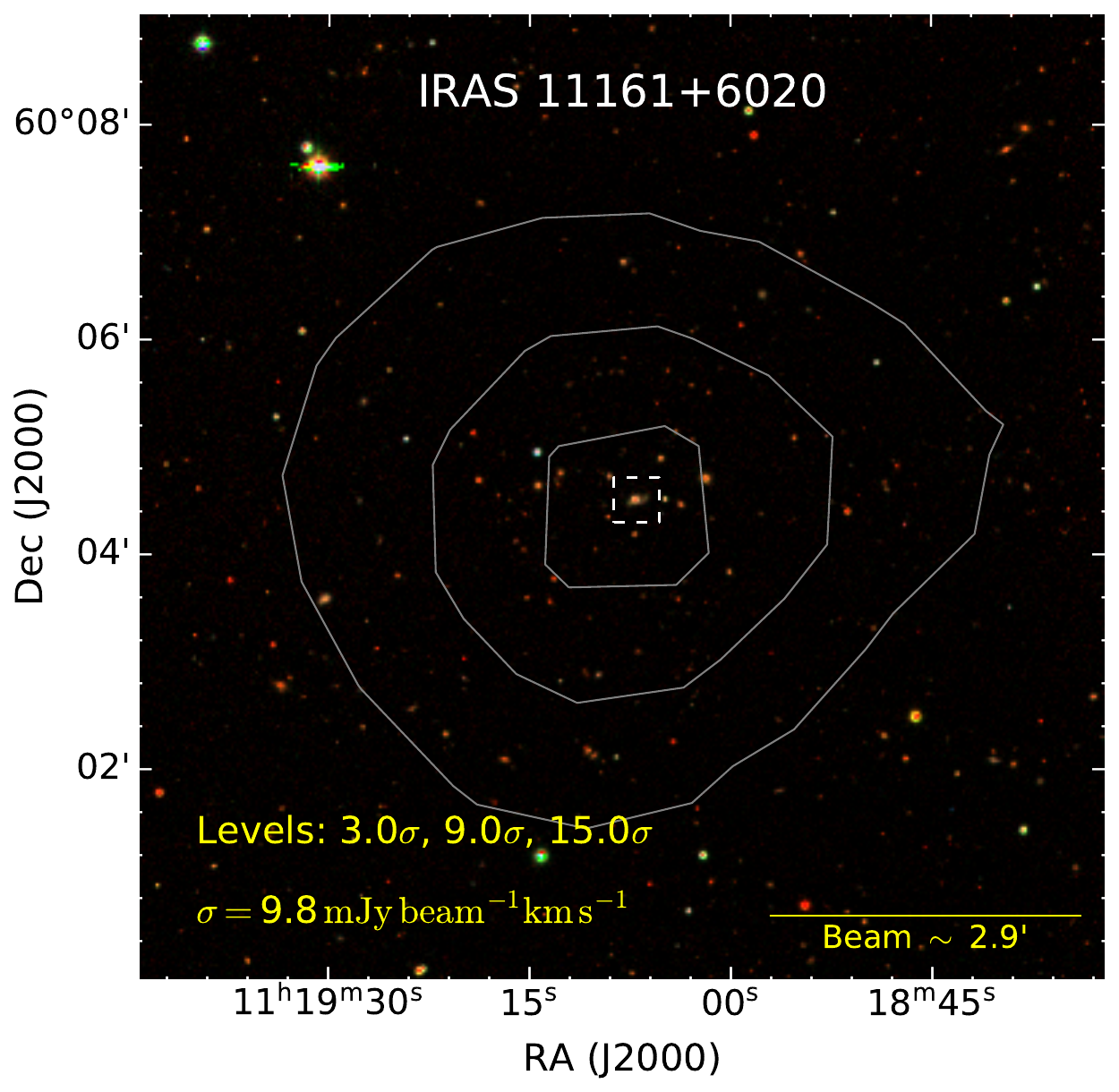}
 \includegraphics[height=0.29\textwidth, angle=0]{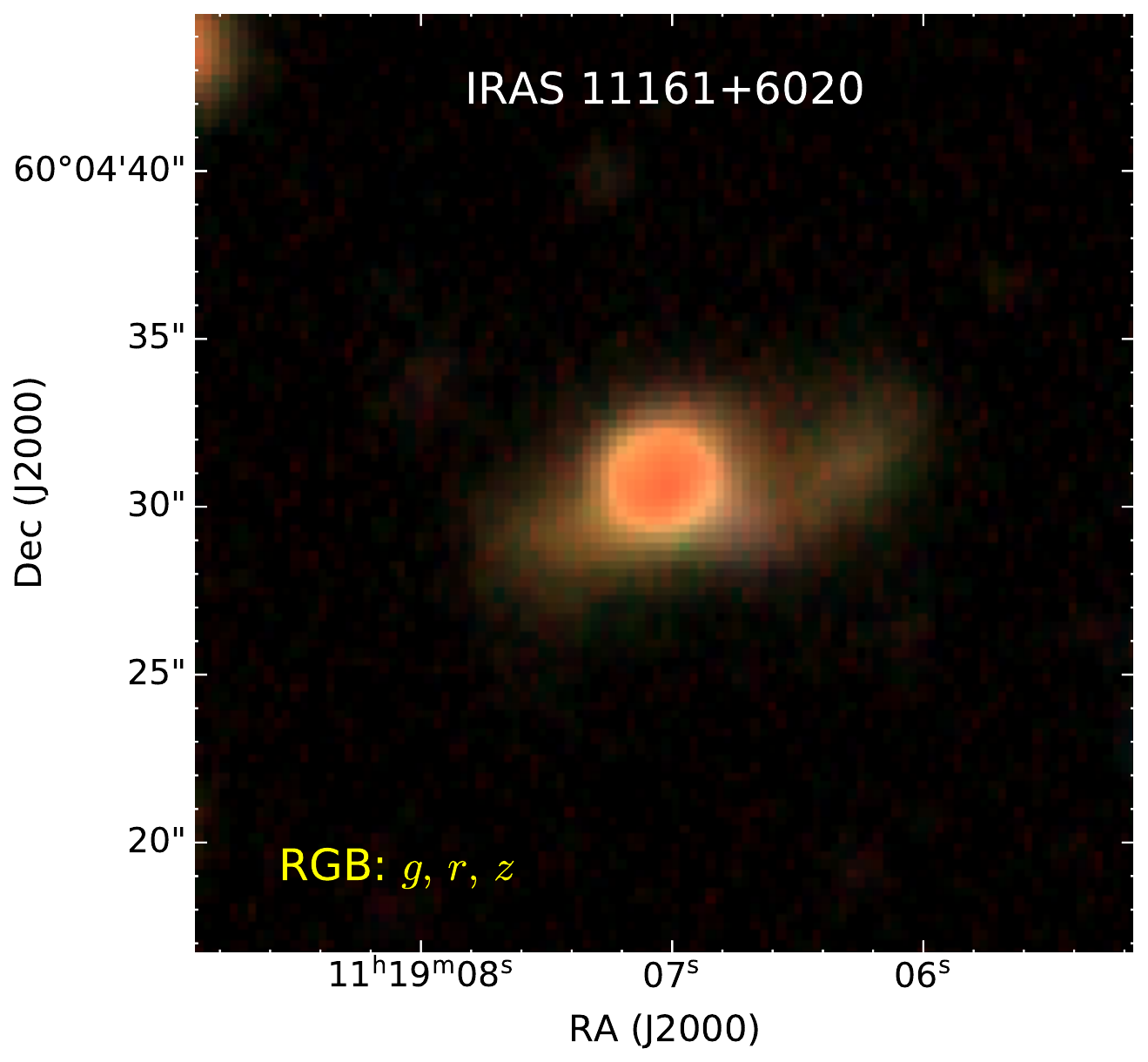}
 \includegraphics[height=0.25\textwidth, angle=0]{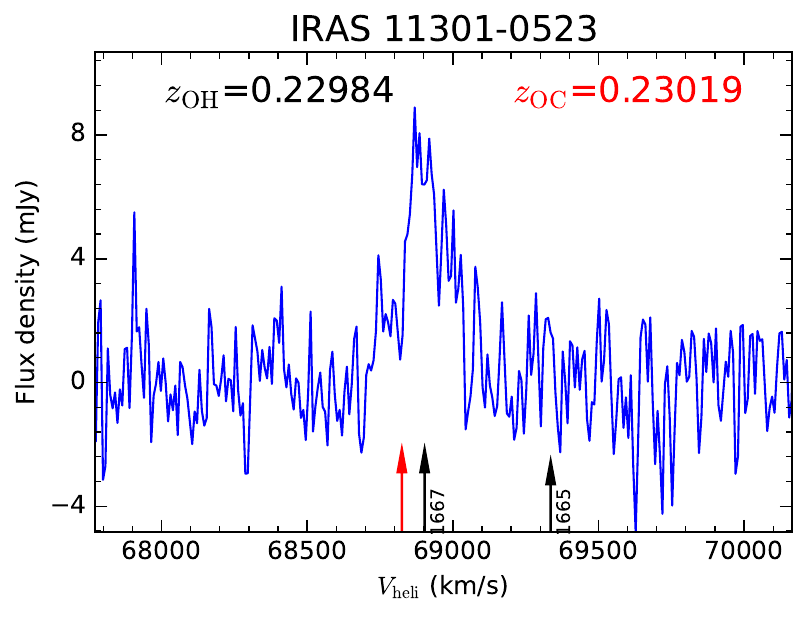}
 \includegraphics[height=0.31\textwidth, angle=0]{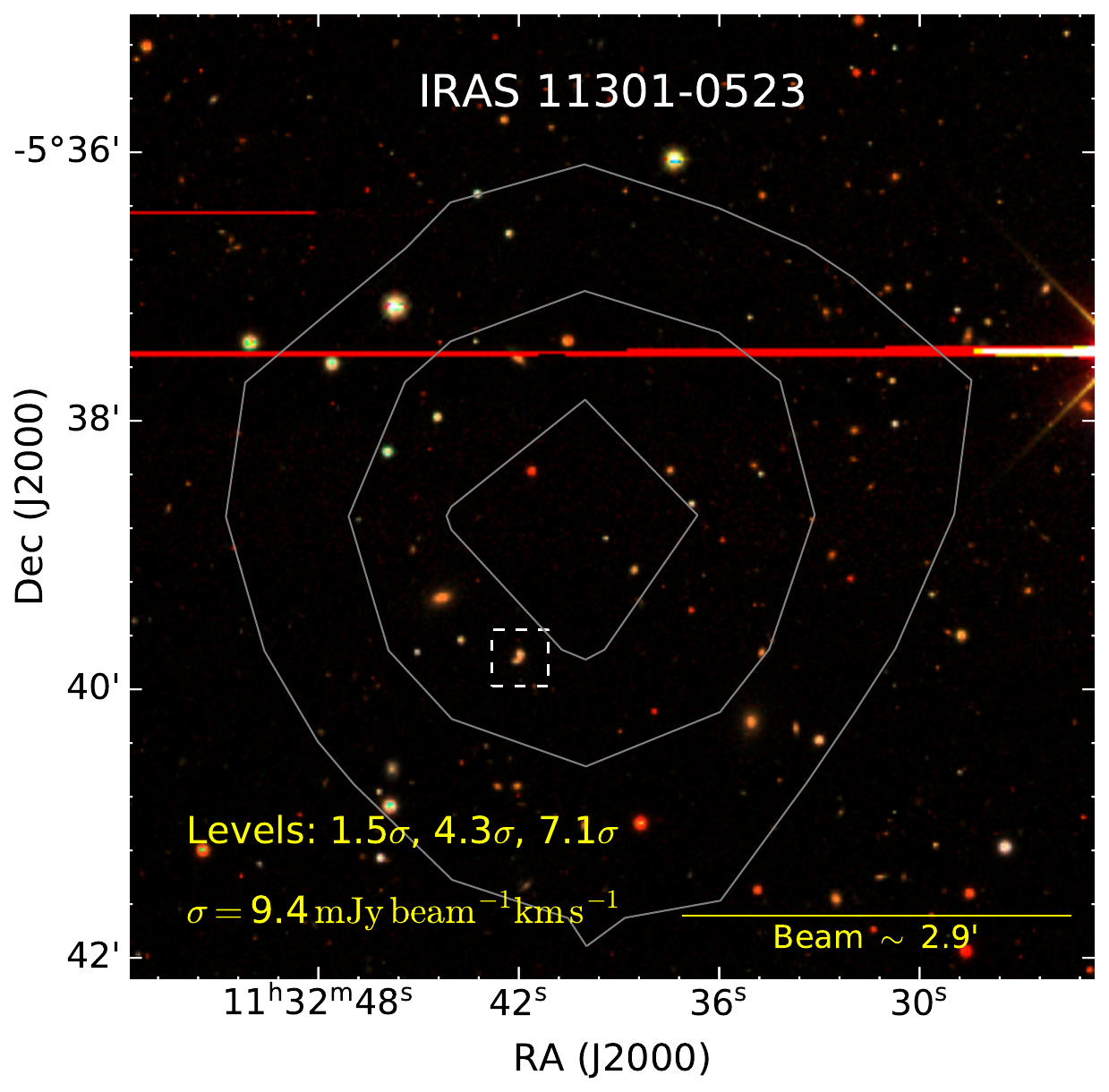}
 \includegraphics[height=0.29\textwidth, angle=0]{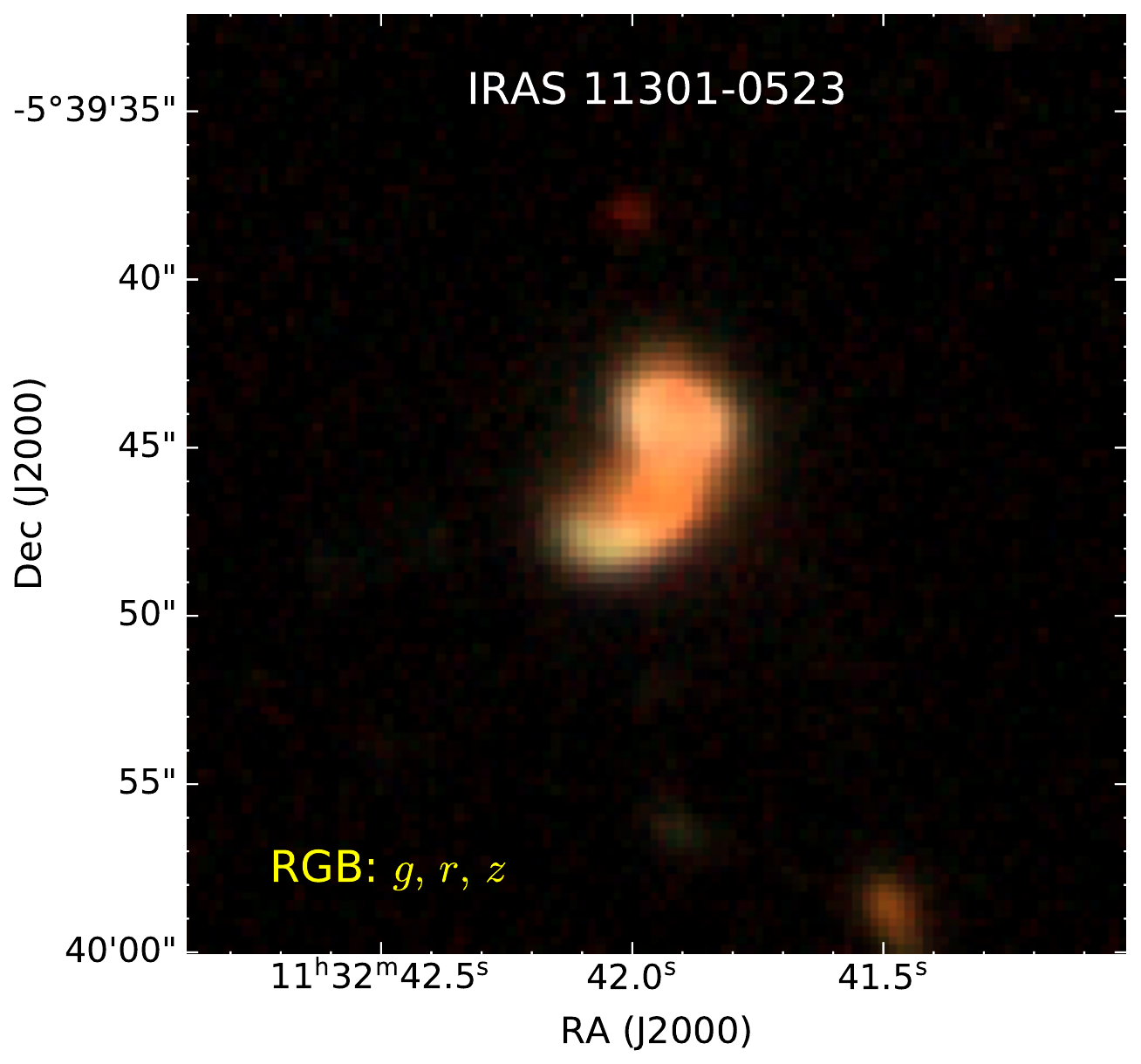}
 \includegraphics[height=0.25\textwidth, angle=0]{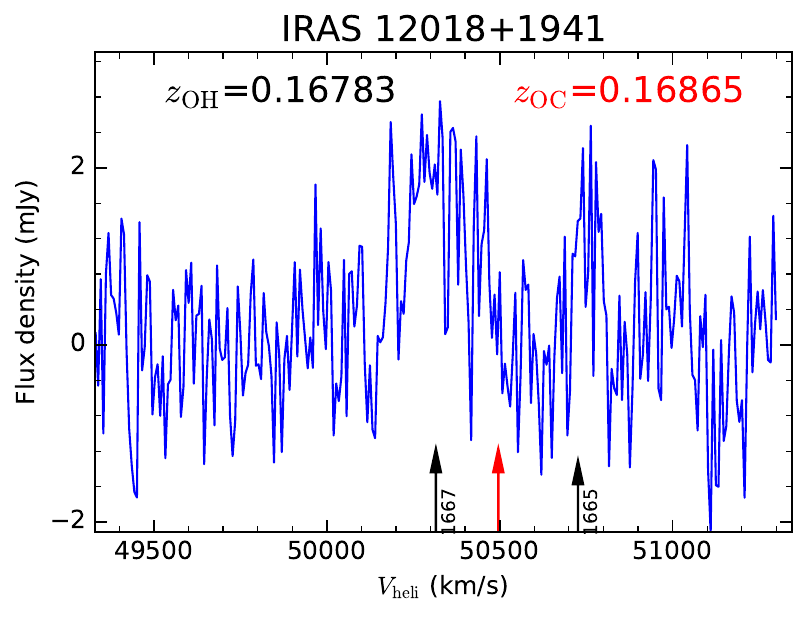}
 \includegraphics[height=0.31\textwidth, angle=0]{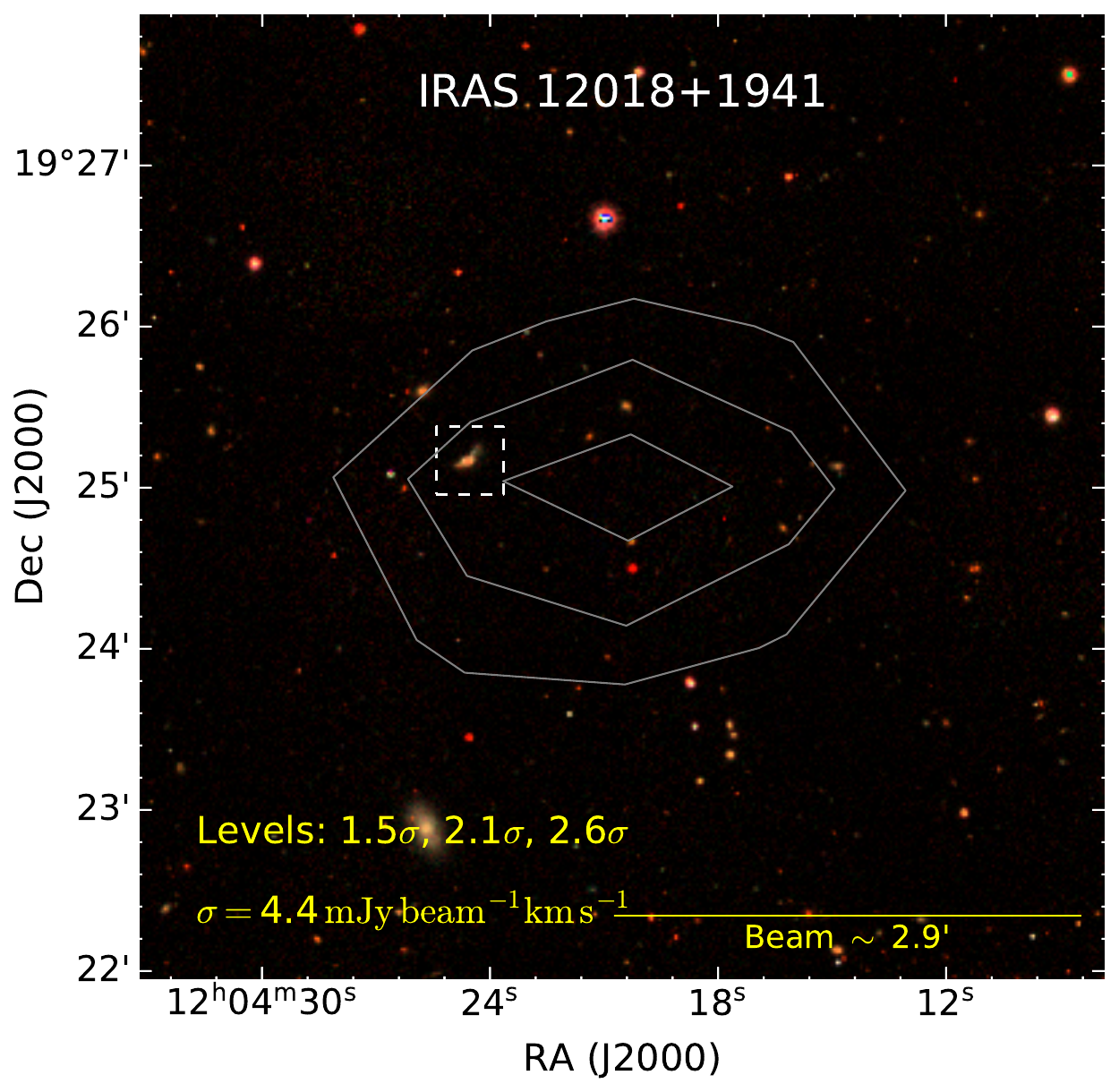}
 \includegraphics[height=0.29\textwidth, angle=0]{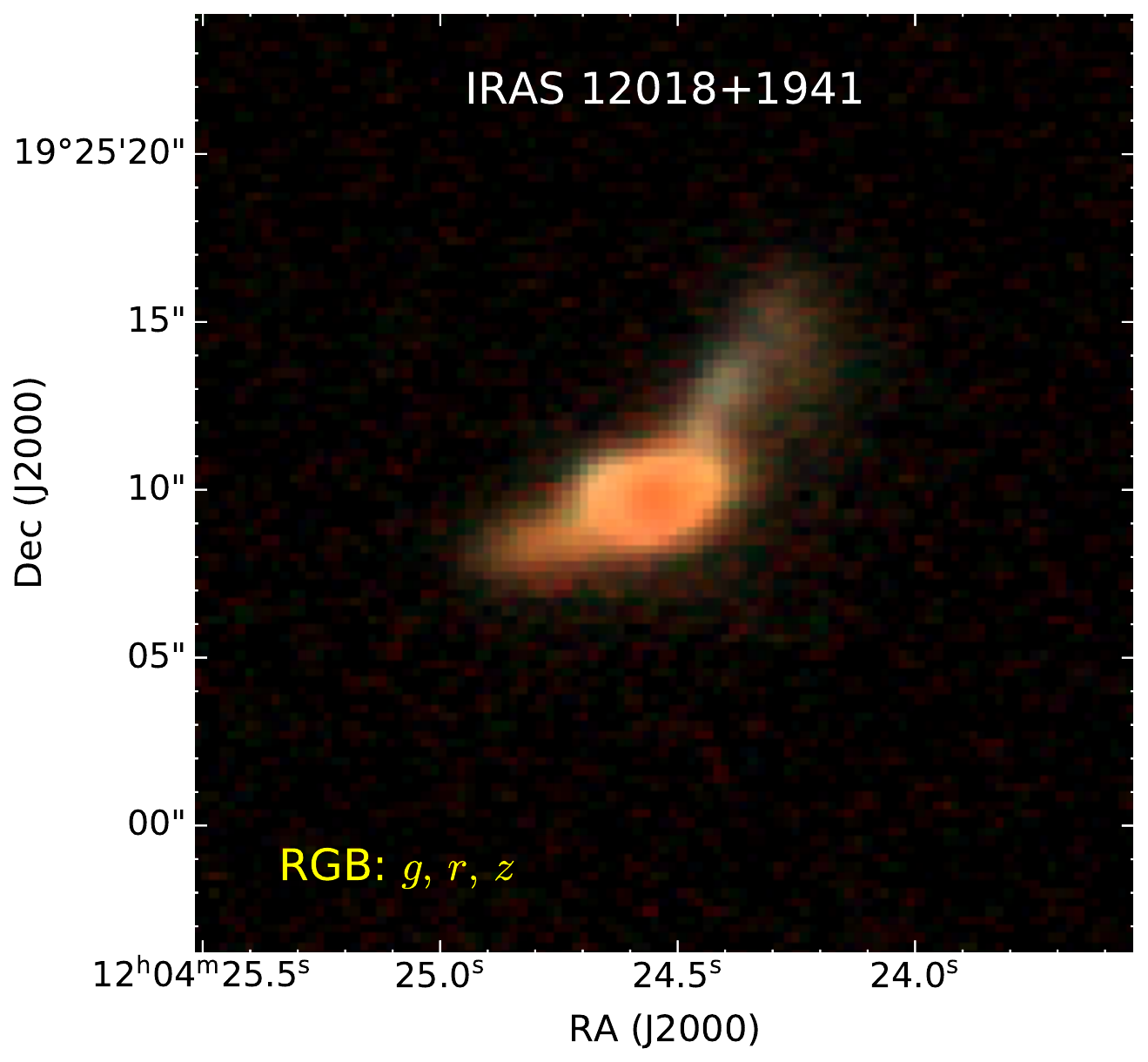}
 \caption{See caption in Figure\,\ref{Fig:IRAS00256-0208}}
 \end{figure*} 

 \begin{figure*}[htp]
 \centering
 \renewcommand{\thefigure}{\arabic{figure} (Continued)}
 \addtocounter{figure}{-1}
 \includegraphics[height=0.25\textwidth, angle=0]{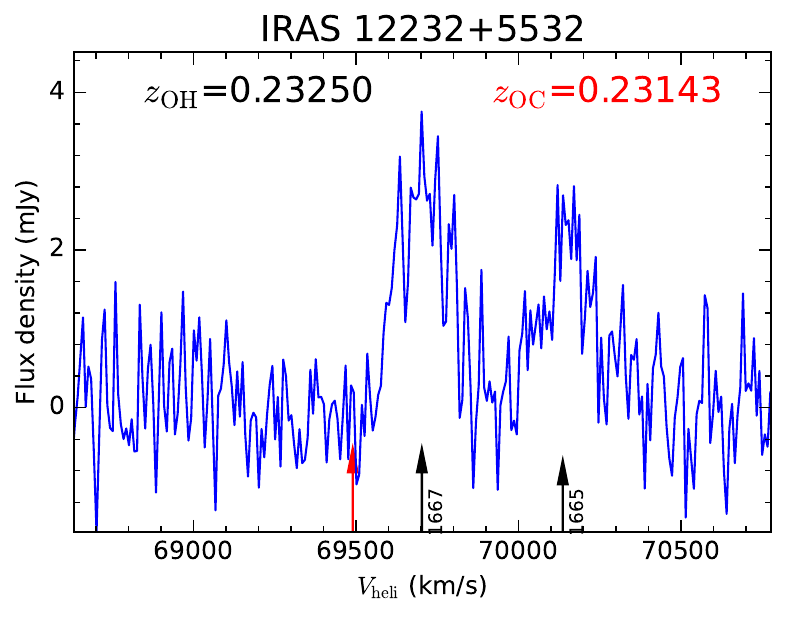}
 \includegraphics[height=0.31\textwidth, angle=0]{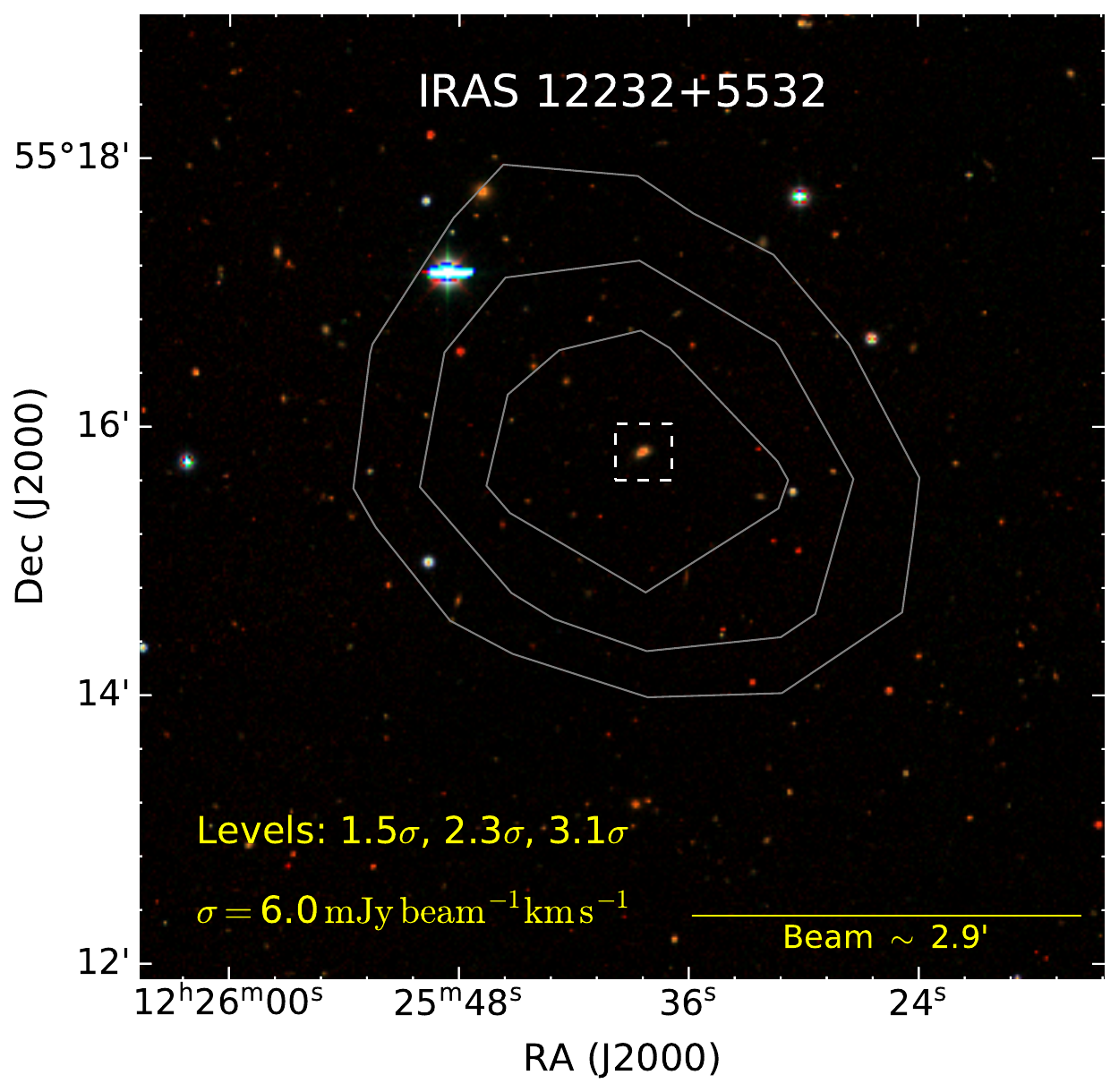}
 \includegraphics[height=0.29\textwidth, angle=0]{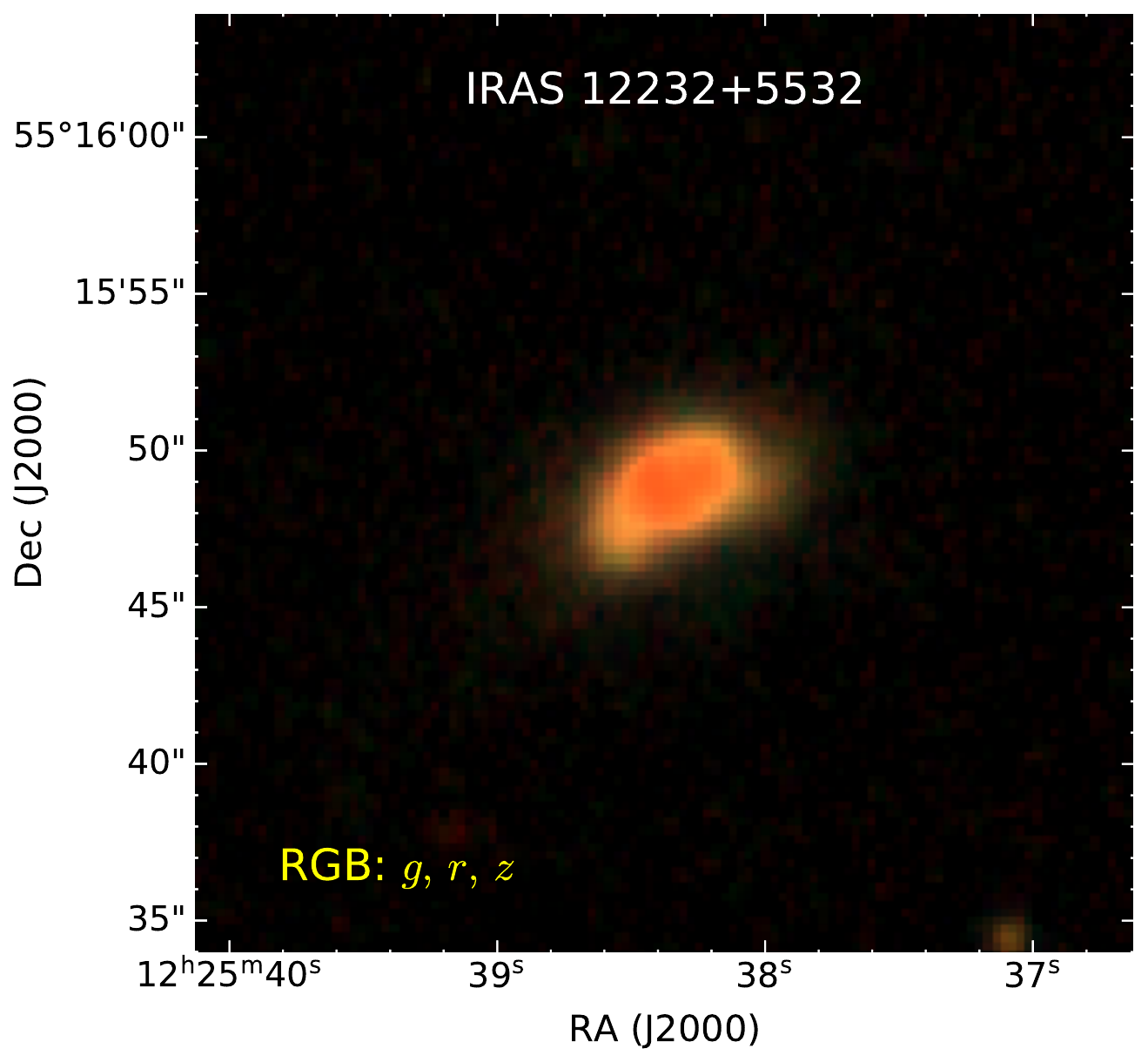}
 \includegraphics[height=0.25\textwidth, angle=0]{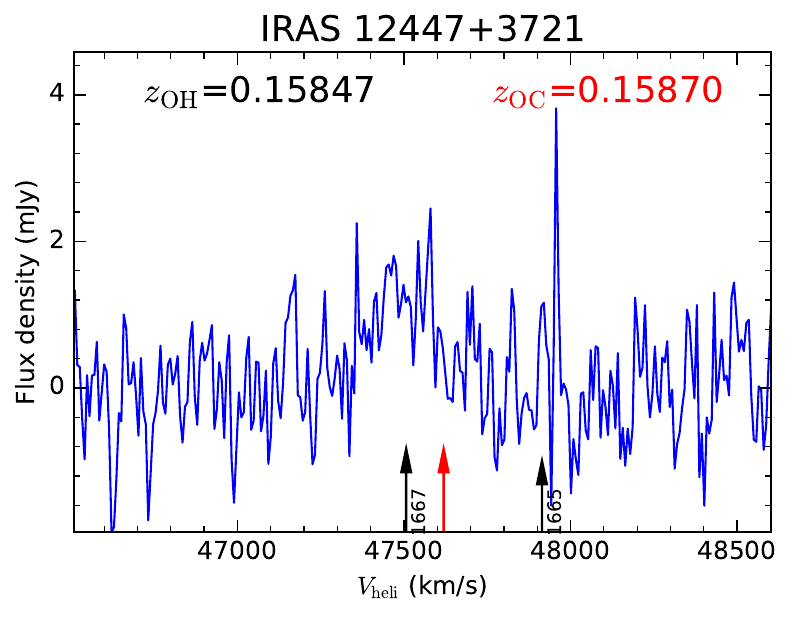}
 \includegraphics[height=0.31\textwidth, angle=0]{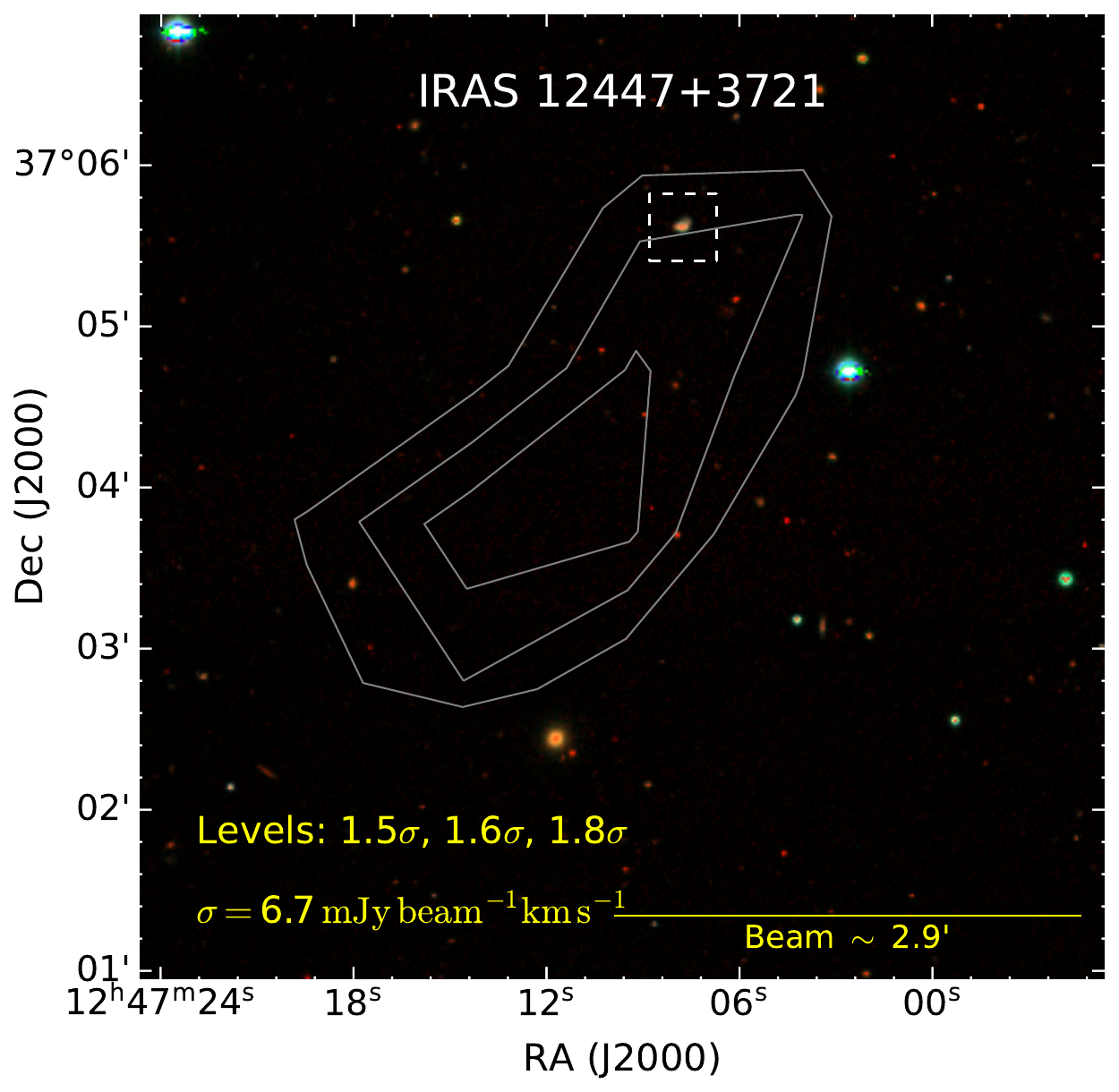}
 \includegraphics[height=0.29\textwidth, angle=0]{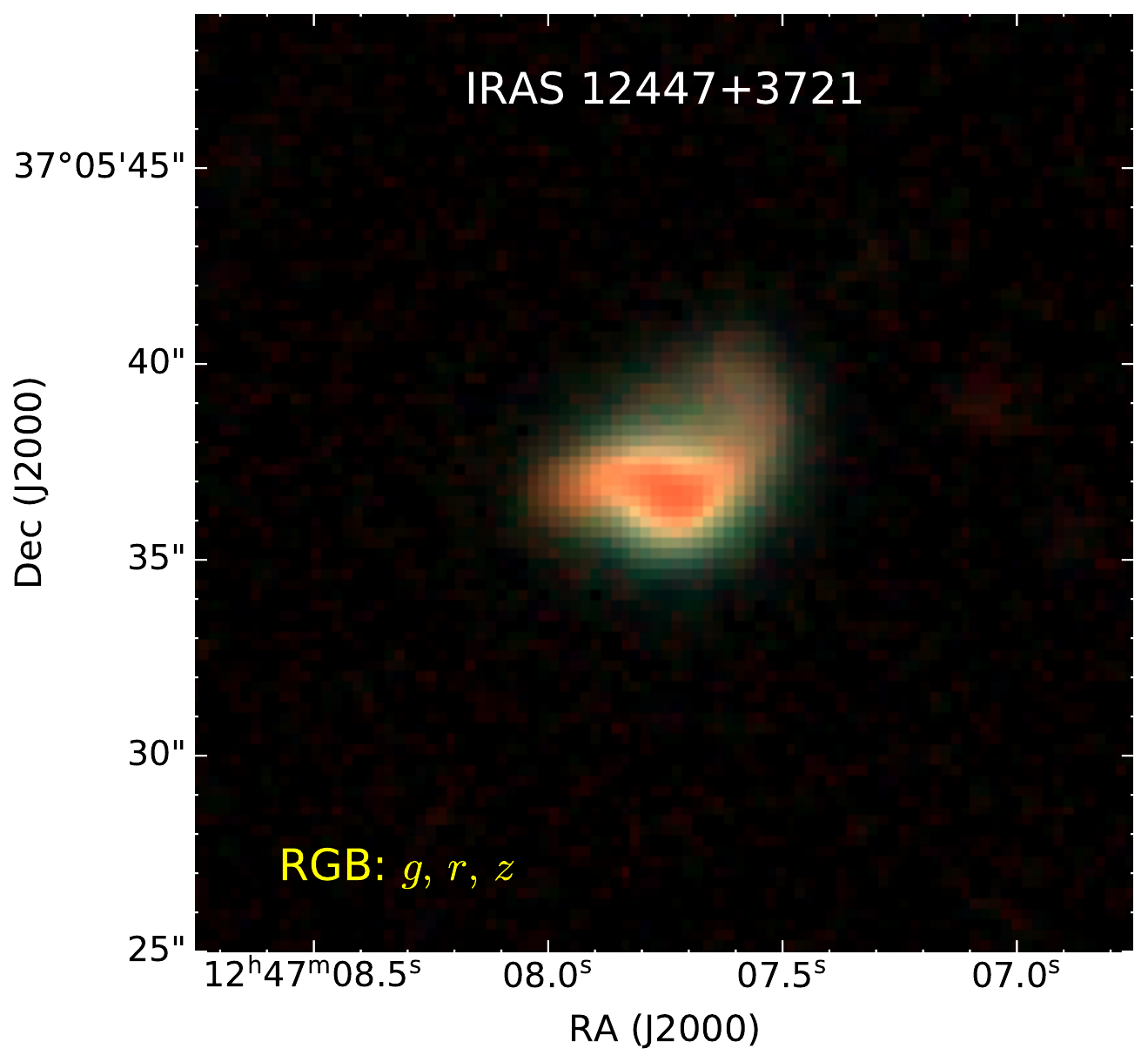}
 \includegraphics[height=0.25\textwidth, angle=0]{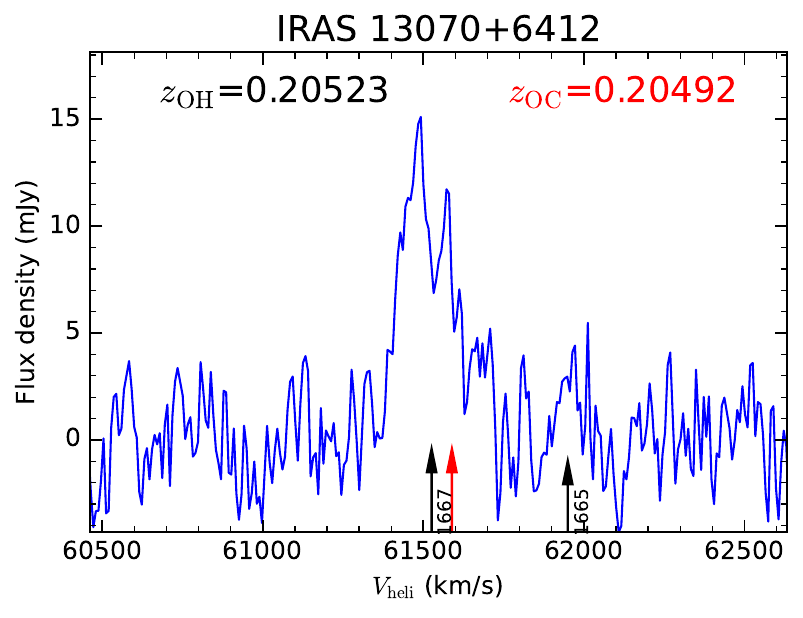}
 \includegraphics[height=0.31\textwidth, angle=0]{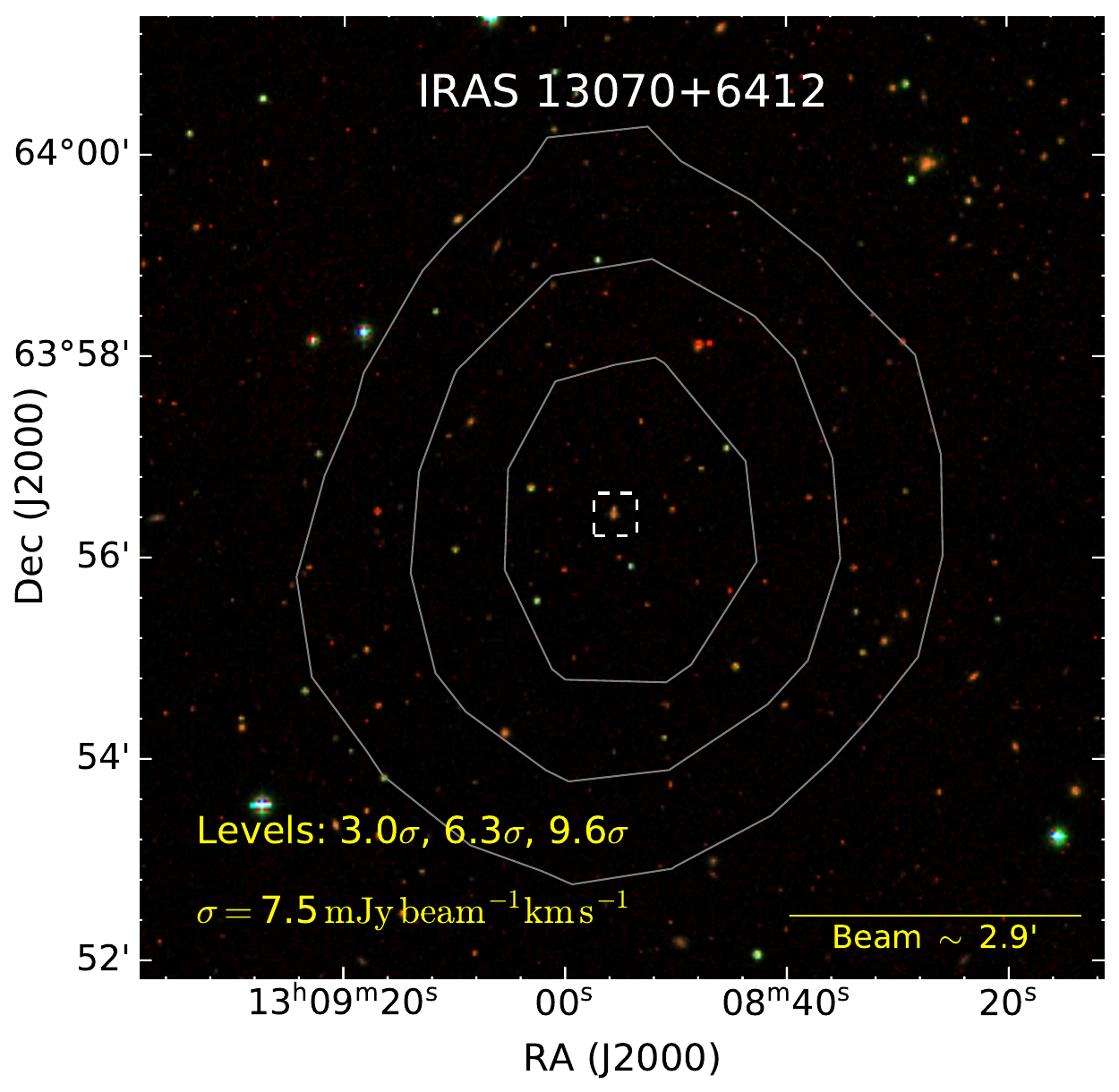}
 \includegraphics[height=0.29\textwidth, angle=0]{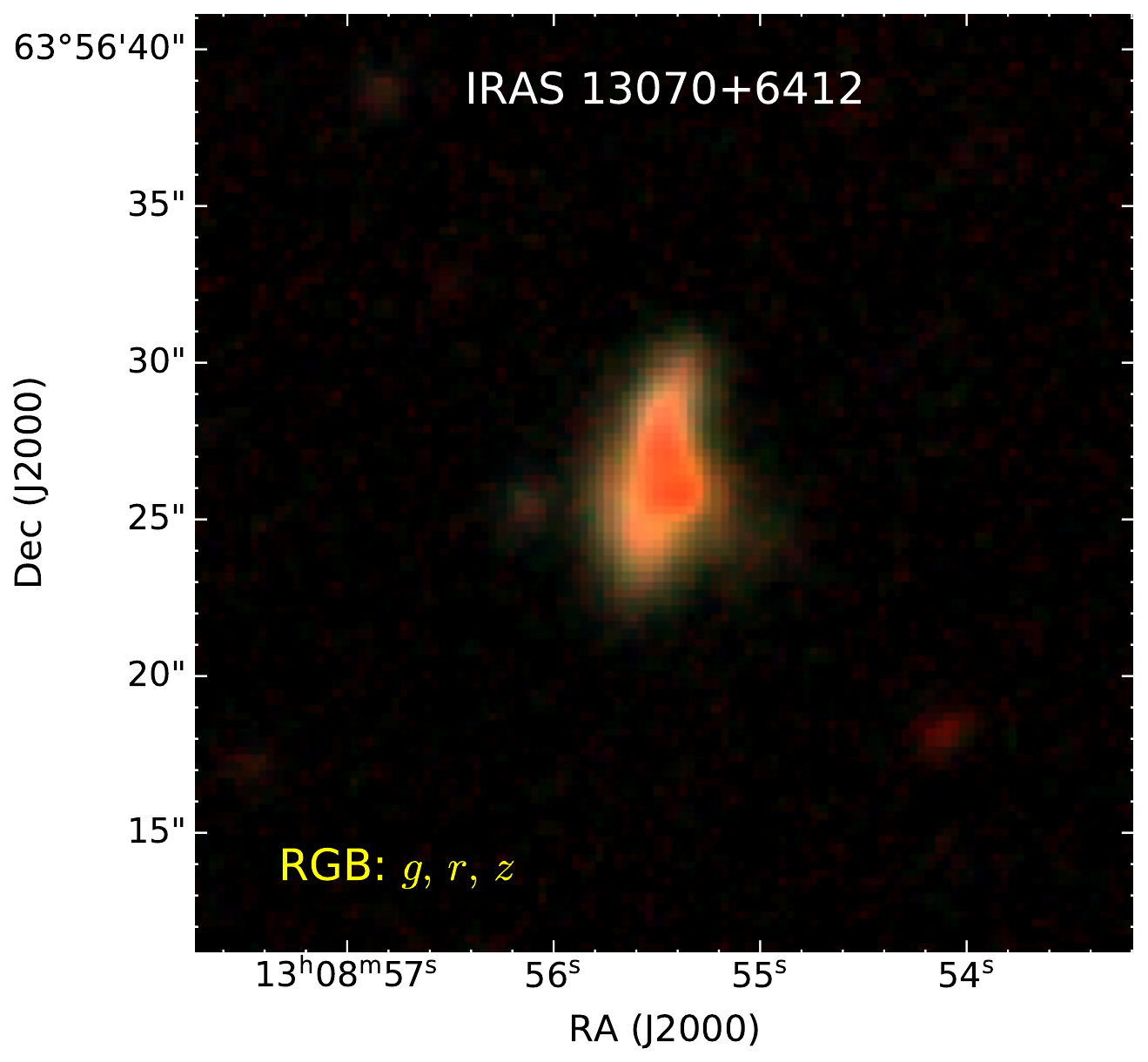}
 \includegraphics[height=0.25\textwidth, angle=0]{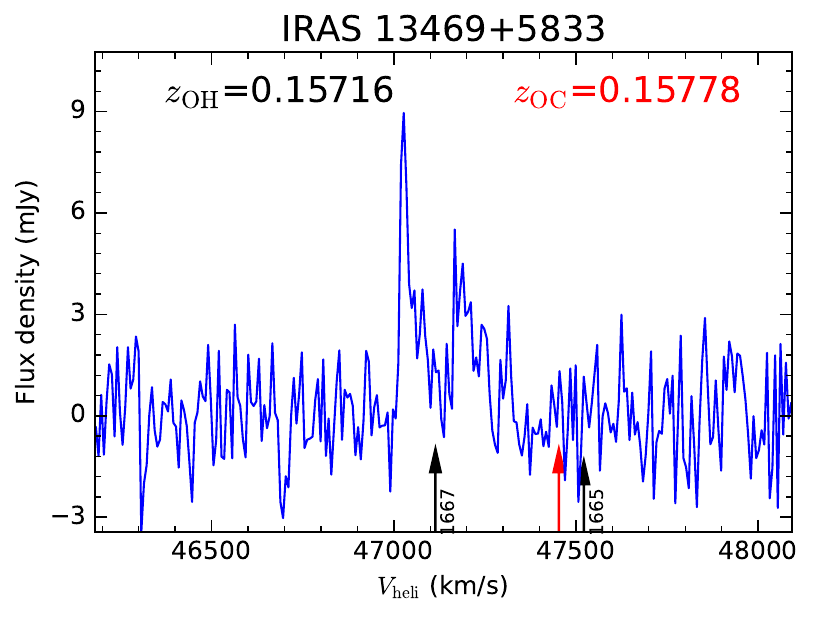}
 \includegraphics[height=0.31\textwidth, angle=0]{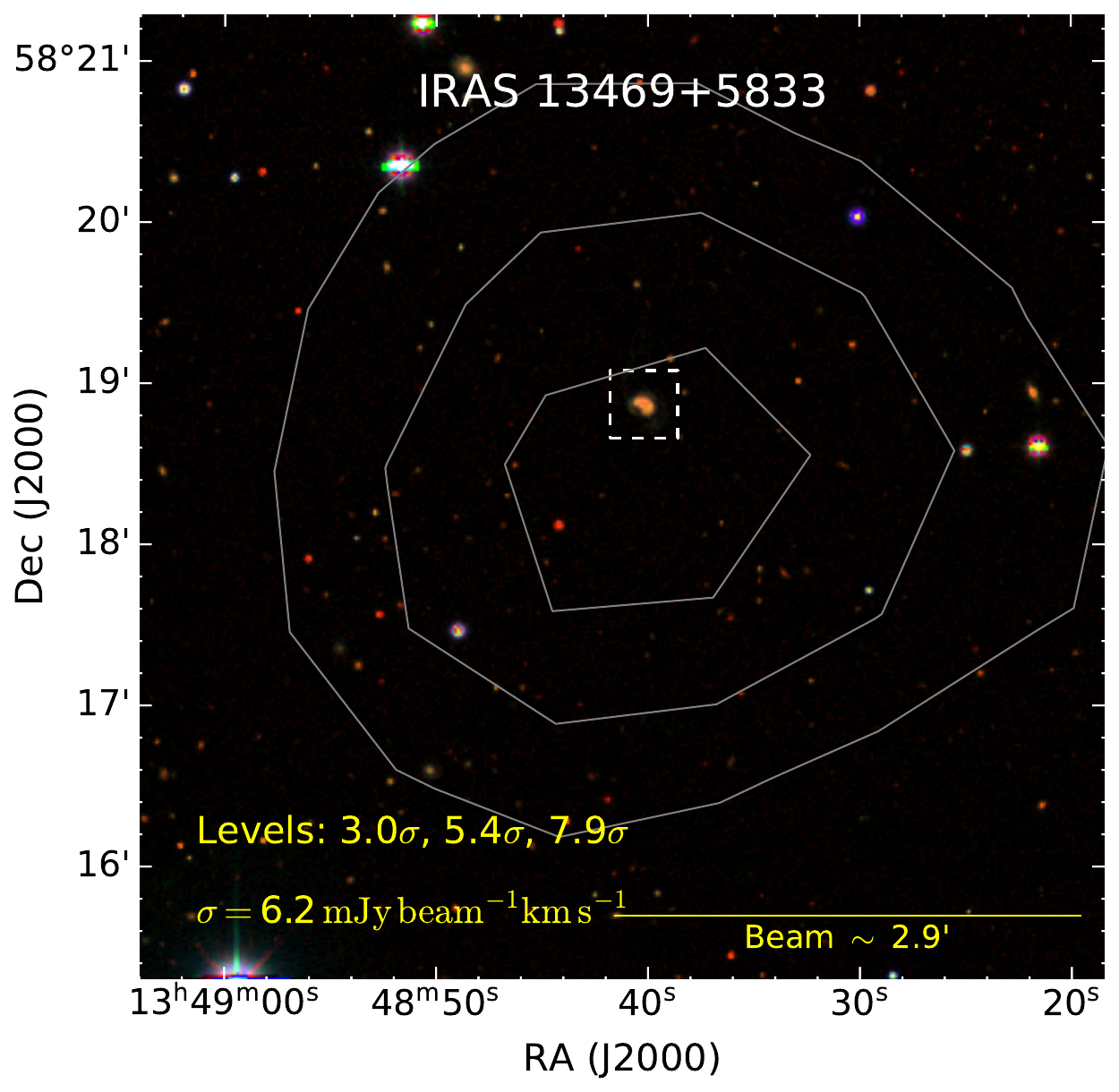}
 \includegraphics[height=0.29\textwidth, angle=0]{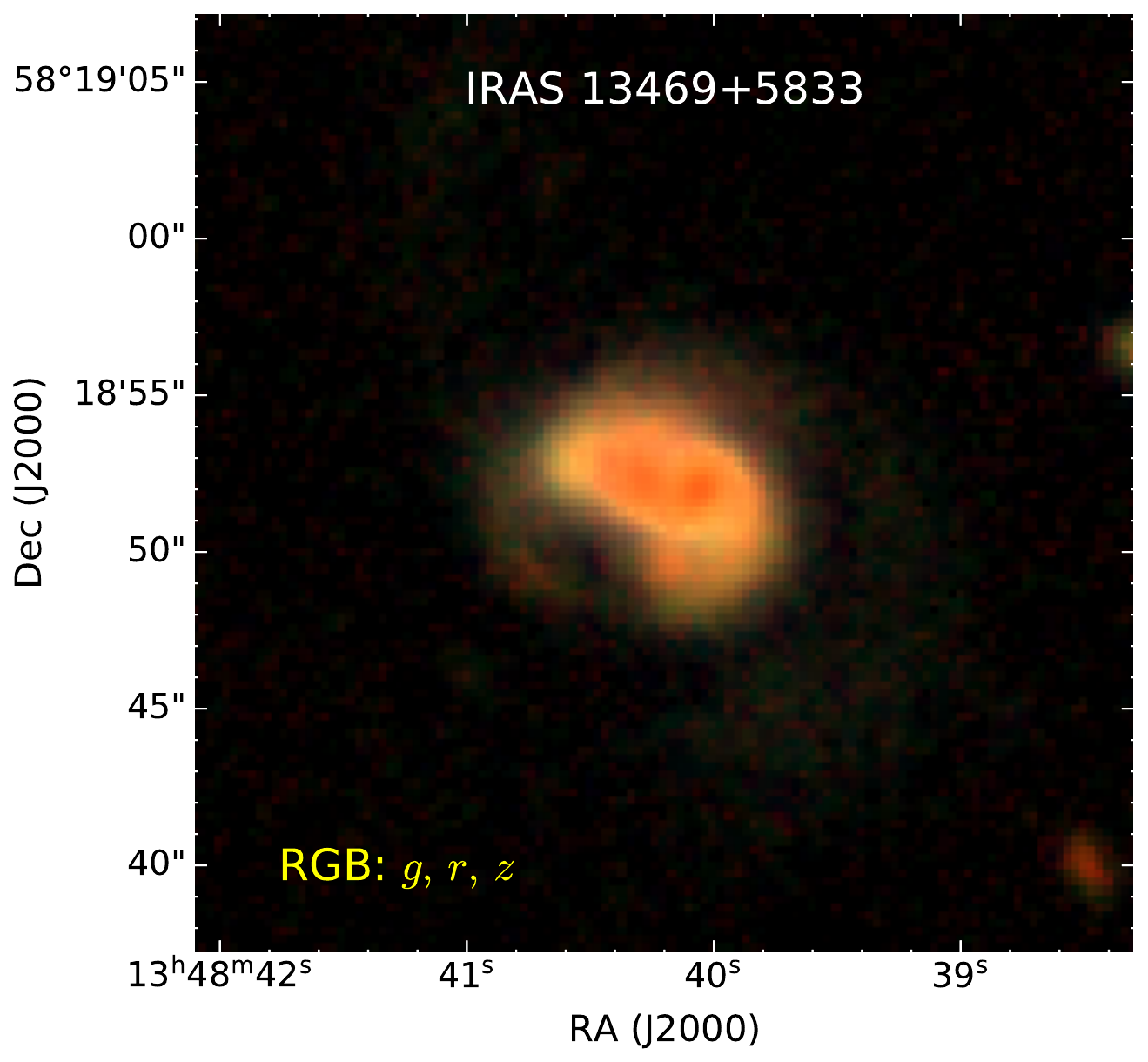}
 \caption{See caption in Figure\,\ref{Fig:IRAS00256-0208}}
 \end{figure*} 

 \begin{figure*}[htp]
 \centering
 \renewcommand{\thefigure}{\arabic{figure} (Continued)}
 \addtocounter{figure}{-1}
 \includegraphics[height=0.25\textwidth, angle=0]{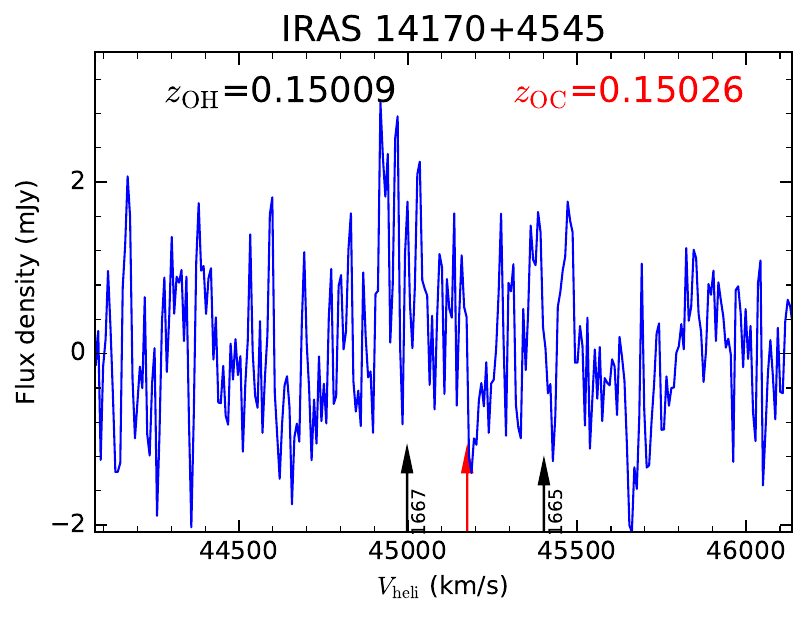}
 \includegraphics[height=0.31\textwidth, angle=0]{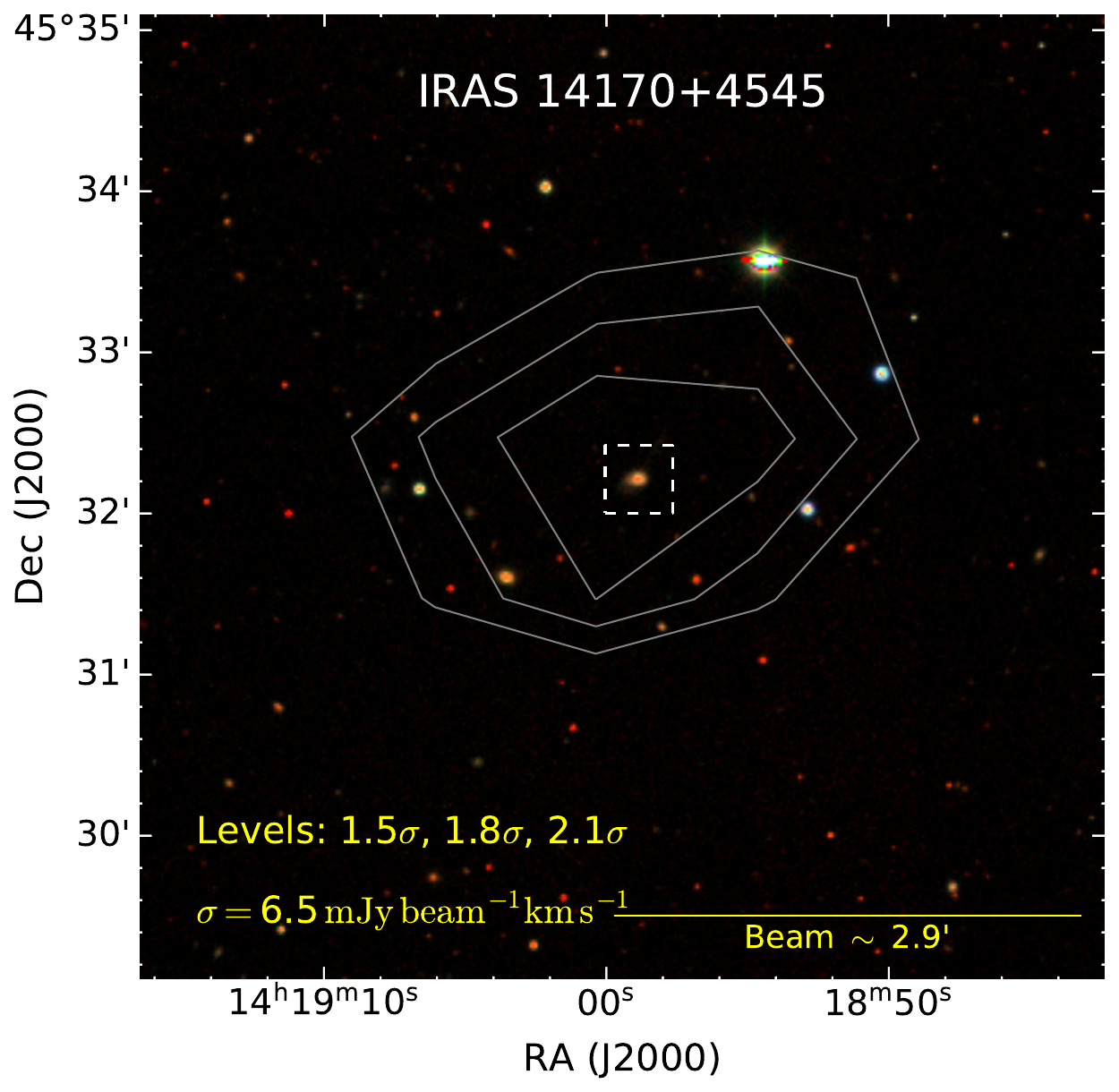}
 \includegraphics[height=0.29\textwidth, angle=0]{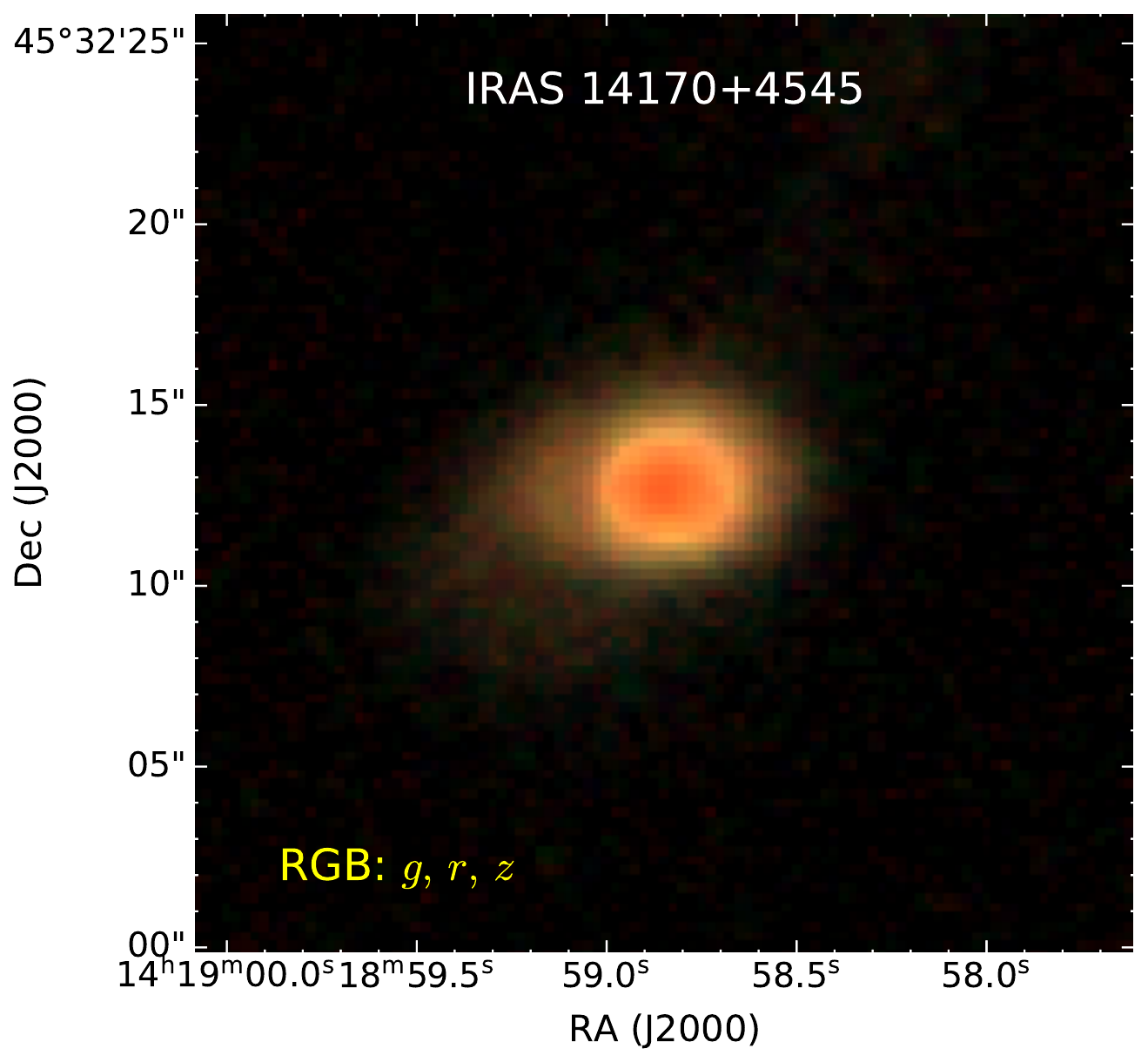}
 \includegraphics[height=0.25\textwidth, angle=0]{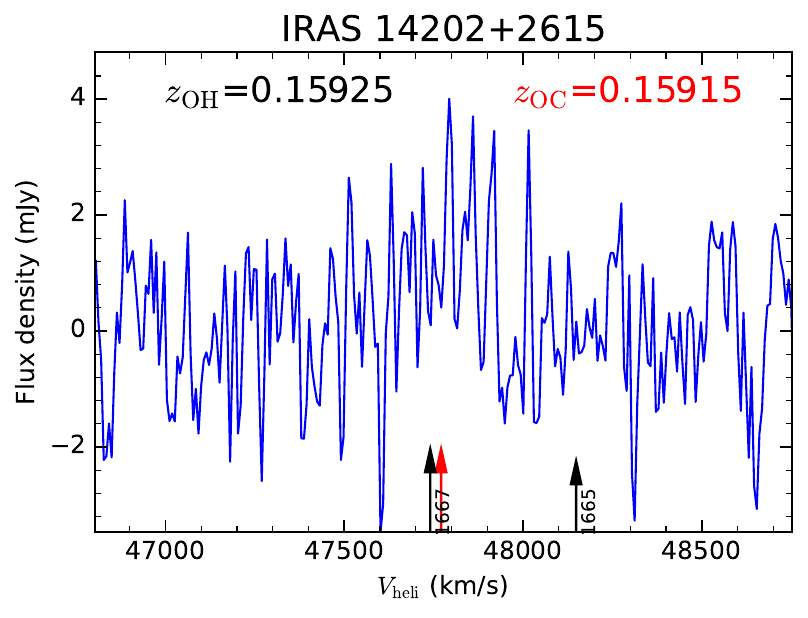}
 \includegraphics[height=0.31\textwidth, angle=0]{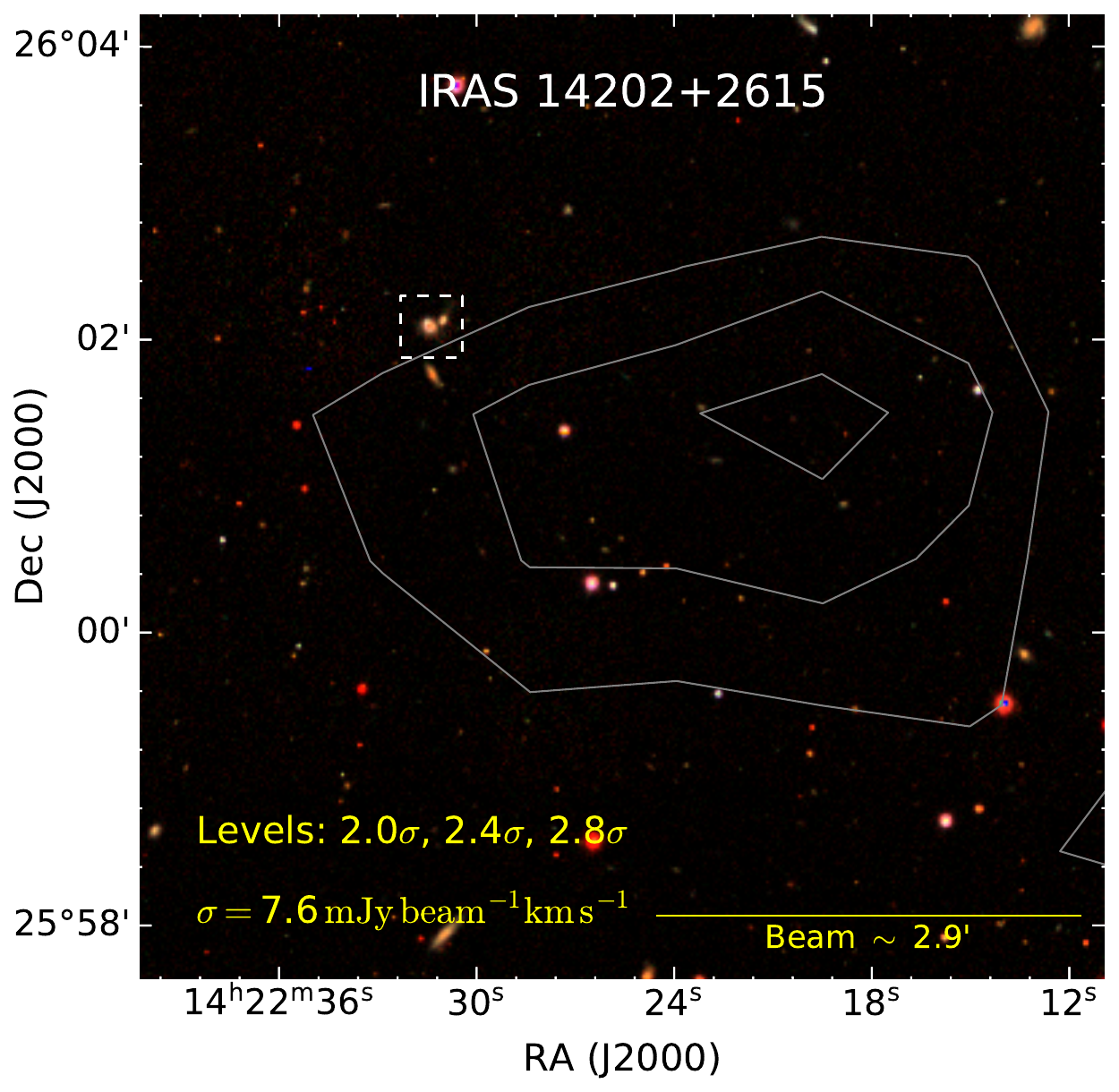}
 \includegraphics[height=0.29\textwidth, angle=0]{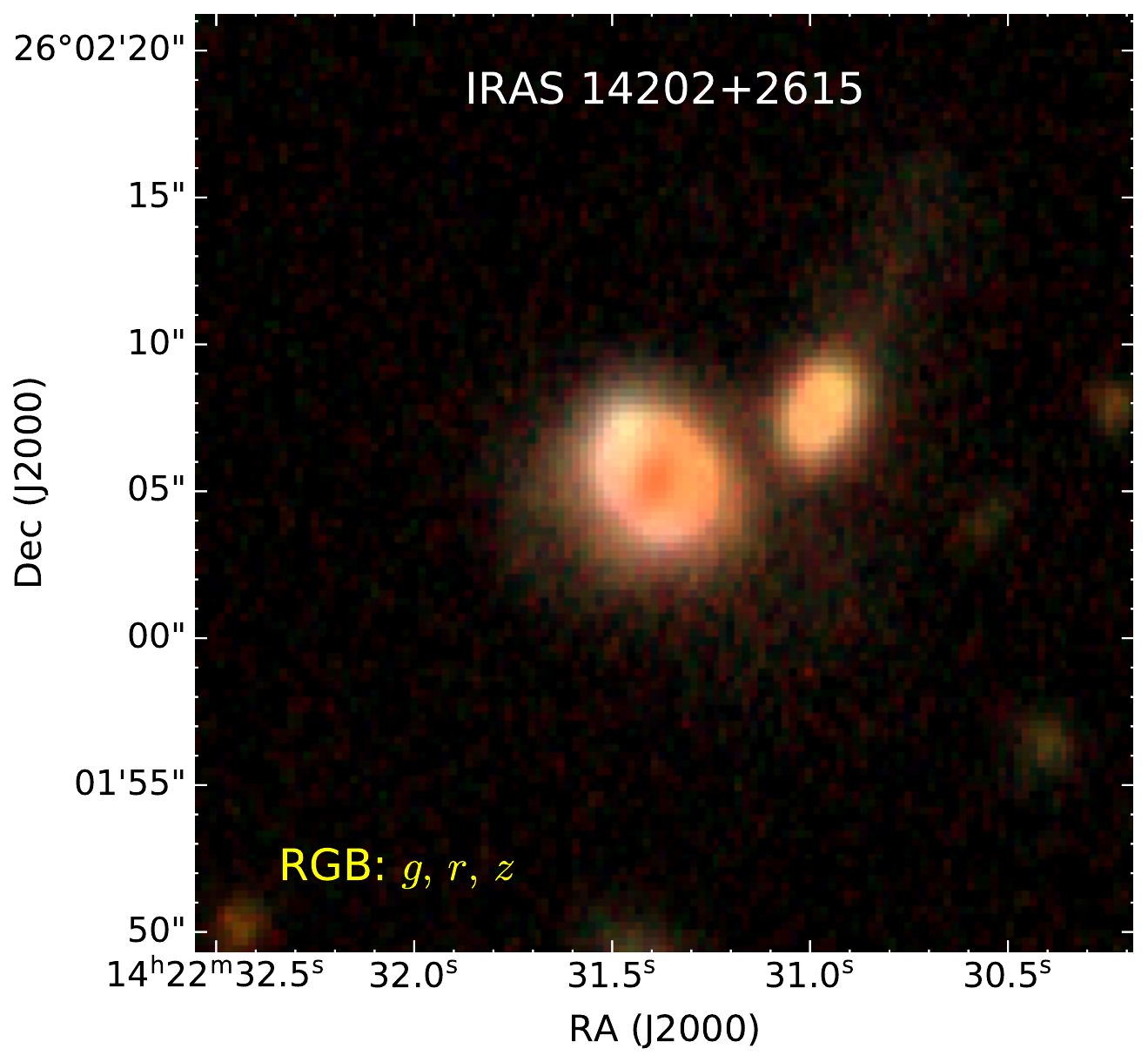}
 \includegraphics[height=0.25\textwidth, angle=0]{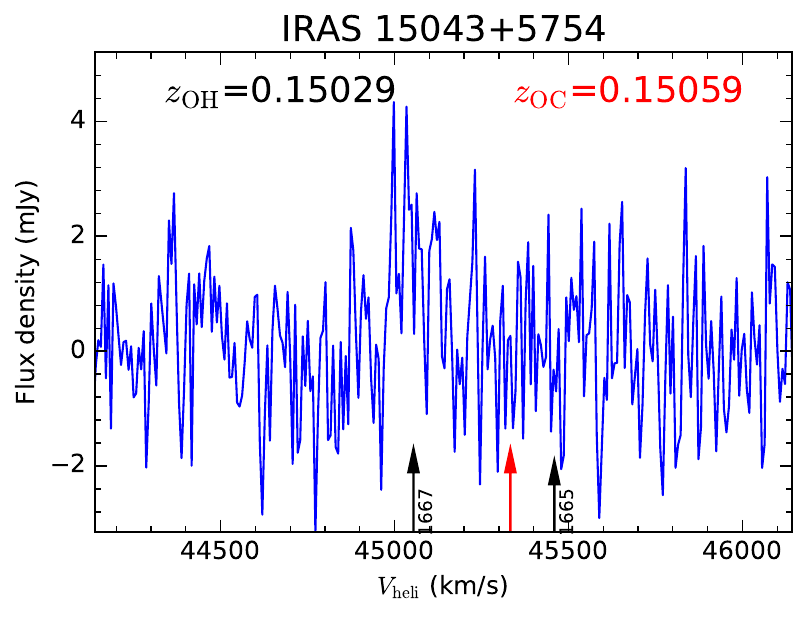}
 \includegraphics[height=0.31\textwidth, angle=0]{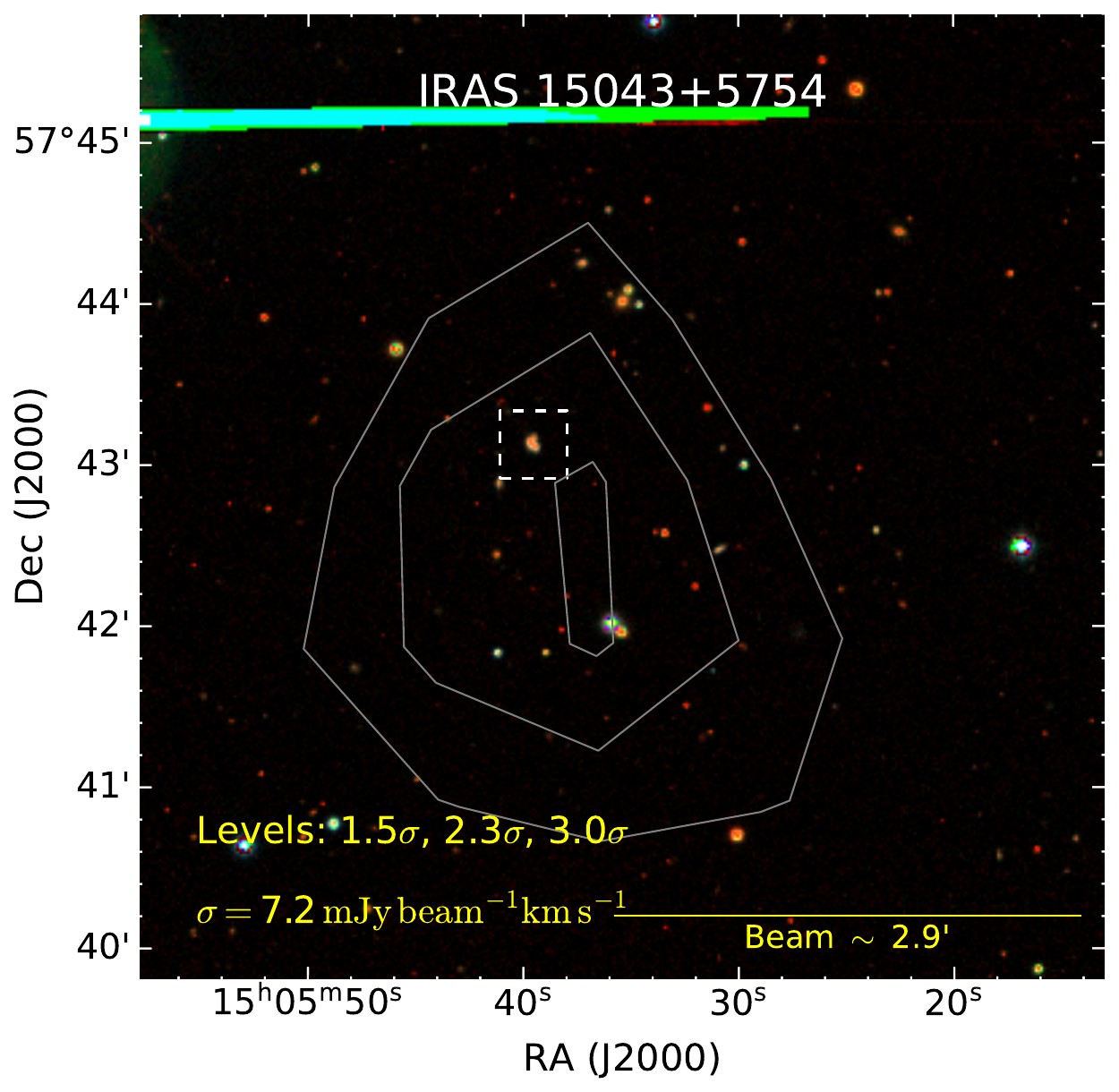}
 \includegraphics[height=0.29\textwidth, angle=0]{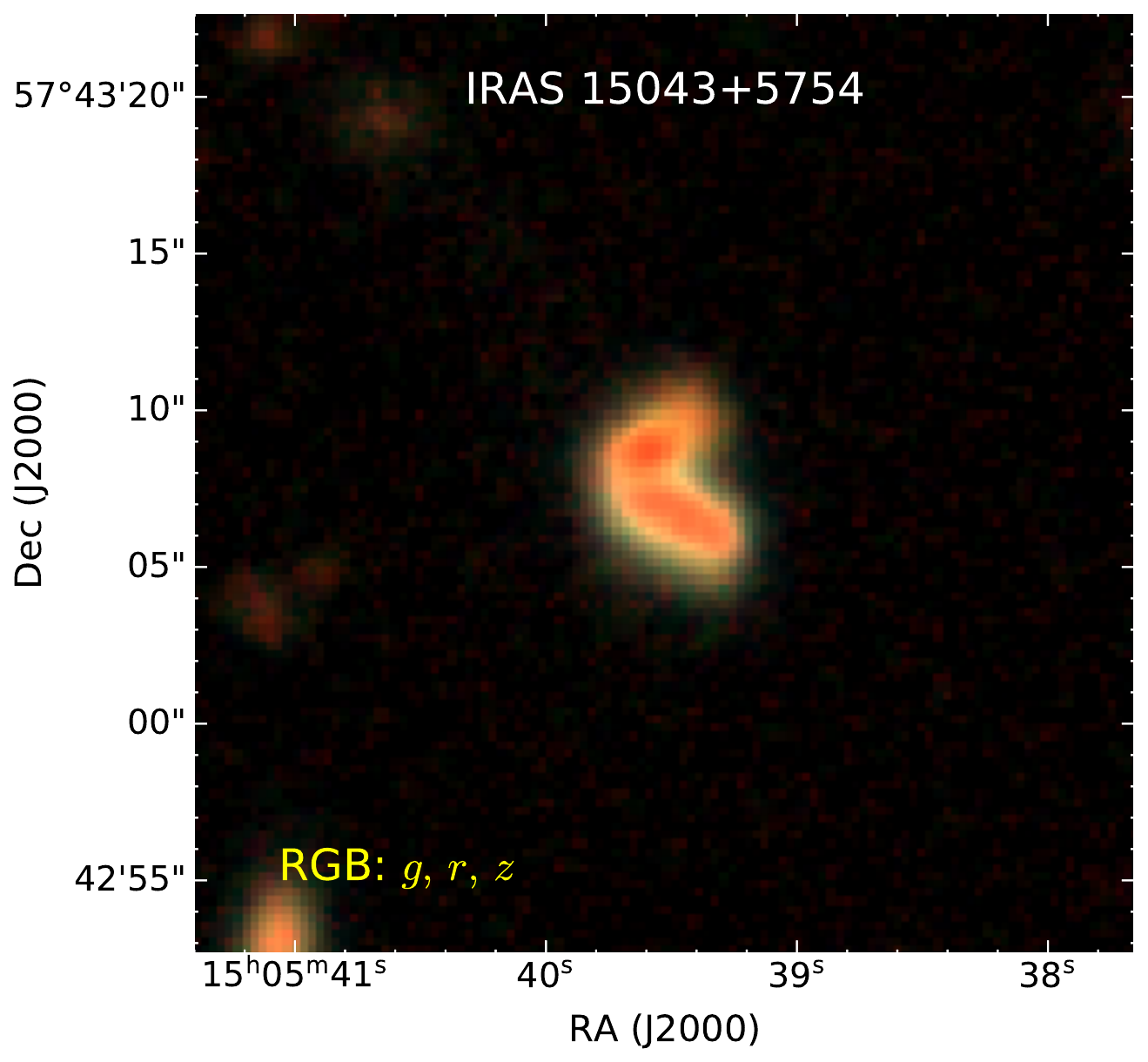}
 \includegraphics[height=0.25\textwidth, angle=0]{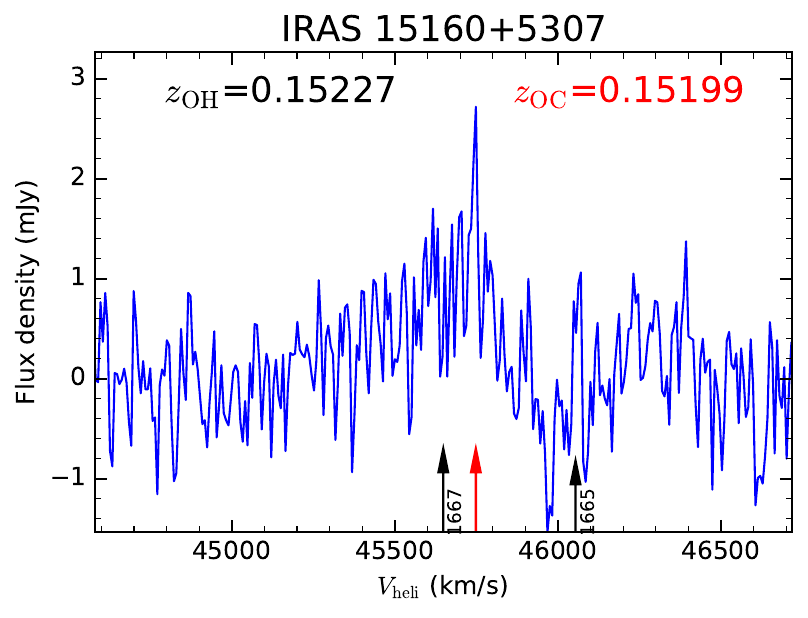}
 \includegraphics[height=0.31\textwidth, angle=0]{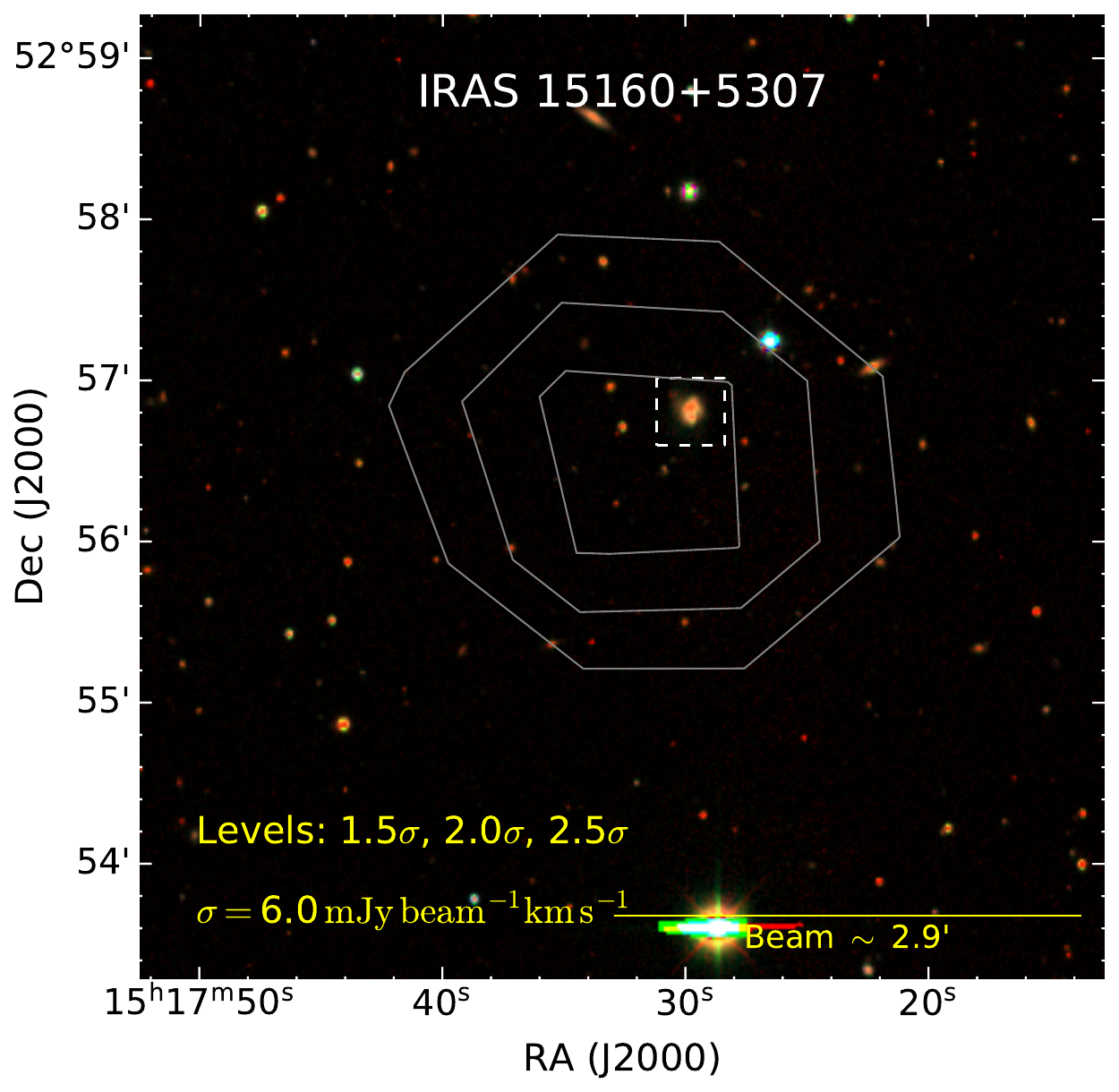}
 \includegraphics[height=0.29\textwidth, angle=0]{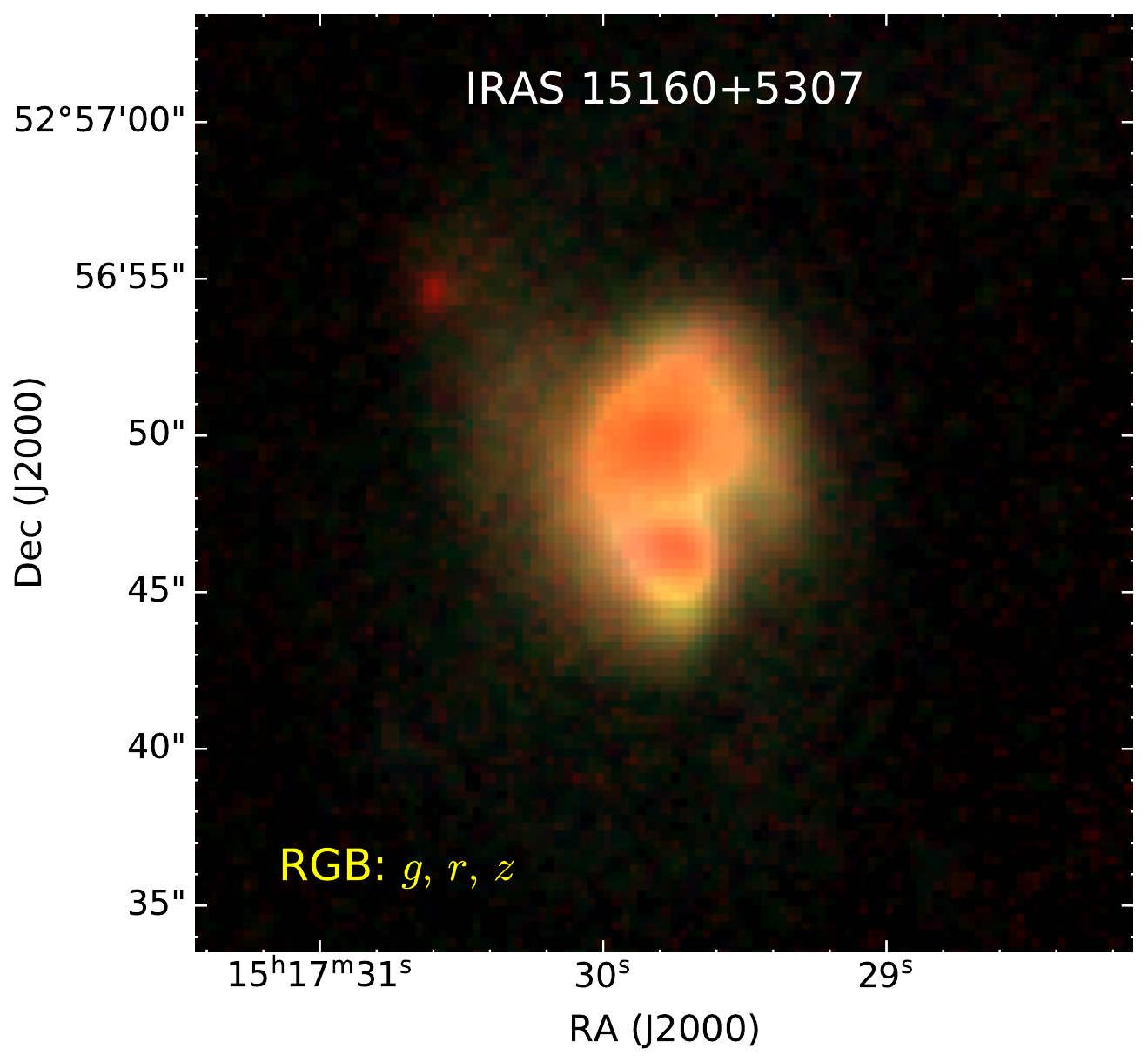}
 \caption{See caption in Figure\,\ref{Fig:IRAS00256-0208}}
 \end{figure*} 

 \begin{figure*}[htp]
 \centering
 \renewcommand{\thefigure}{\arabic{figure} (Continued)}
 \addtocounter{figure}{-1}
 \includegraphics[height=0.25\textwidth, angle=0]{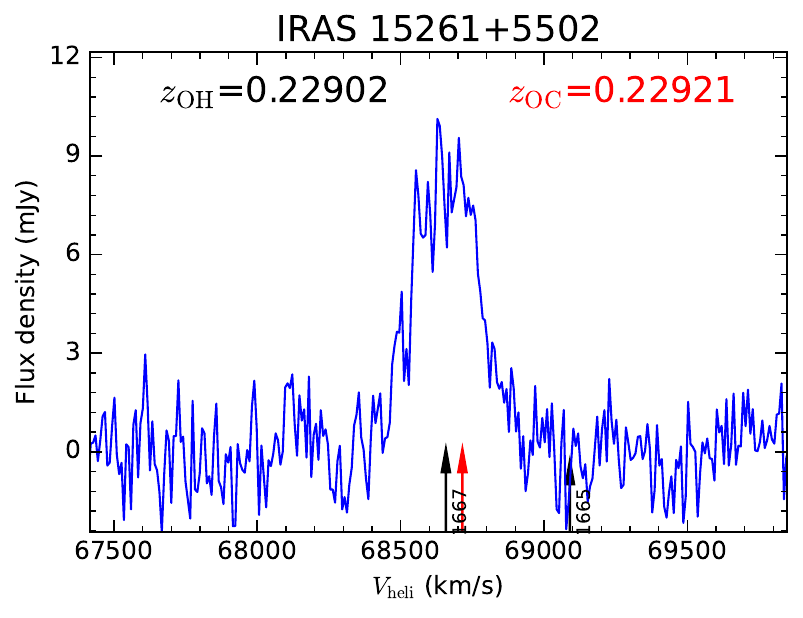}
 \includegraphics[height=0.31\textwidth, angle=0]{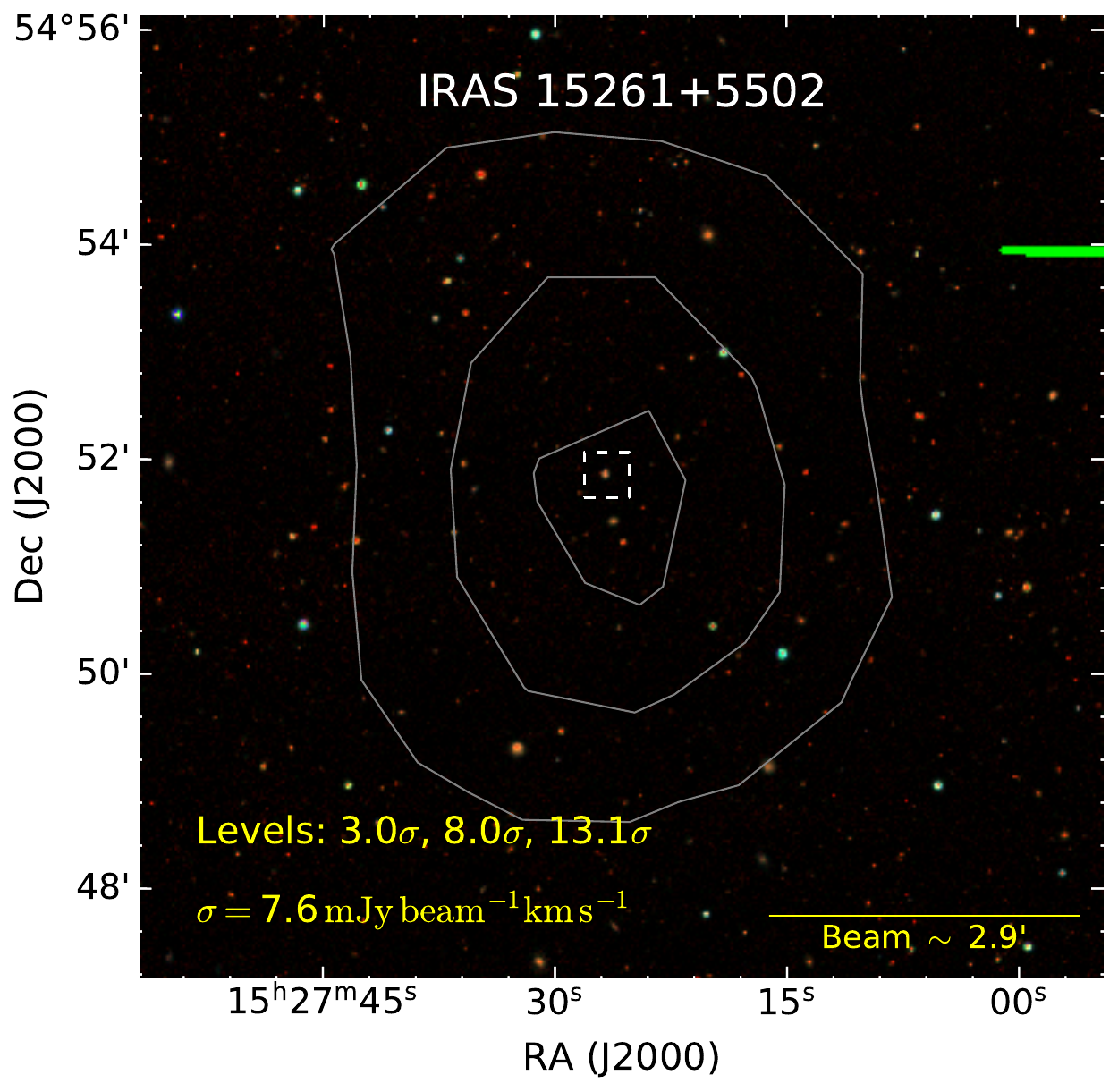}
 \includegraphics[height=0.29\textwidth, angle=0]{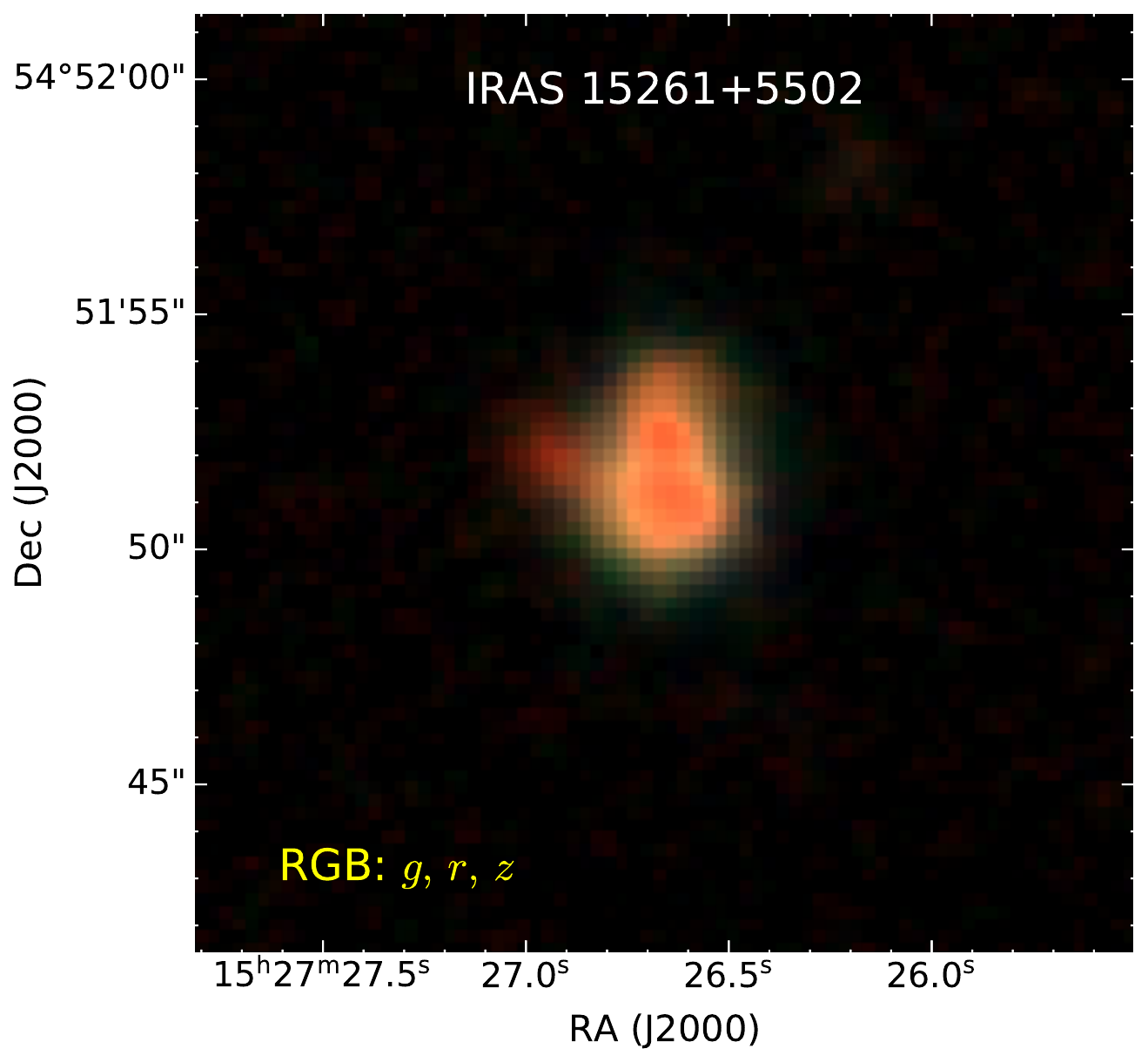}
 \includegraphics[height=0.25\textwidth, angle=0]{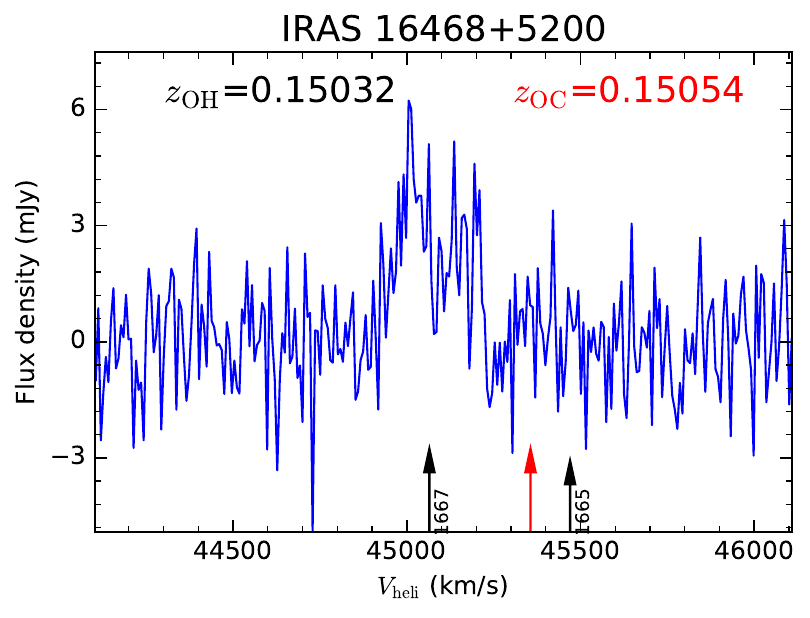}
 \includegraphics[height=0.31\textwidth, angle=0]{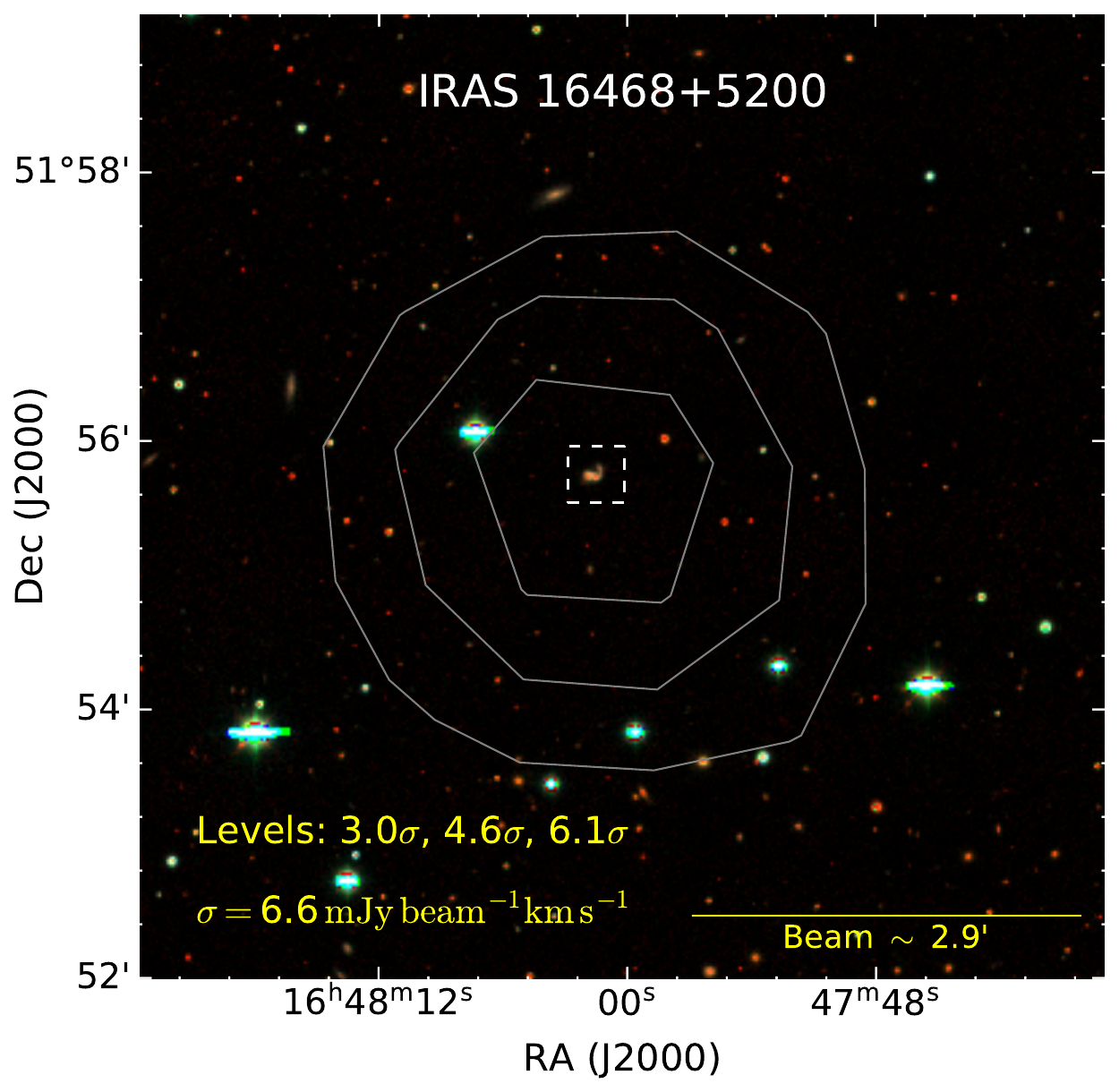}
 \includegraphics[height=0.29\textwidth, angle=0]{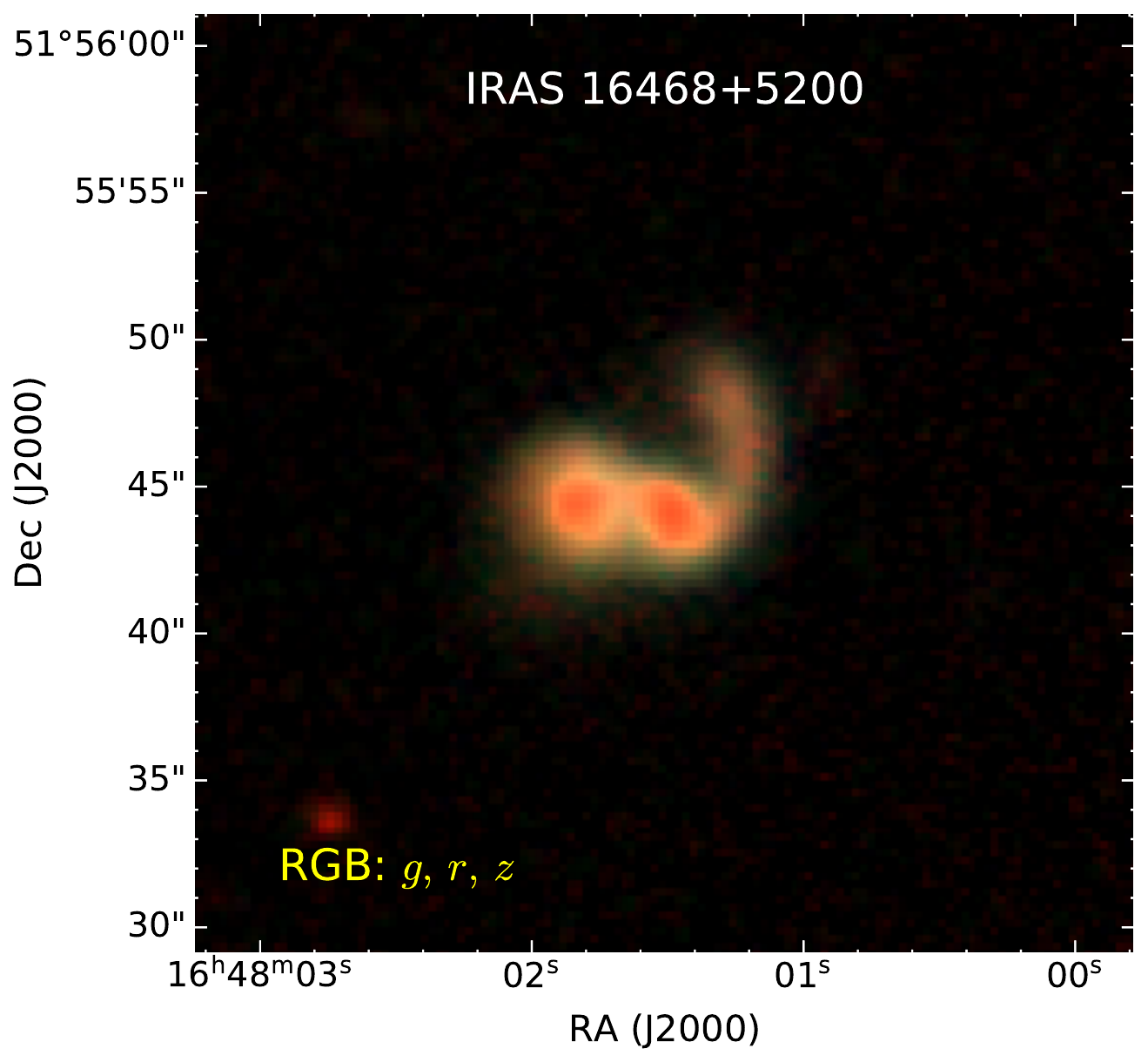}
 \includegraphics[height=0.25\textwidth, angle=0]{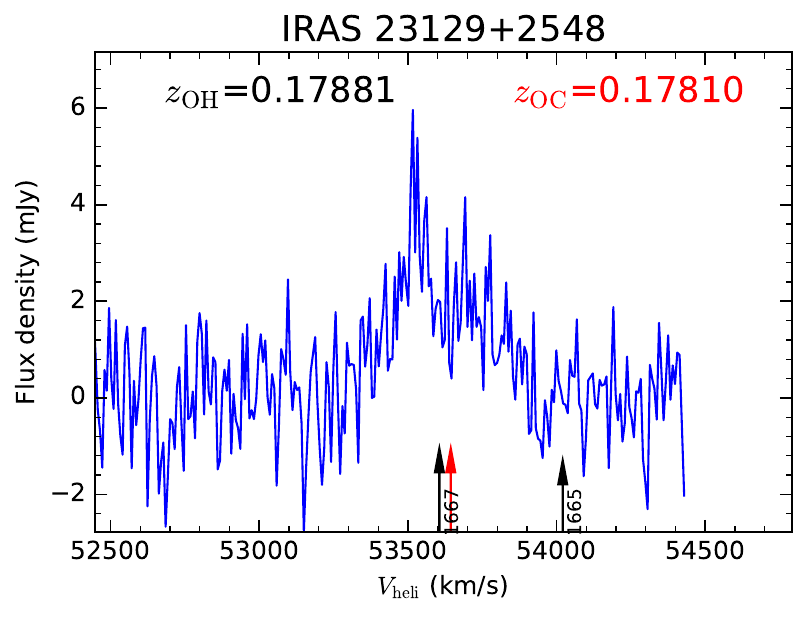}
 \includegraphics[height=0.31\textwidth, angle=0]{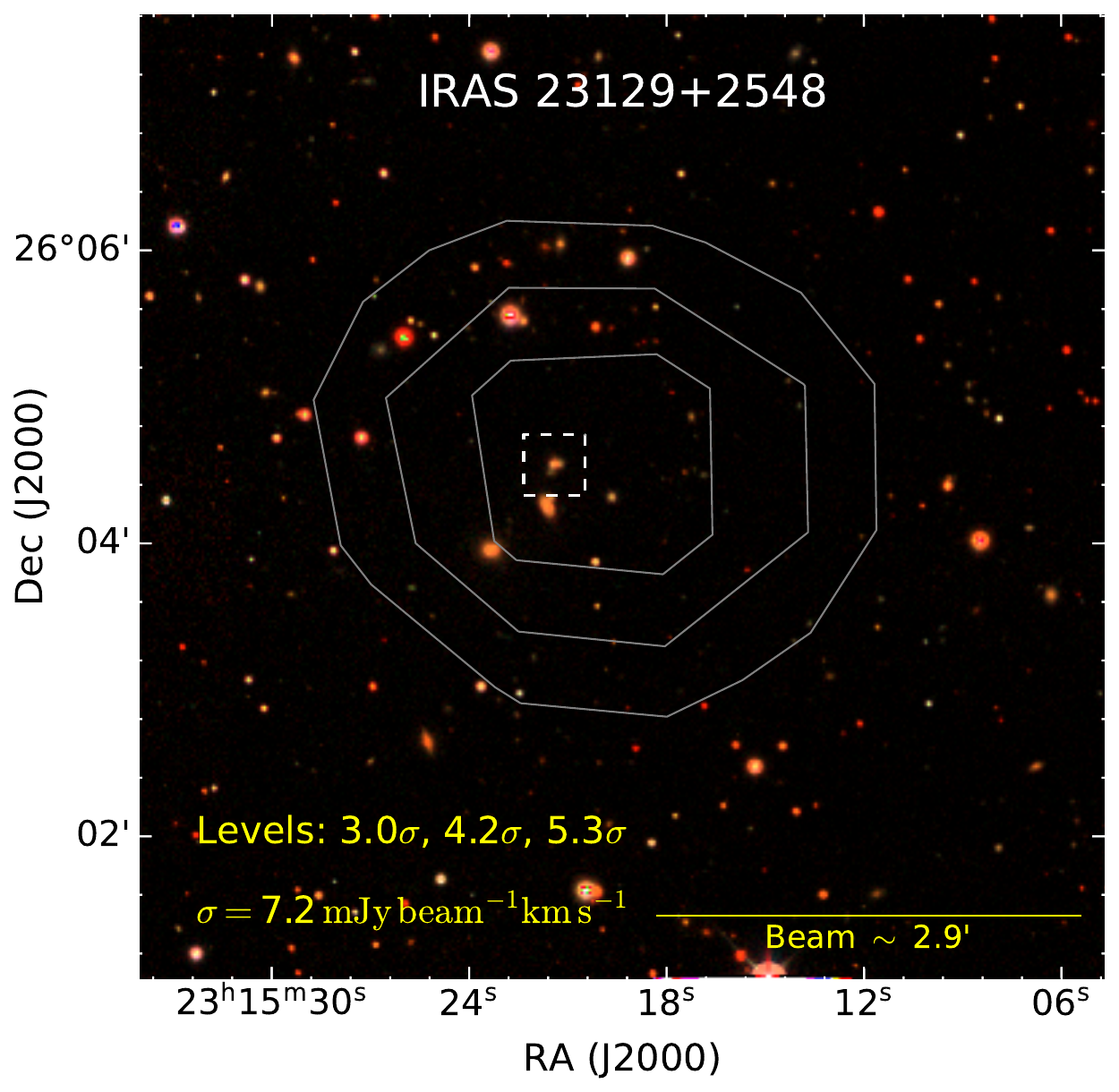}
 \includegraphics[height=0.29\textwidth, angle=0]{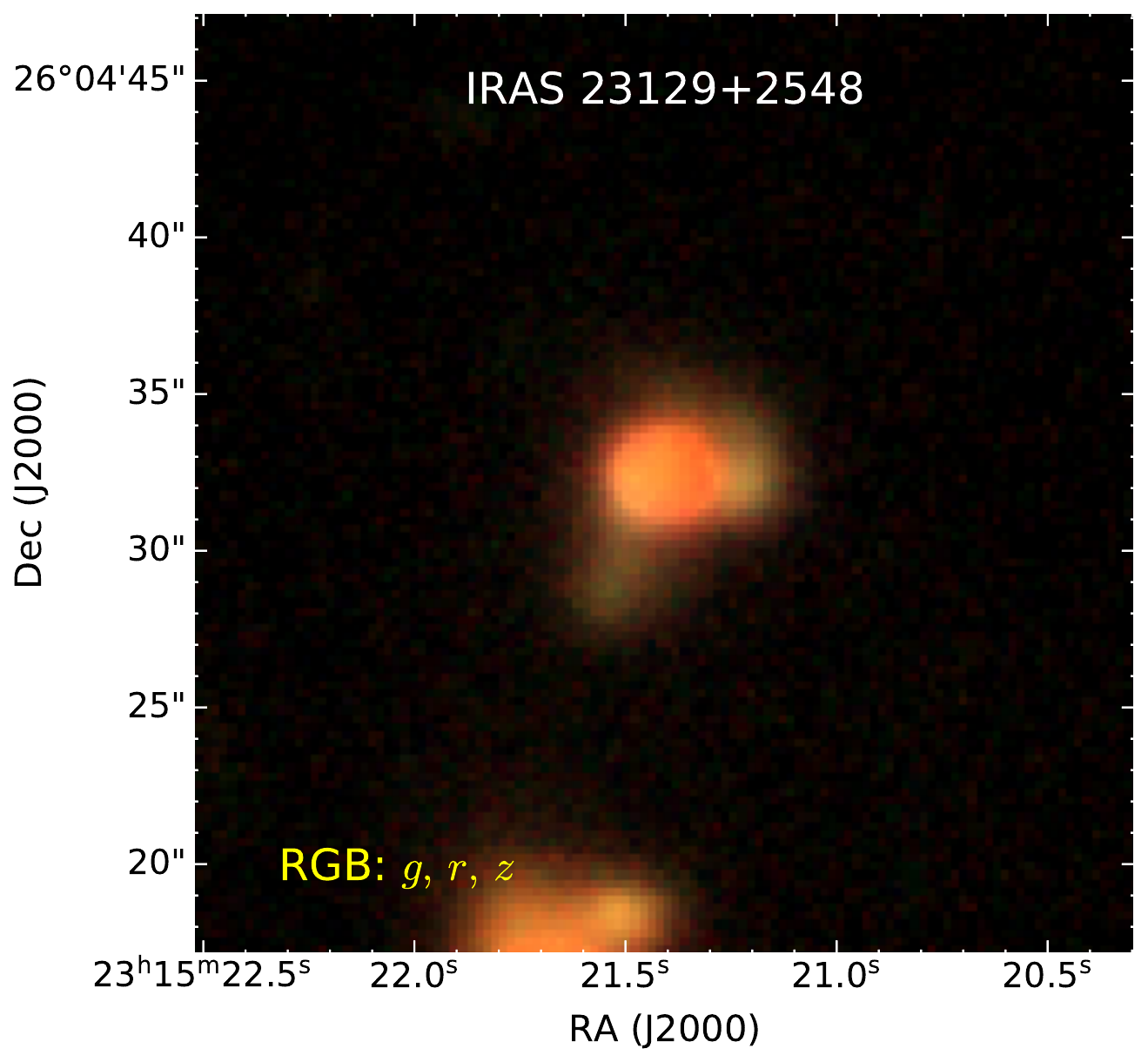}
 \caption{See caption in Figure\,\ref{Fig:IRAS00256-0208}}
 \end{figure*}

 \clearpage

\end{document}